\documentclass[traditabstract]{aa}  

\usepackage{psfig}

\begin{document}

\title{Unveiling the nature of {\it INTEGRAL} objects through optical 
spectroscopy\thanks{Based on observations collected at the following 
observatories: Cerro Tololo Interamerican Observatory (Chile);
Observatorio del Roque de los Muchachos of the Instituto de 
Astrof\'{\i}sica de Canarias (Canary Islands, Spain); Astronomical 
Observatory of Bologna in Loiano (Italy); Astronomical Observatory of 
Asiago (Italy); Observatorio Astron\'omico Nacional (San Pedro M\'artir, 
Mexico); South African Astronomical Observatory (Sutherland, South Africa); 
and Australian Astronomical Observatory (Siding Spring, Australia).}}

\subtitle{X. A new multi-year, multi-observatory campaign}

\author{N. Masetti\inst{1}, 
P. Parisi\inst{2}, 
E. Palazzi\inst{1},
E. Jim\'enez-Bail\'on\inst{3}, 
V. Chavushyan\inst{4}, 
V. McBride\inst{5,6},
A.F. Rojas\inst{7}, 
L. Steward\inst{5,6},
L. Bassani\inst{1},
A. Bazzano\inst{2}, 
A.J. Bird\inst{8},
P.A. Charles\inst{8},
G. Galaz\inst{7},
R. Landi\inst{1},
A. Malizia\inst{1},
E. Mason\inst{9},
D. Minniti\inst{7,10},
L. Morelli\inst{11,12},
F. Schiavone\inst{1},
J.B. Stephen\inst{1} and
P. Ubertini\inst{2}
}

\institute{
INAF -- Istituto di Astrofisica Spaziale e Fisica Cosmica di 
Bologna, Via Gobetti 101, I-40129 Bologna, Italy
\and
INAF -- Istituto di Astrofisica e Planetologia Spaziali, Via Fosso del 
Cavaliere 100, I-00133 Rome, Italy
\and
Instituto de Astronom\'{\i}a, Universidad Nacional Aut\'onoma de M\'exico,
Apartado Postal 70-264, 04510 M\'exico D.F., M\'exico
\and
Instituto Nacional de Astrof\'{i}sica, \'Optica y Electr\'onica,
Apartado Postal 51-216, 72000 Puebla, M\'exico
\and
South African Astronomical Observatory, P.O. Box 9, Observatory 7935, 
South Africa
\and
Department of Astronomy, University of Cape Town, Private Bag X3, 
Rondebosch 7701, South Africa
\and
Departamento de Astronom\'{i}a y Astrof\'{i}sica, Pontificia Universidad 
Cat\'olica de Chile, Casilla 306, Santiago 22, Chile
\and
School of Physics \& Astronomy, University of Southampton, Highfield, 
Southampton, SO17 1BJ, United Kingdom  
\and
Space Telescope Science Institute, 3700 San Martin Drive, Baltimore, MD 
21218, USA
\and
Specola Vaticana, V-00120 Citt\`a del Vaticano
\and
Dipartimento di Fisica ed Astronomia ``G. Galilei", Universit\`a di Padova,
vicolo dell'Osservatorio 3, I-35122 Padova, Italy
\and
INAF-Osservatorio Astronomico di Padova, Vicolo dell'Osservatorio 5, 
I-35122 Padua, Italy
}

\offprints{N. Masetti (\texttt{masetti@iasfbo.inaf.it)}}
\date{Received 6 June 2013; accepted 20 June 2013}

\abstract{Within the framework of our program (running since 2004) of 
identification of hard X--ray {\it INTEGRAL} sources through optical 
spectroscopy, we present the results concerning the nature of 33 
high-energy objects. The data were acquired with the use of six telescopes 
of different sizes and from one on-line archive. The results indicate that 
the majority of these objects (23 out of 33) are active galactic nuclei 
(AGNs), whereas 10 are sources in the local Universe with eight of which 
in the Galaxy and two in the Small Magellanic Cloud (SMC). Among the 
identified AGNs, 13 are of Type 1 (i.e., with broad emission lines), eight 
are of Type 2 (with narrow emissions only), and two are X--ray bright, 
optically normal galaxies with no apparent nuclear activity in the 
optical. Six of these AGNs lie at high redshift ($z >$ 0.5). Concerning 
local objects, we found that five of them are Galactic cataclysmic 
variables, three are high-mass X--ray binaries (two of which lying in the 
SMC), one is a low-mass X--ray binary, and one is classified as a flare 
star that is likely of RS CVn type. The main optical properties and 
inferred physical characteristics of these sources are presented and 
discussed.}


\titlerunning{The nature of 33 more {\it INTEGRAL} sources}
\authorrunning{N. Masetti et al.}

\maketitle

\section{Introduction}

\begin{figure*}
\hspace{-.1cm}
\centering{\mbox{\psfig{file=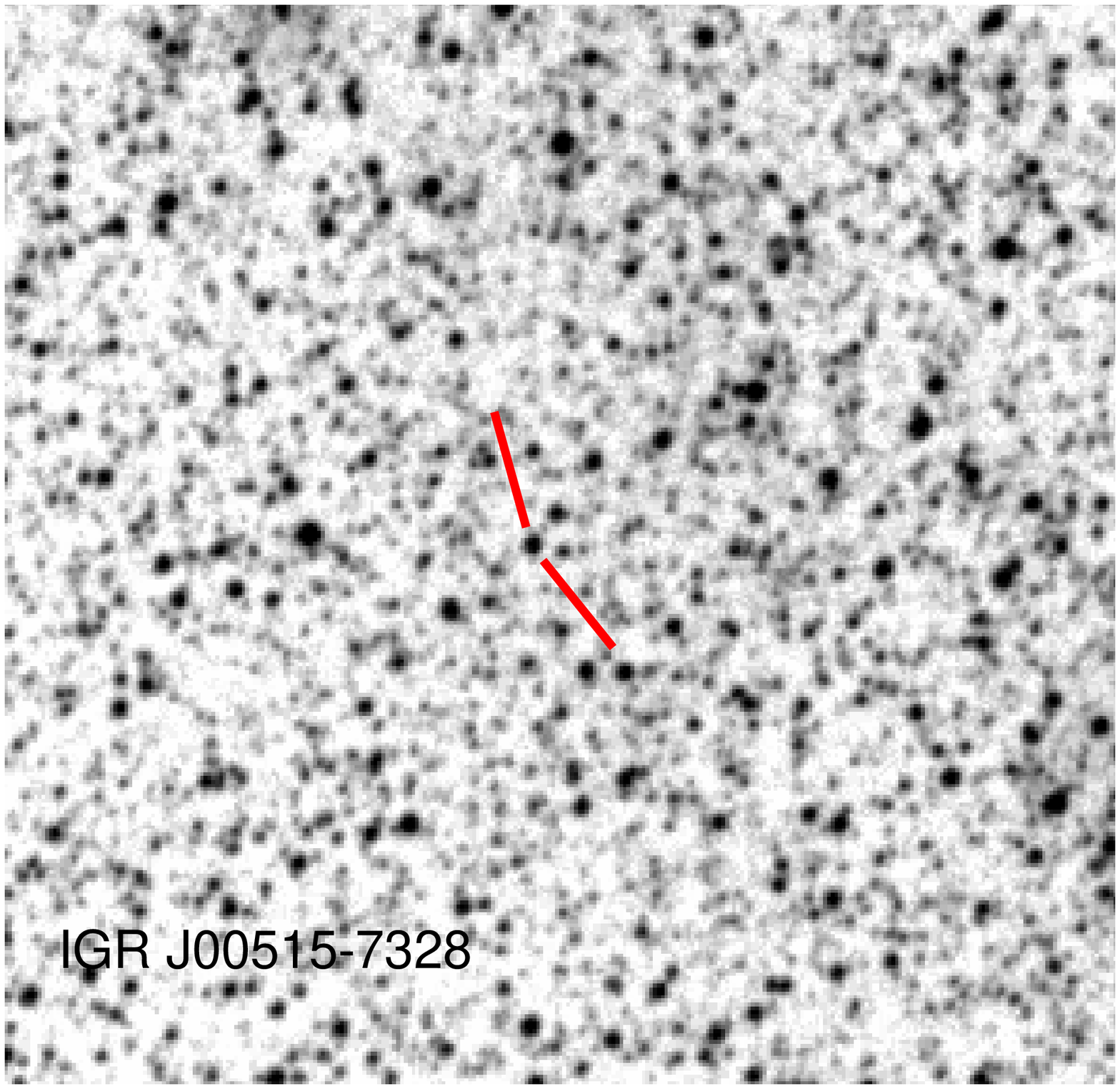,width=5.9cm}}}
\centering{\mbox{\psfig{file=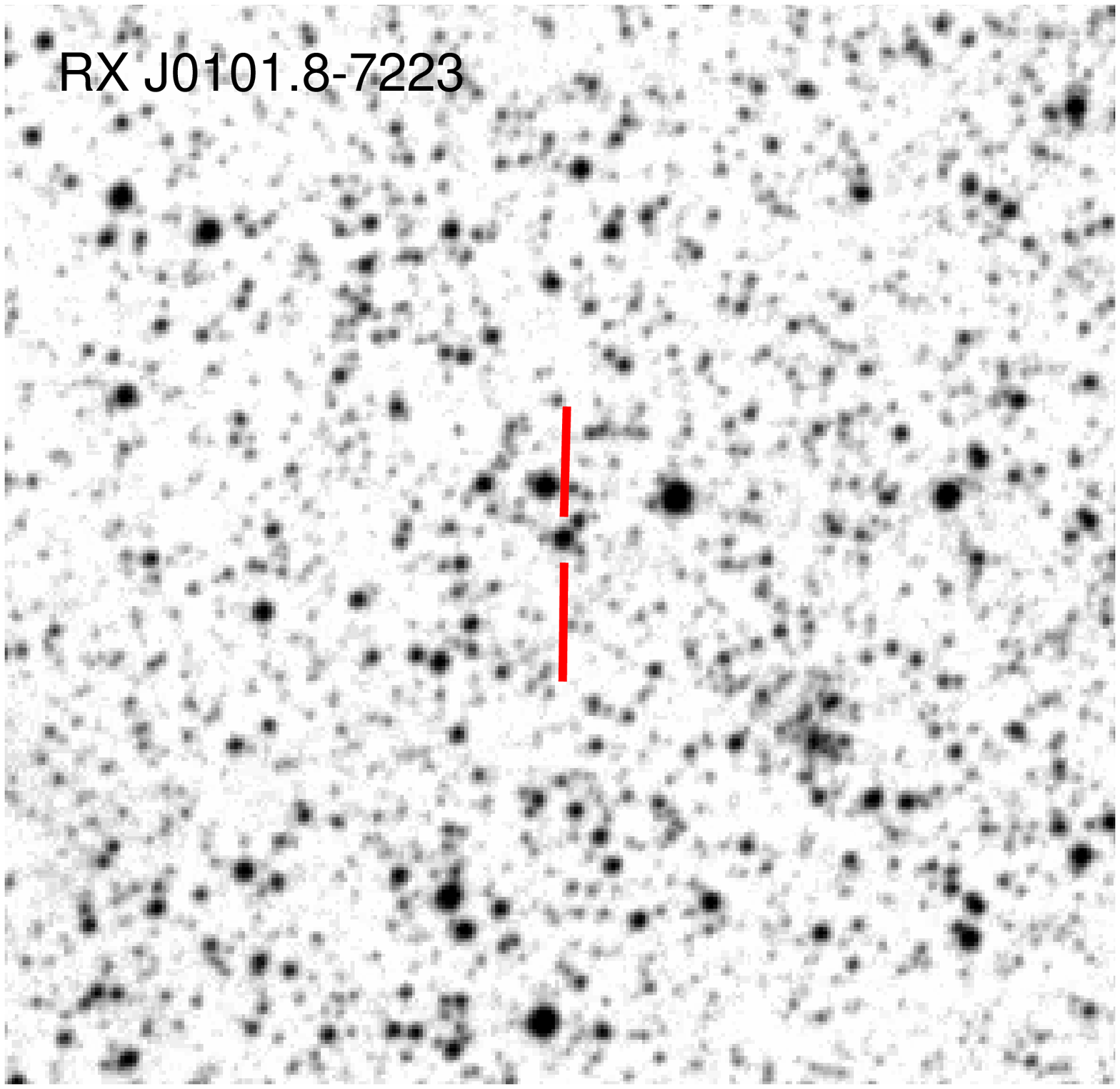,width=5.9cm}}}
\centering{\mbox{\psfig{file=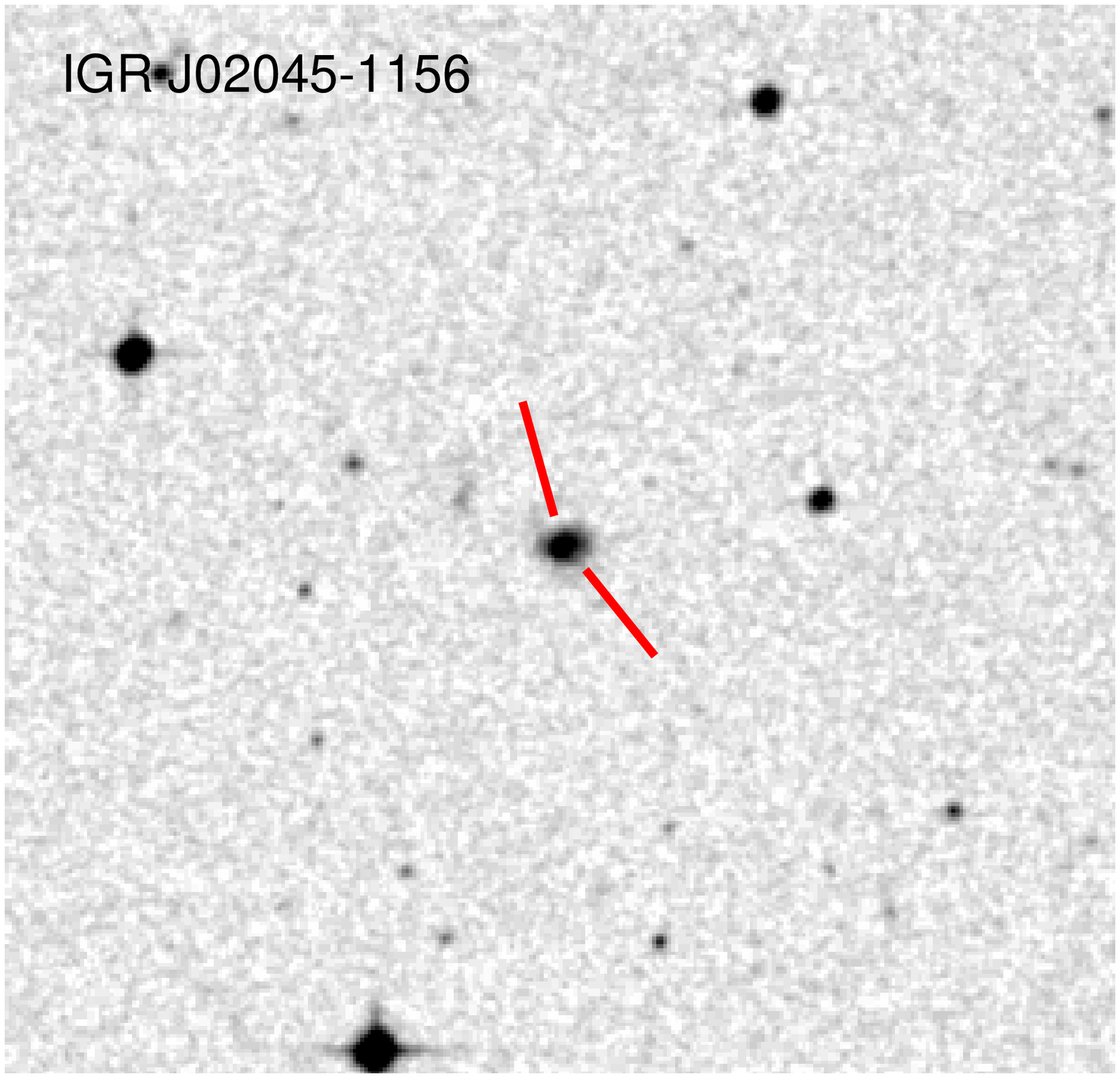,width=5.9cm}}}
\centering{\mbox{\psfig{file=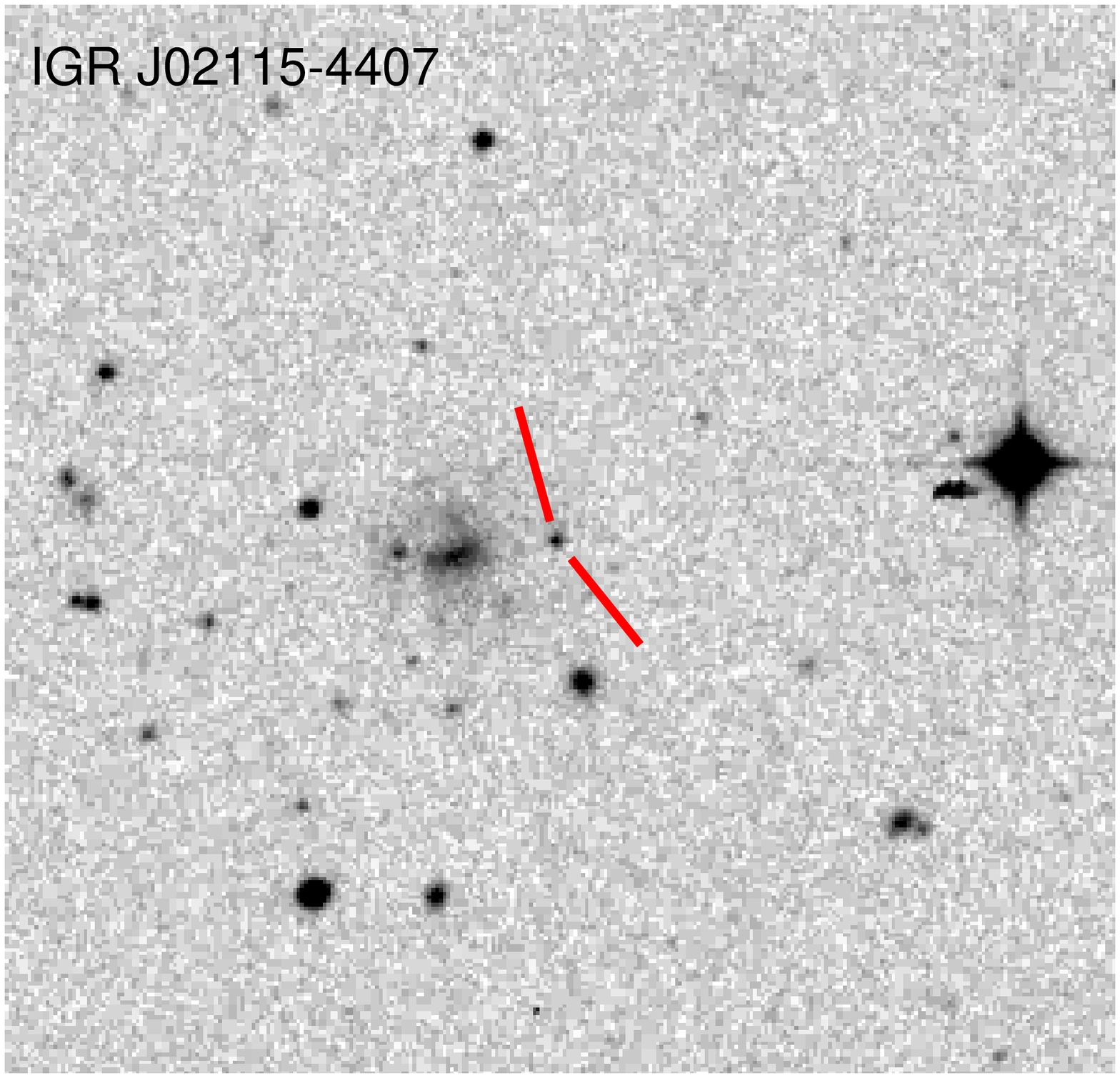,width=5.9cm}}}
\centering{\mbox{\psfig{file=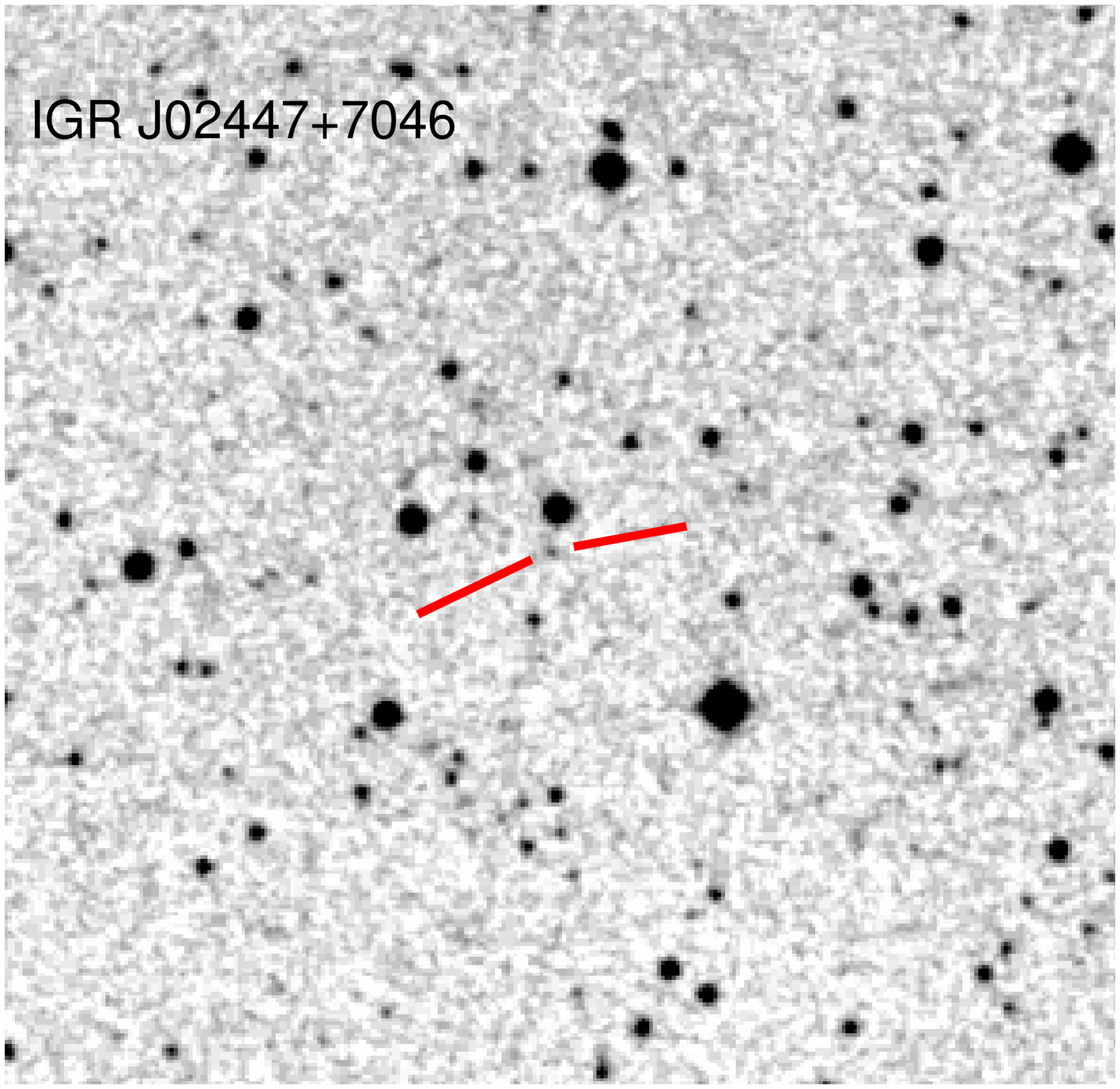,width=5.9cm}}}
\centering{\mbox{\psfig{file=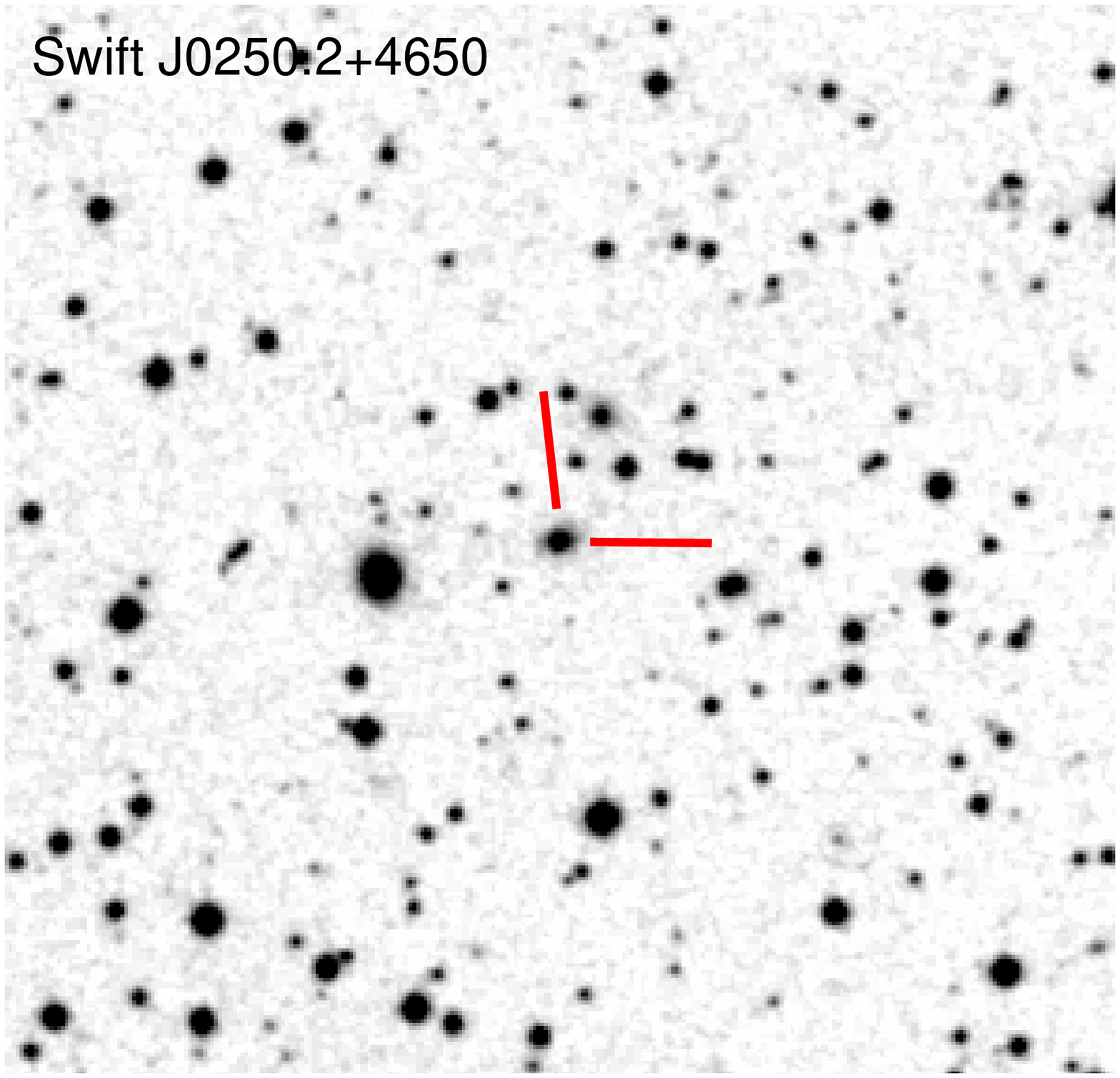,width=5.9cm}}}
\centering{\mbox{\psfig{file=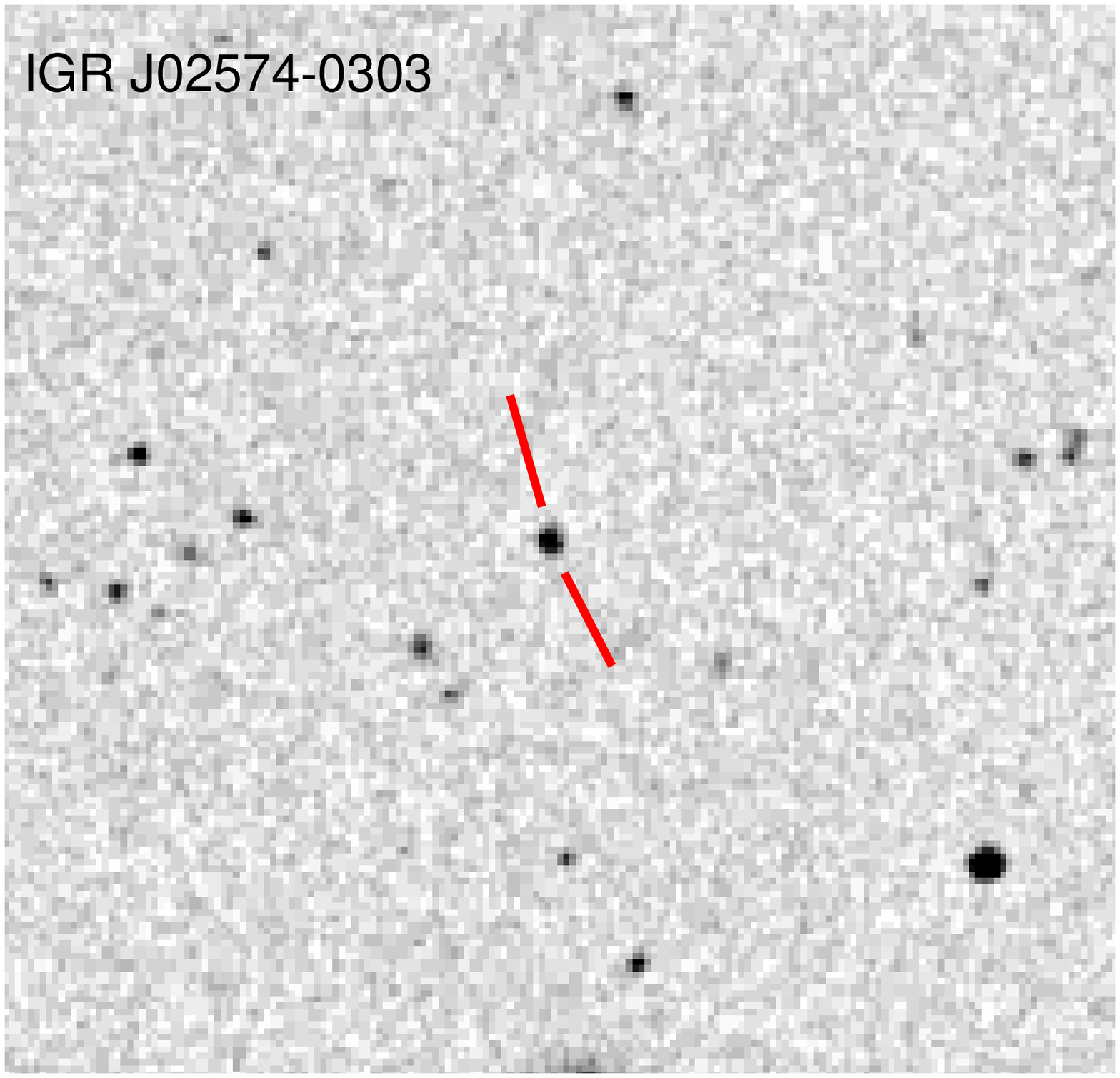,width=5.9cm}}}
\centering{\mbox{\psfig{file=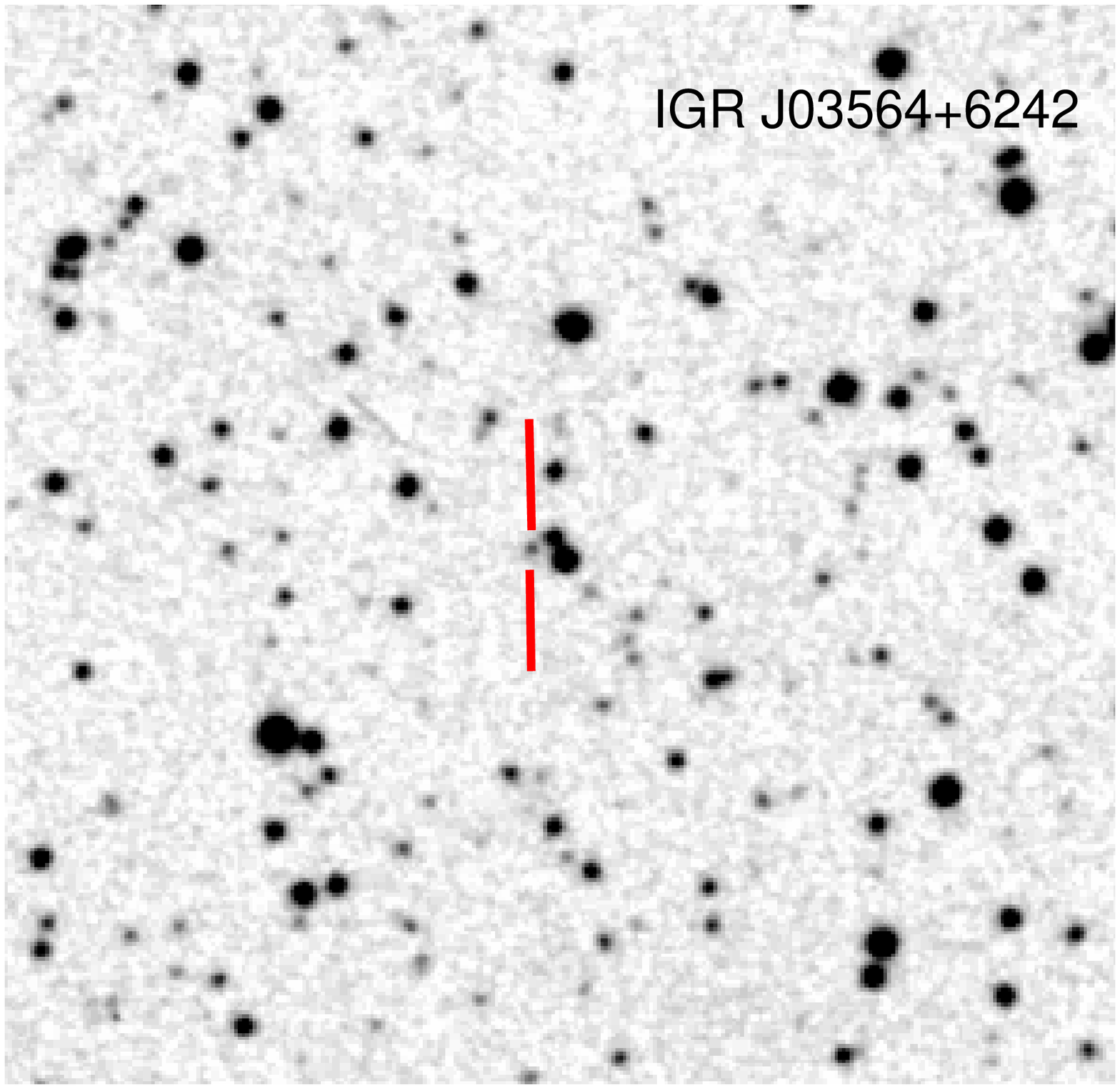,width=5.9cm}}}
\centering{\mbox{\psfig{file=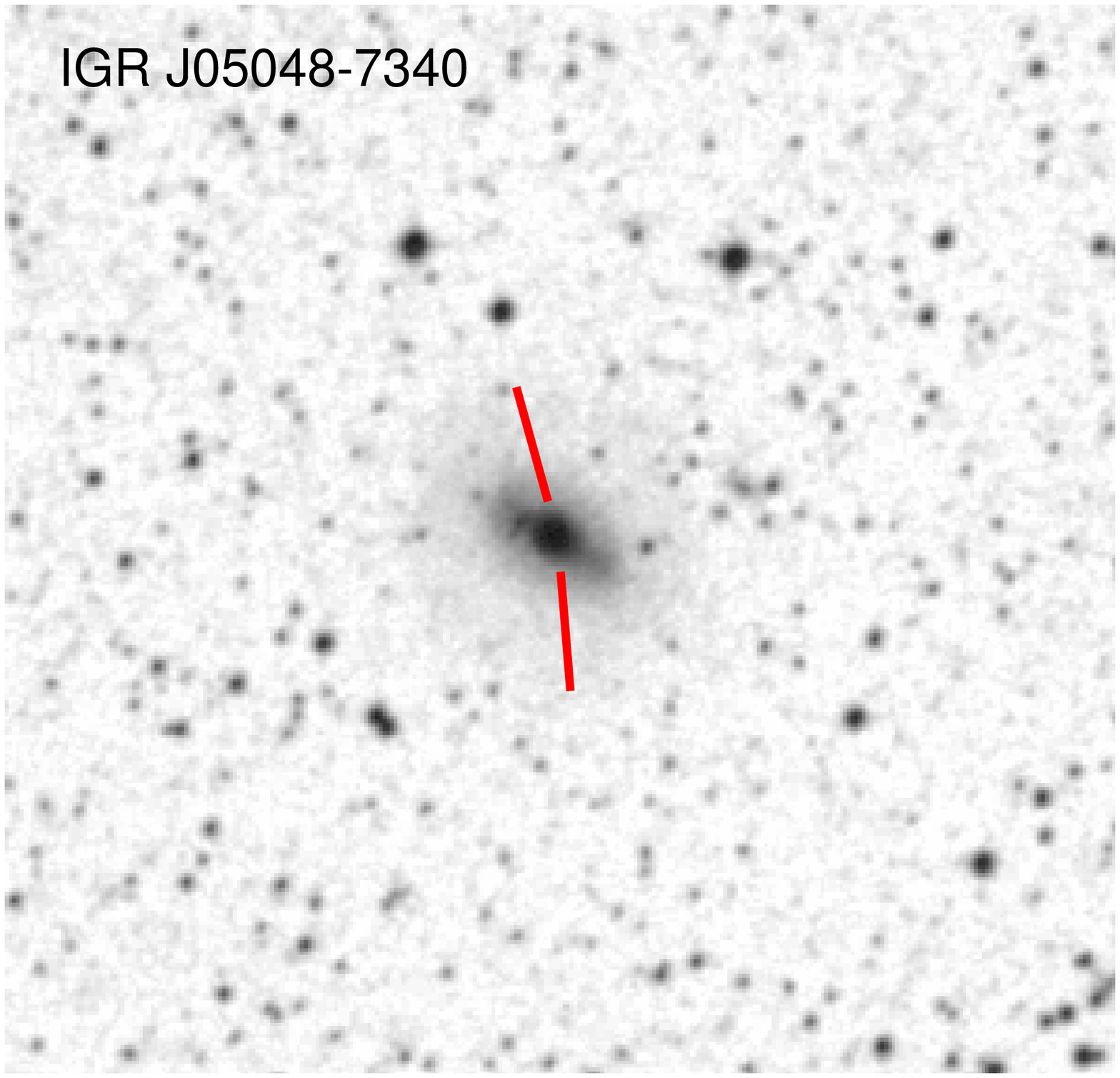,width=5.9cm}}}
\caption{Optical images of the fields of nine {\it INTEGRAL} hard 
X--ray sources selected in this paper for optical spectroscopic 
follow-up (see Table 1). The object name is indicated in each panel.
The proposed optical counterparts are indicated with tick marks. Field 
sizes are 5$'$$\times$5$'$. Nearly all mages were extracted from the 
DSS-II-Red survey; the one for IGR J02574$-$0303 (bottom left panel) was 
instead obtained from the DSS-I-Blue survey. In all cases, north is up 
and east to the left.}
\end{figure*}

\begin{figure*}
\hspace{-.1cm}
\centering{\mbox{\psfig{file=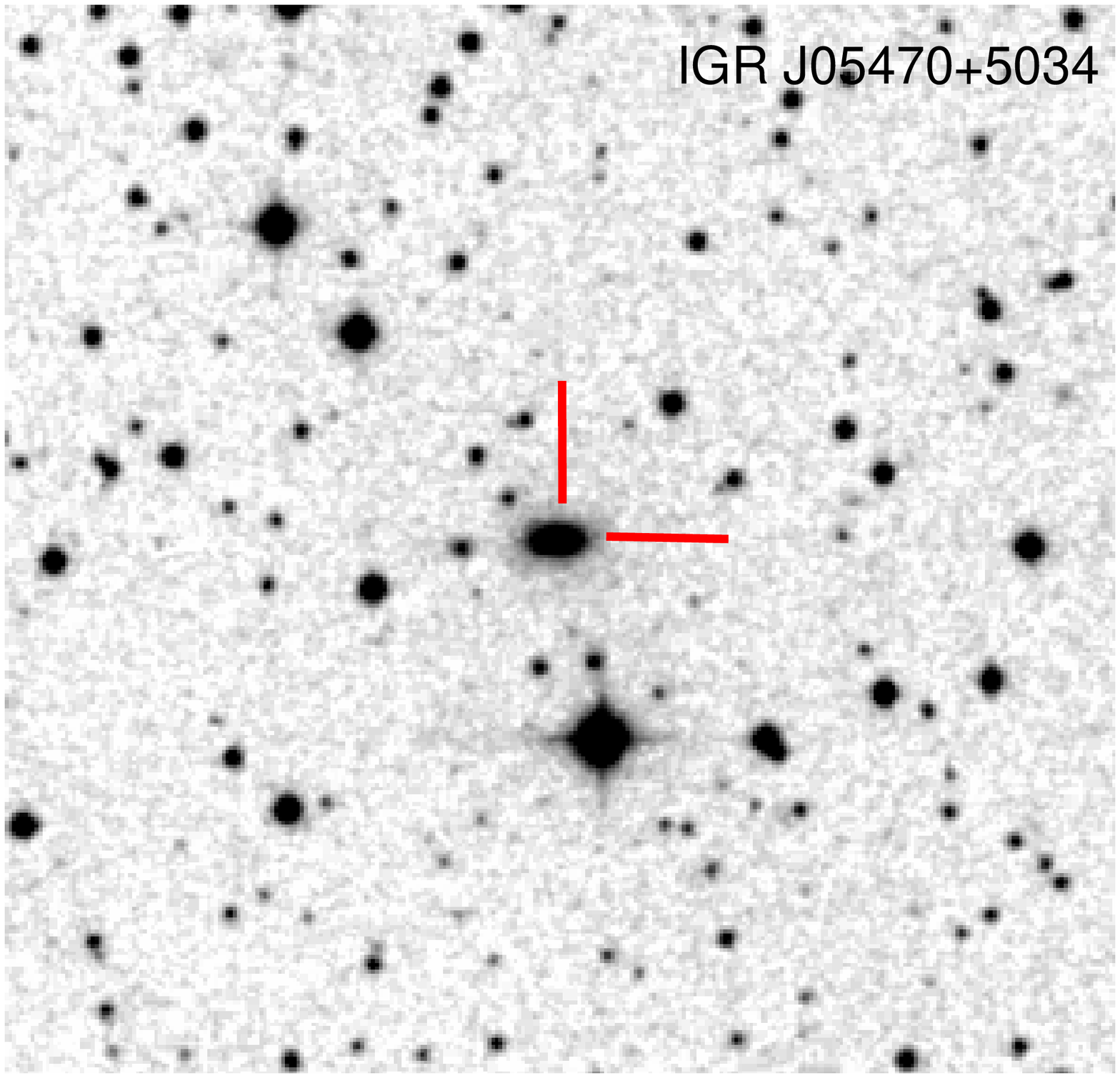,width=5.9cm}}}
\centering{\mbox{\psfig{file=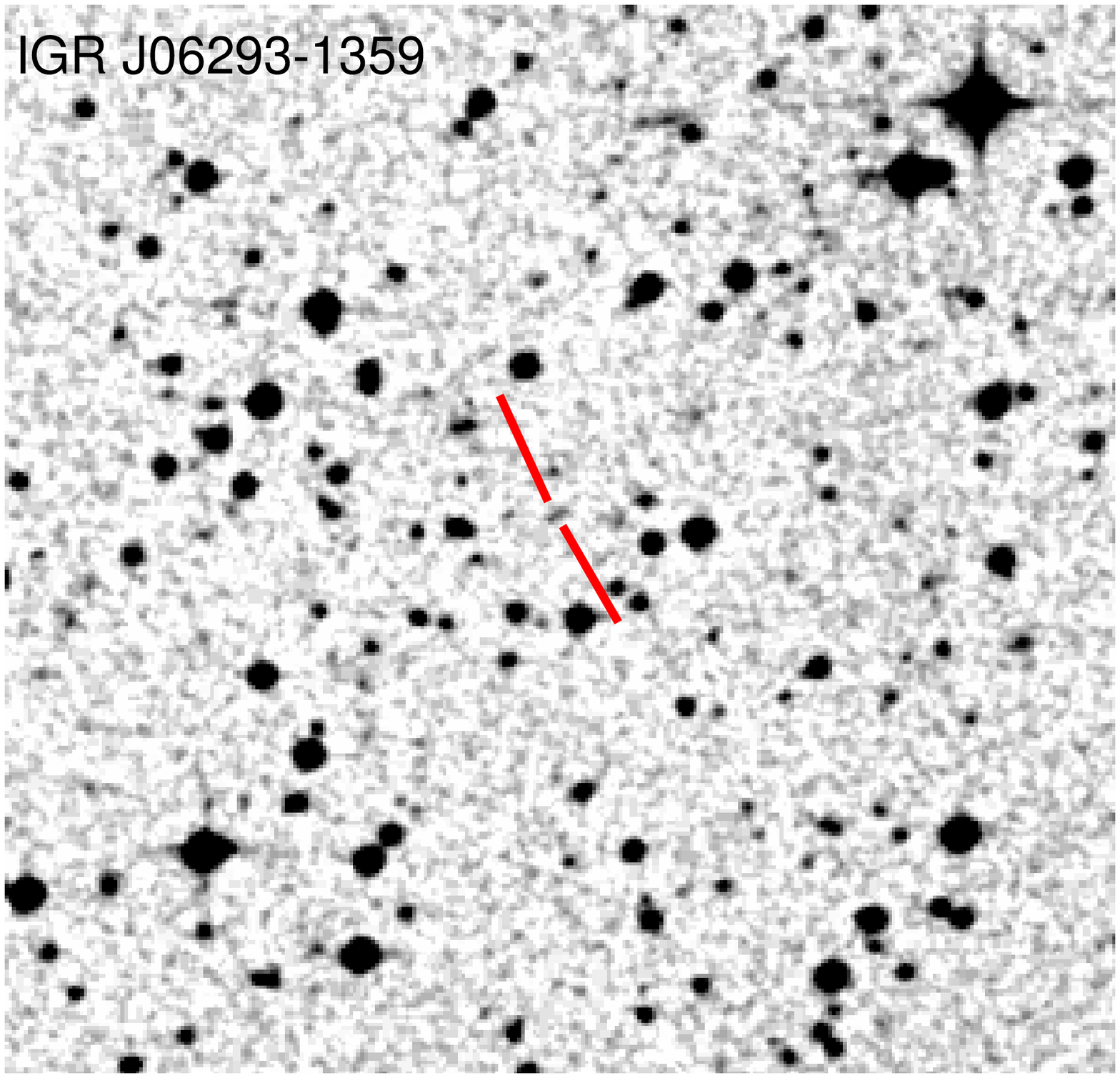,width=5.9cm}}}
\centering{\mbox{\psfig{file=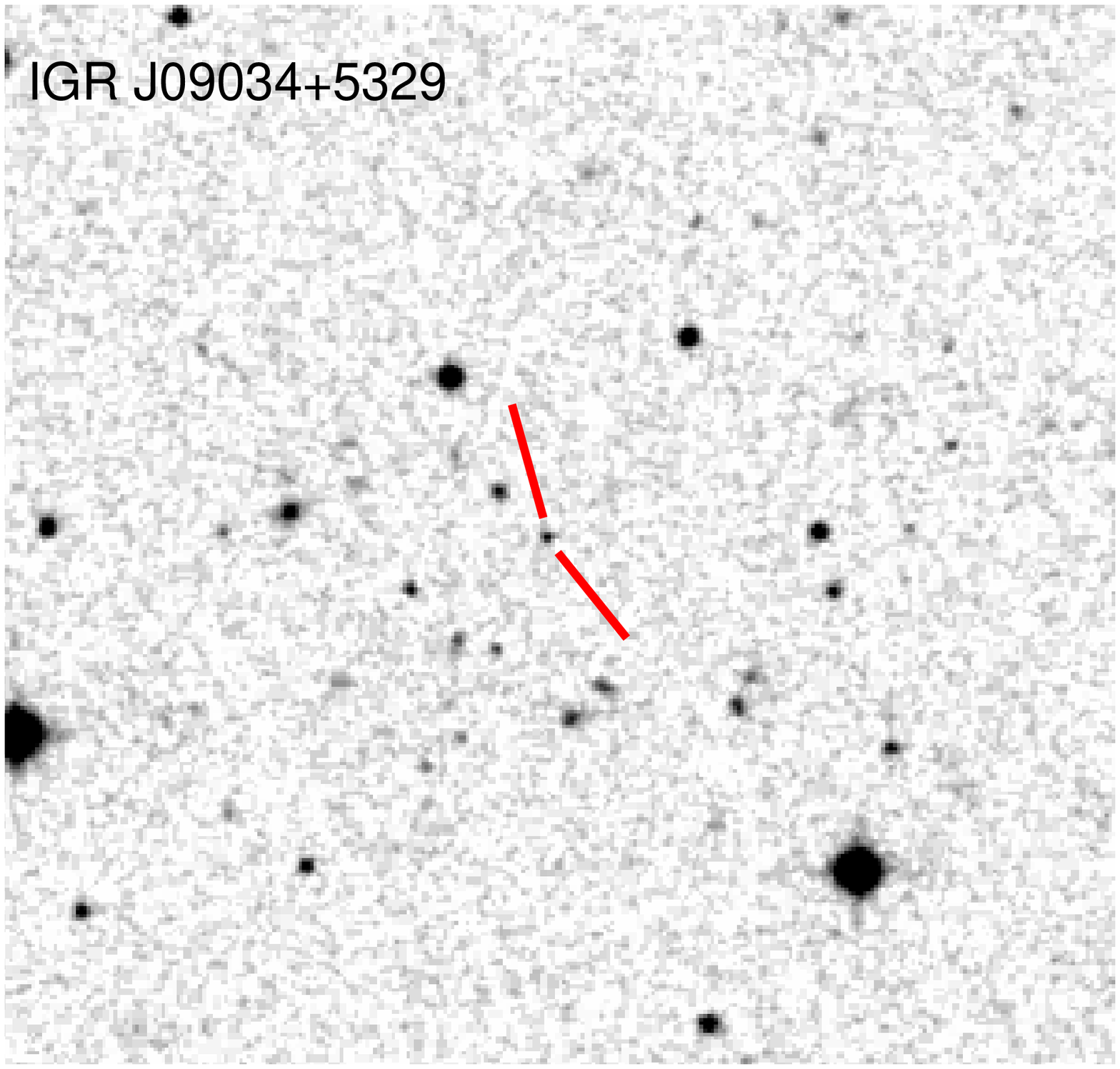,width=5.9cm}}}
\centering{\mbox{\psfig{file=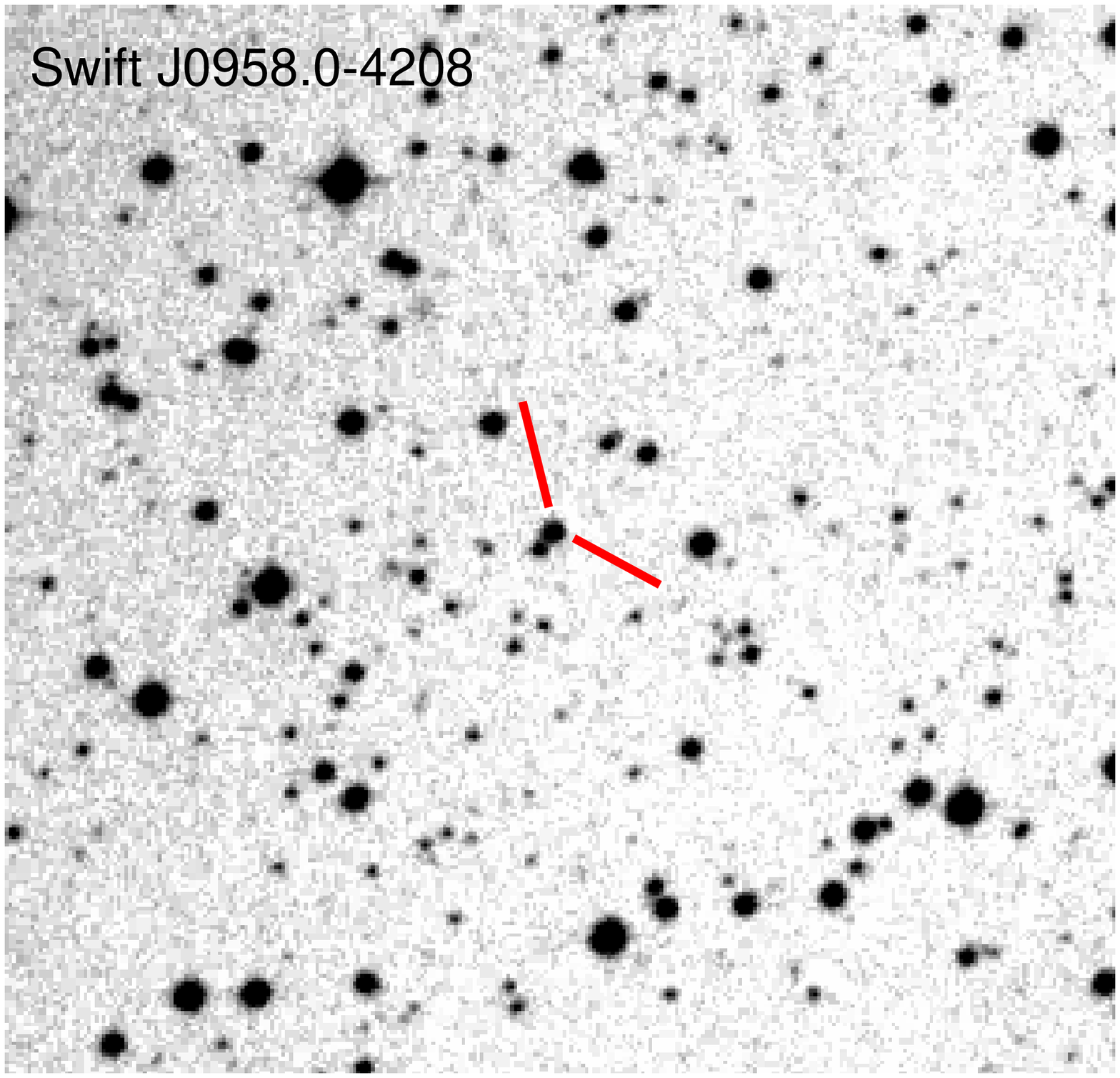,width=5.9cm}}}
\centering{\mbox{\psfig{file=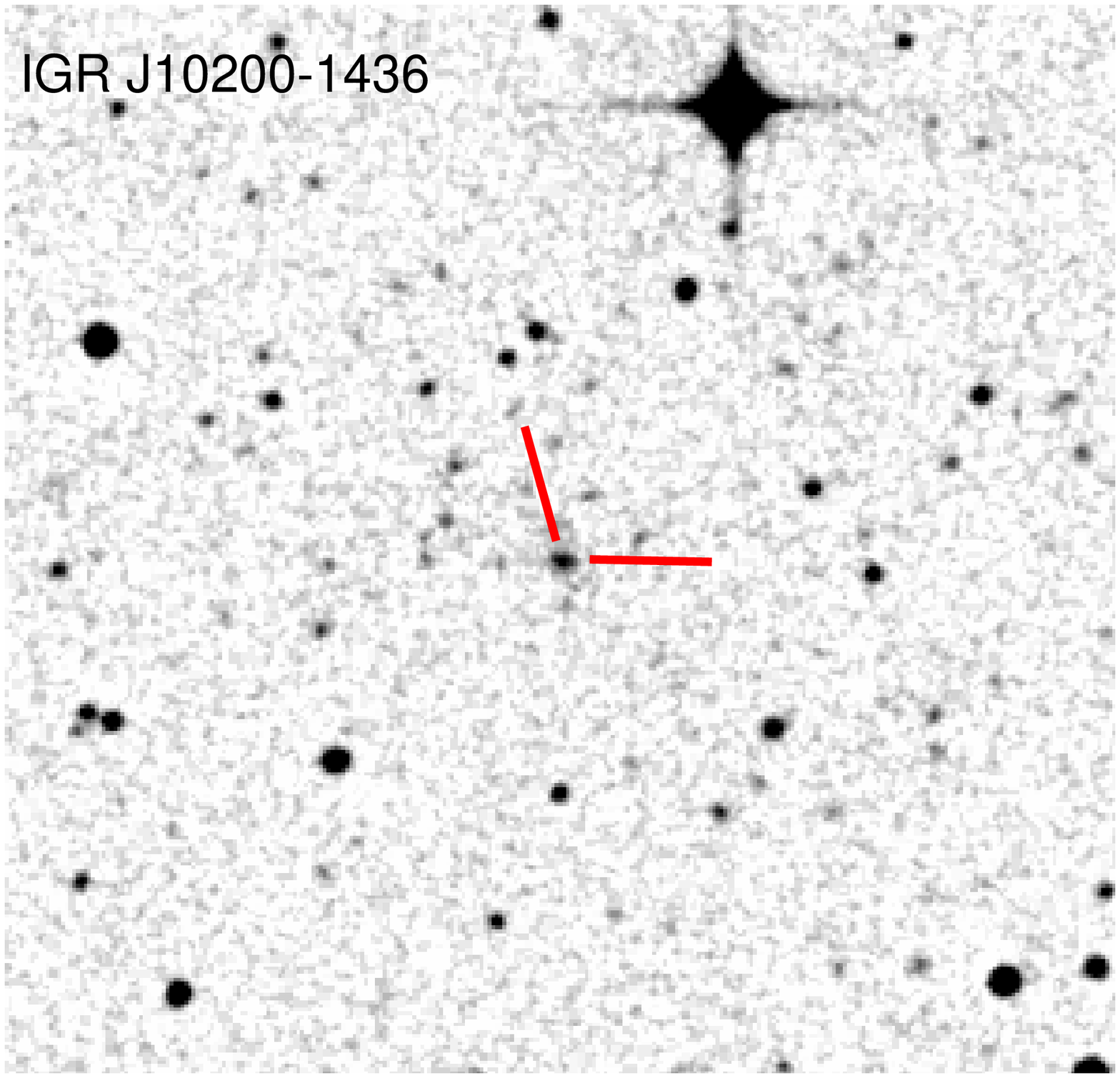,width=5.9cm}}}
\centering{\mbox{\psfig{file=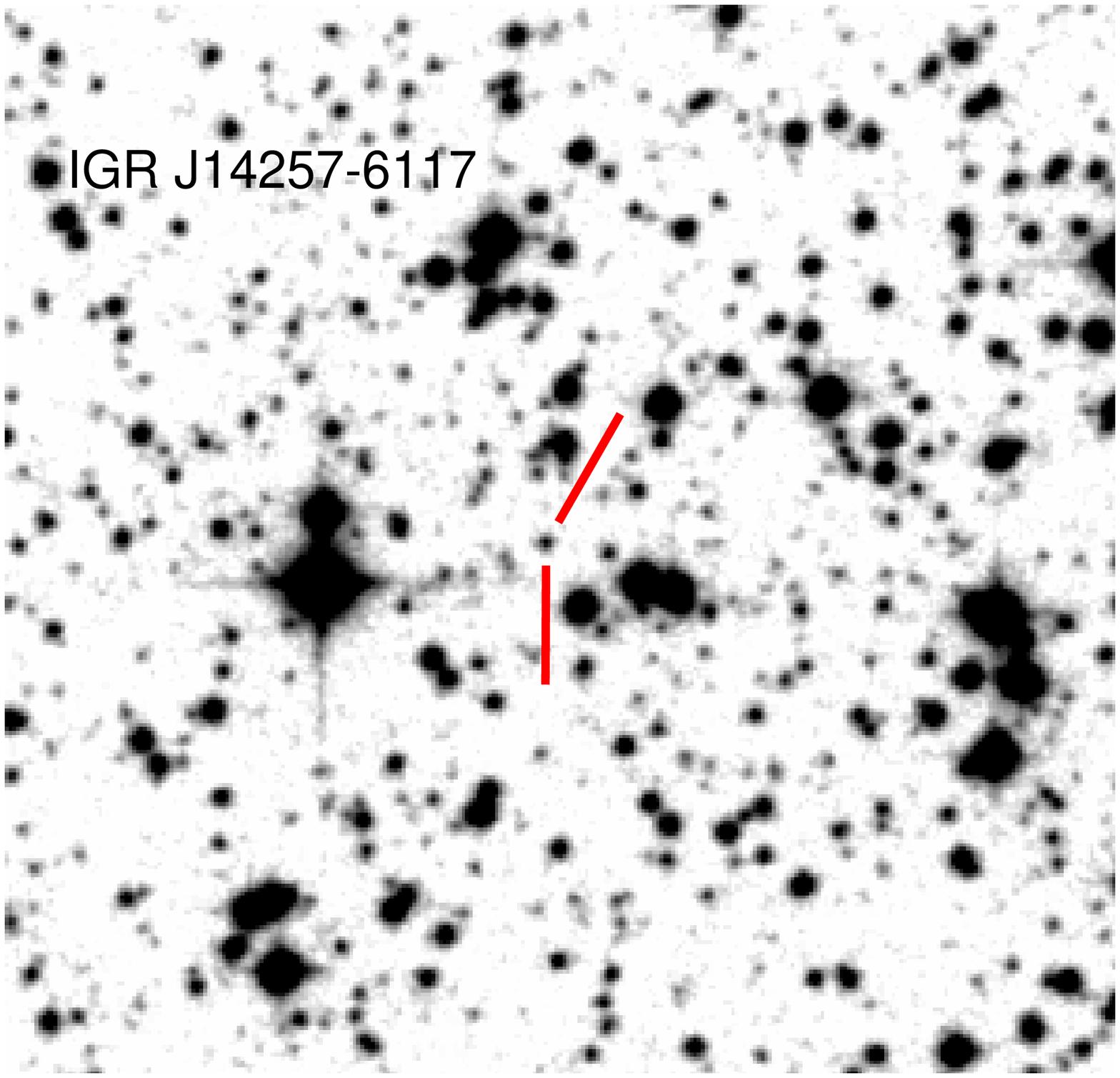,width=5.9cm}}}
\centering{\mbox{\psfig{file=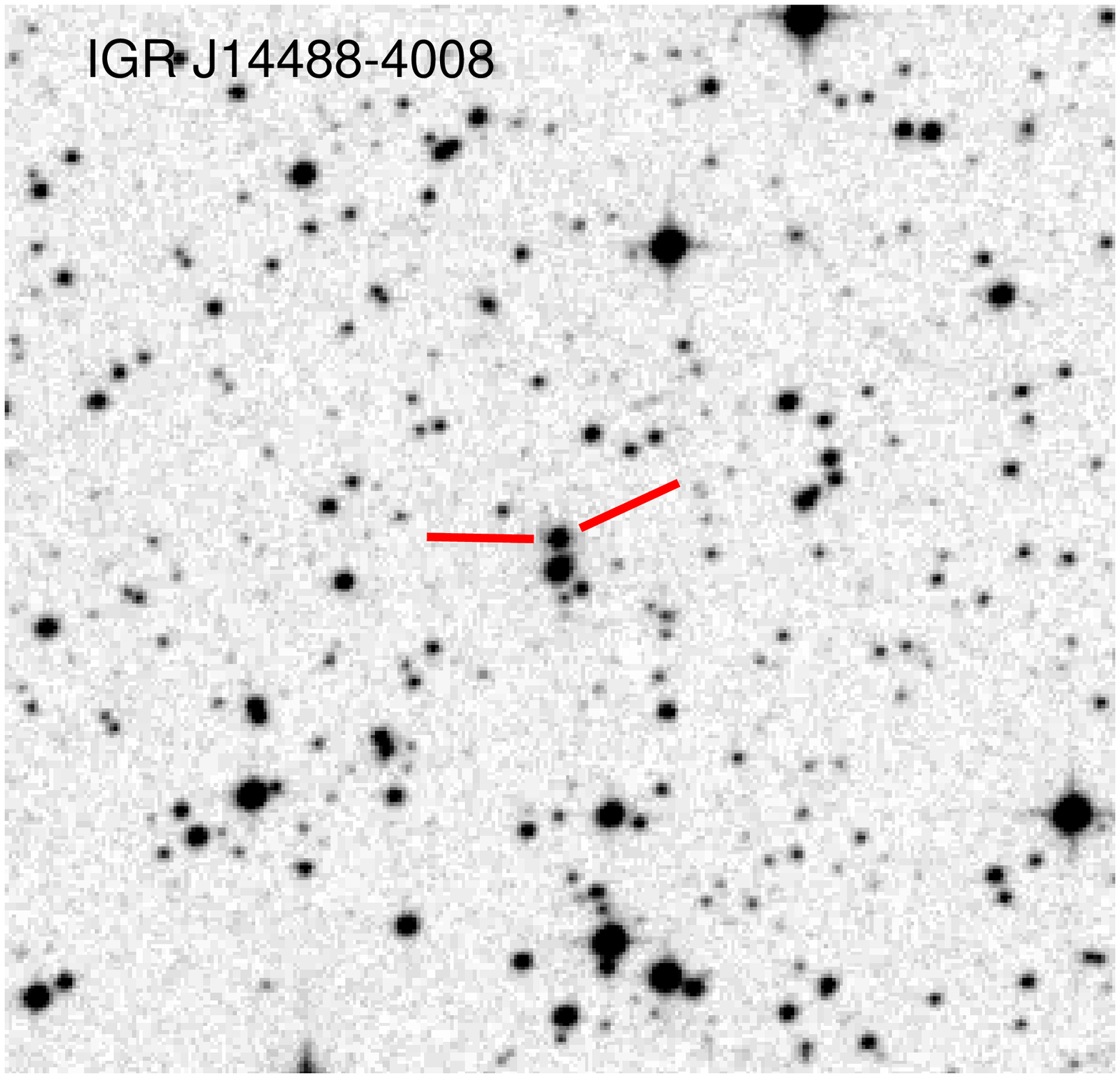,width=5.9cm}}}
\centering{\mbox{\psfig{file=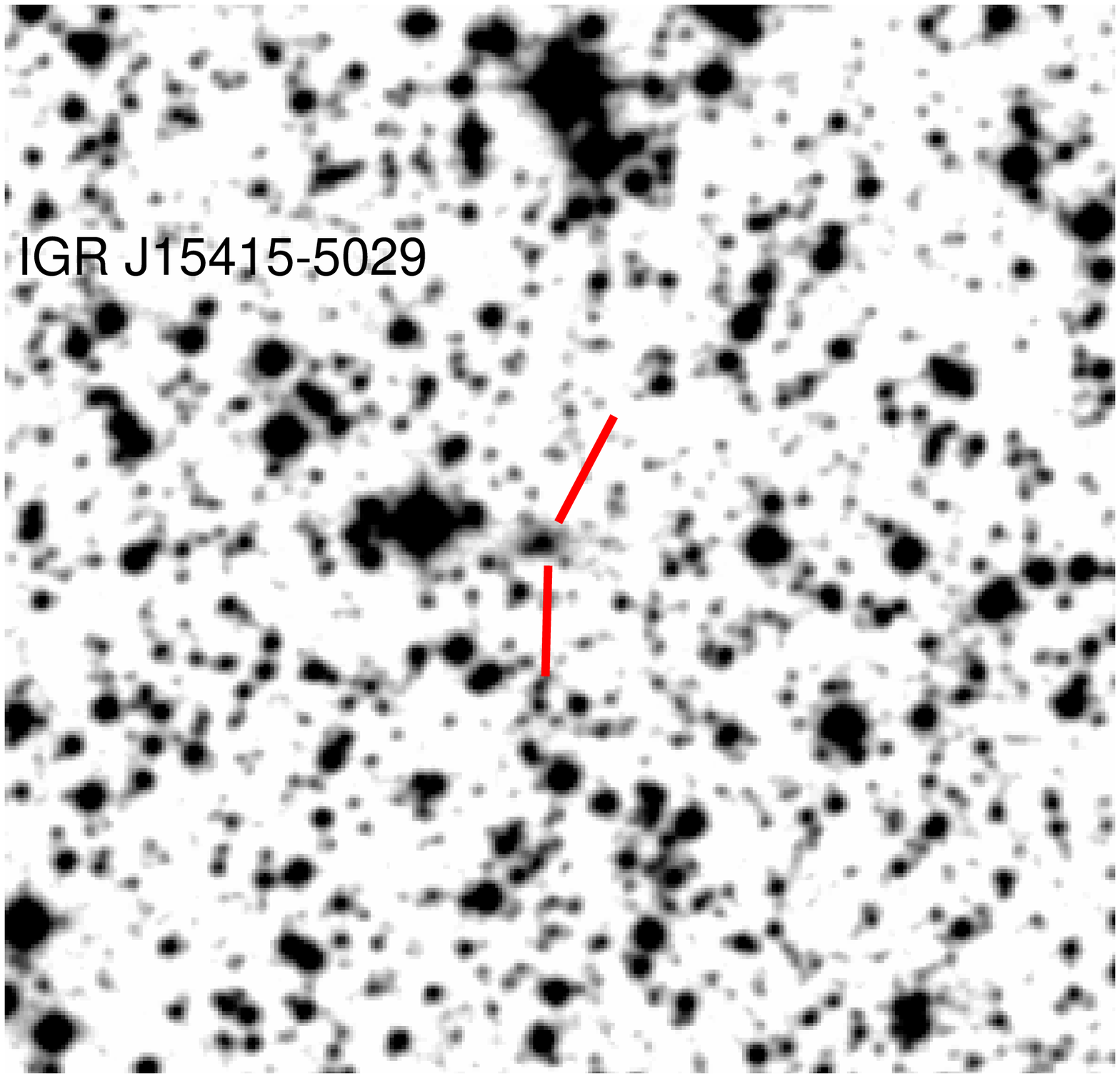,width=5.9cm}}}
\centering{\mbox{\psfig{file=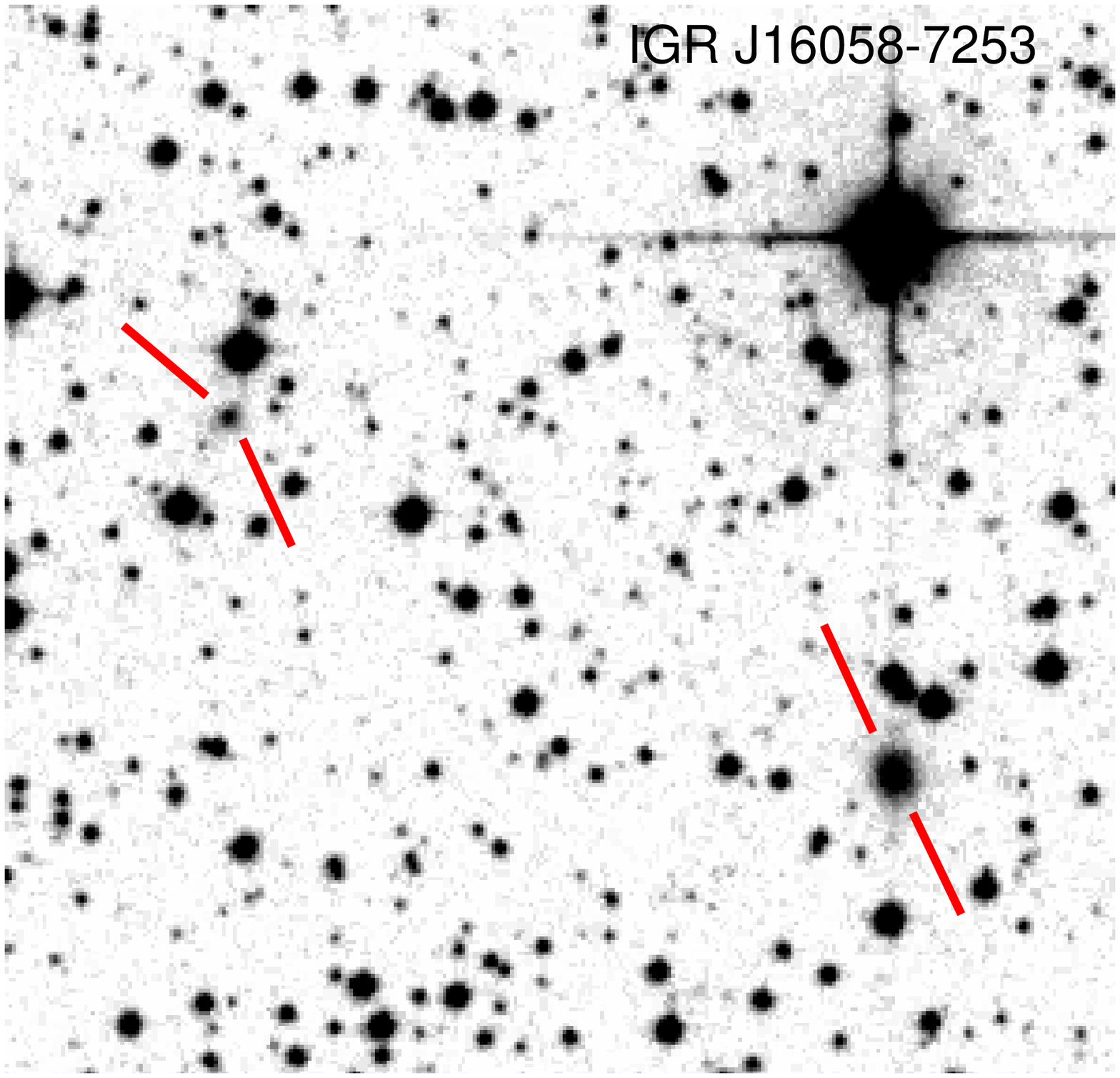,width=5.9cm}}}
\caption{Similar to Fig. 1 but for nine more {\it INTEGRAL} sources of our
sample (see Table 1). For IGR J16058$-$7253 (bottom right panel), we 
indicate the two optical objects likely responsible for the total hard 
X--ray emission detected for this source (see Landi et al. 2012b and Sect. 
4.1 of this paper).}
\end{figure*}

\begin{figure*}
\hspace{-.1cm}
\centering{\mbox{\psfig{file=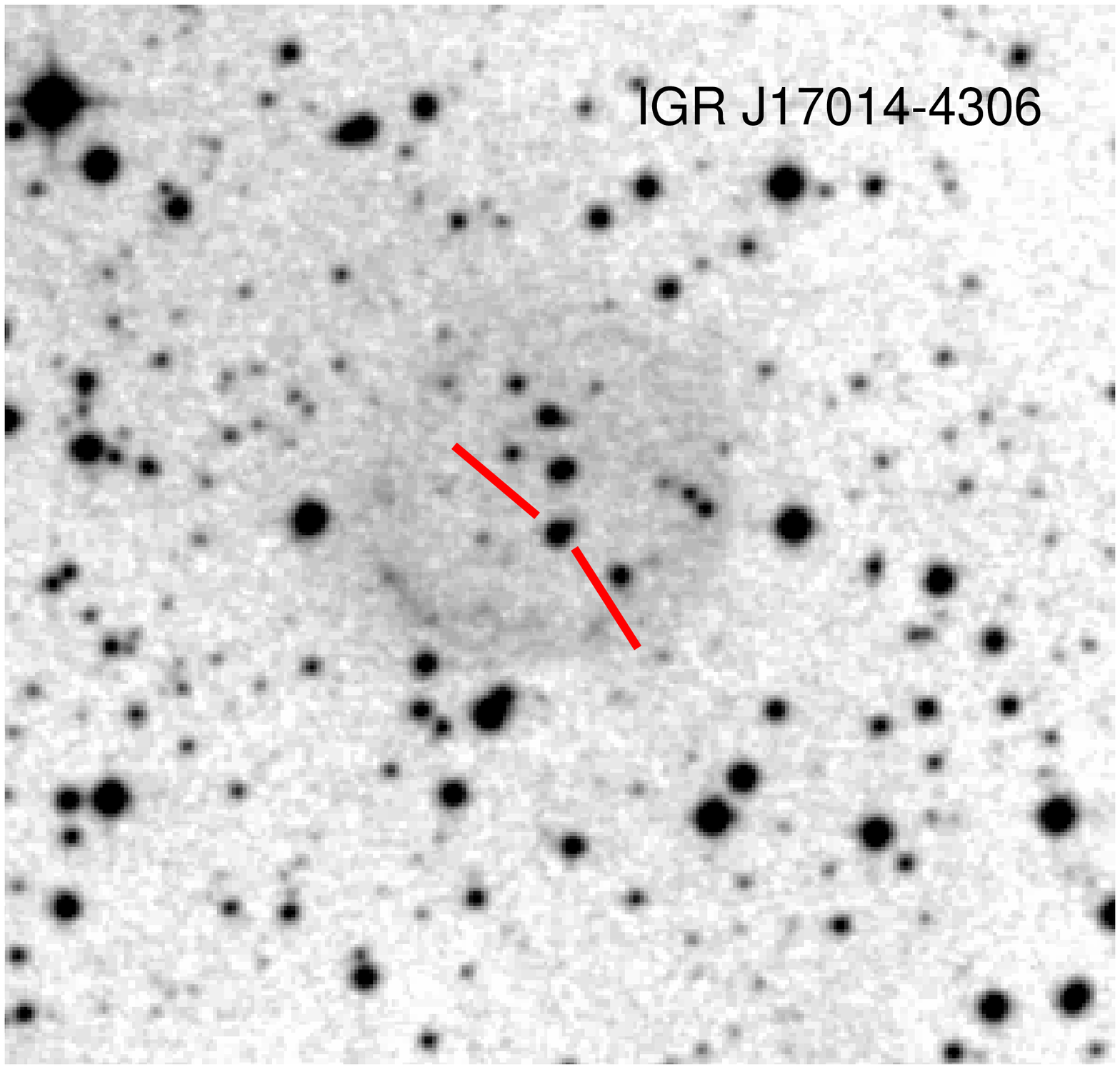,width=5.9cm}}}
\centering{\mbox{\psfig{file=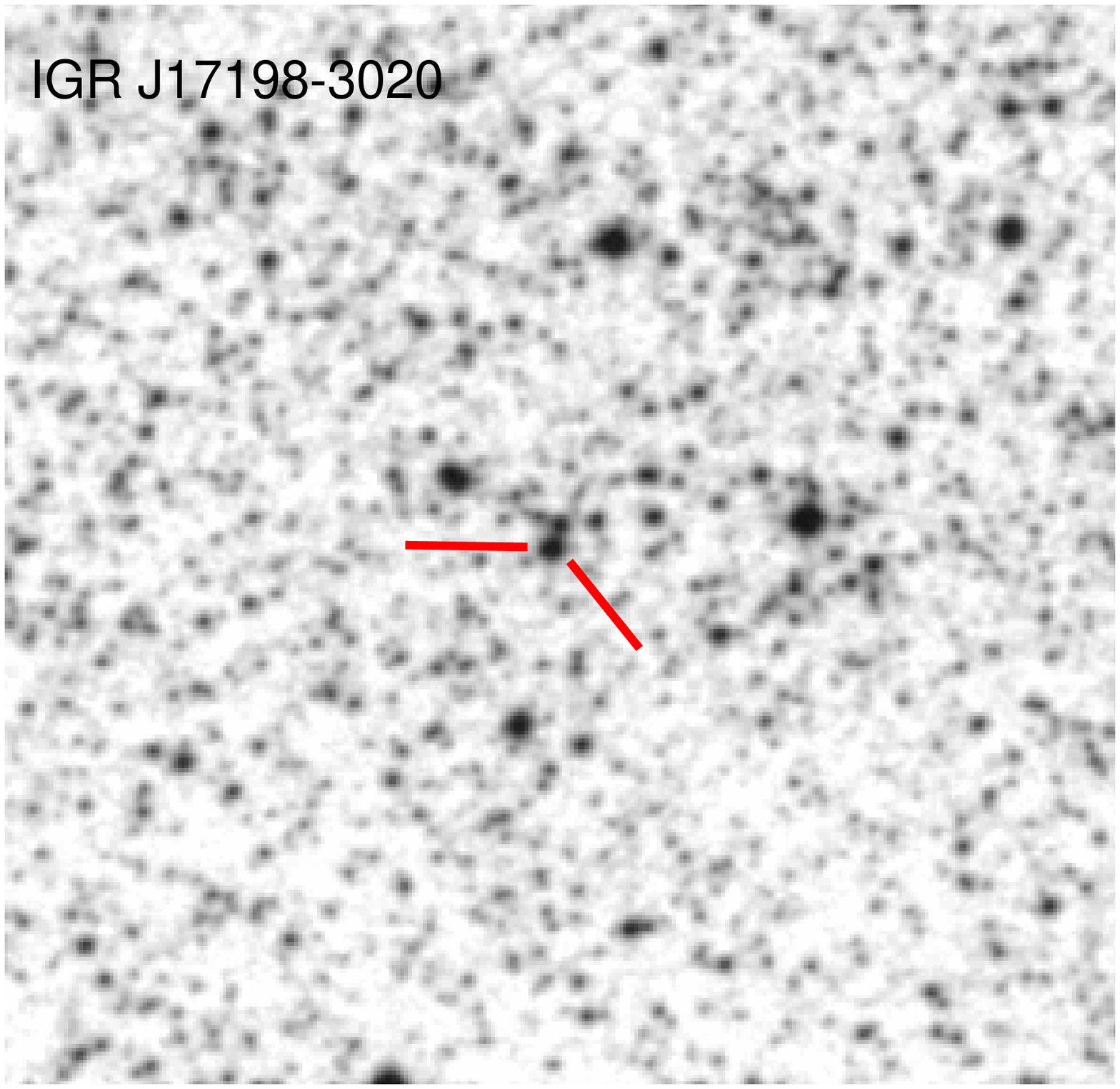,width=5.9cm}}}
\centering{\mbox{\psfig{file=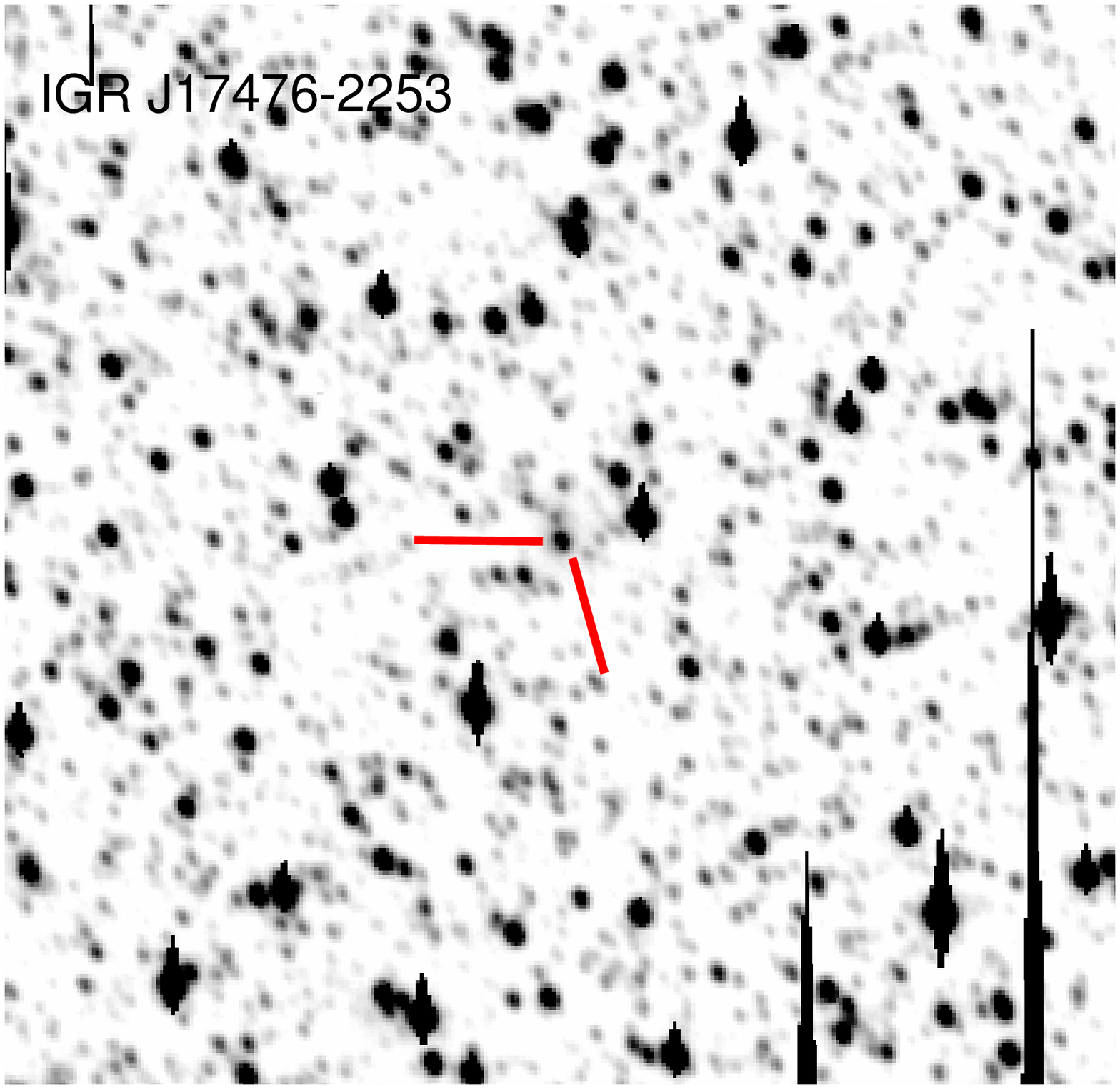,width=5.9cm}}}
\centering{\mbox{\psfig{file=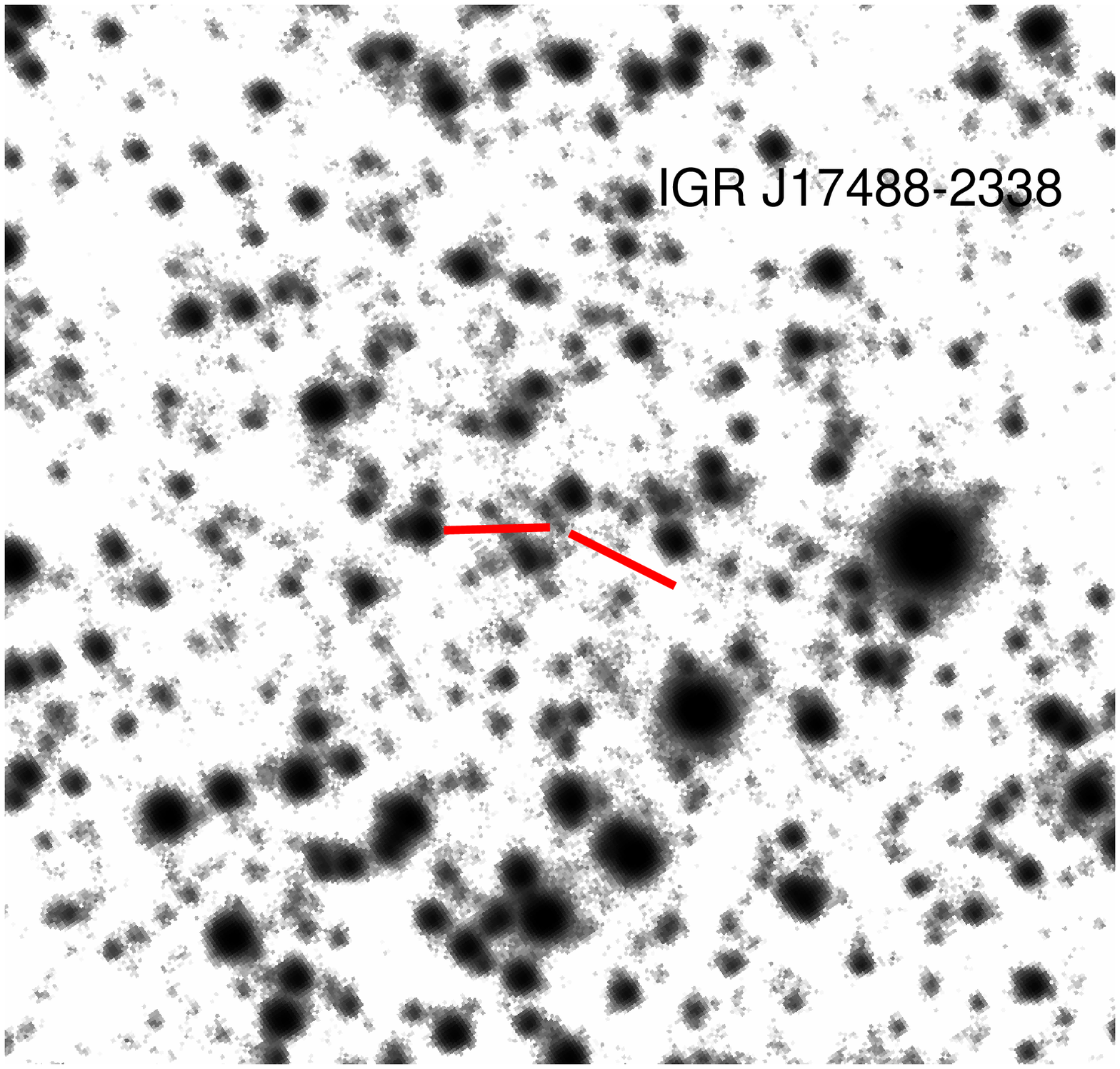,width=5.9cm}}}
\centering{\mbox{\psfig{file=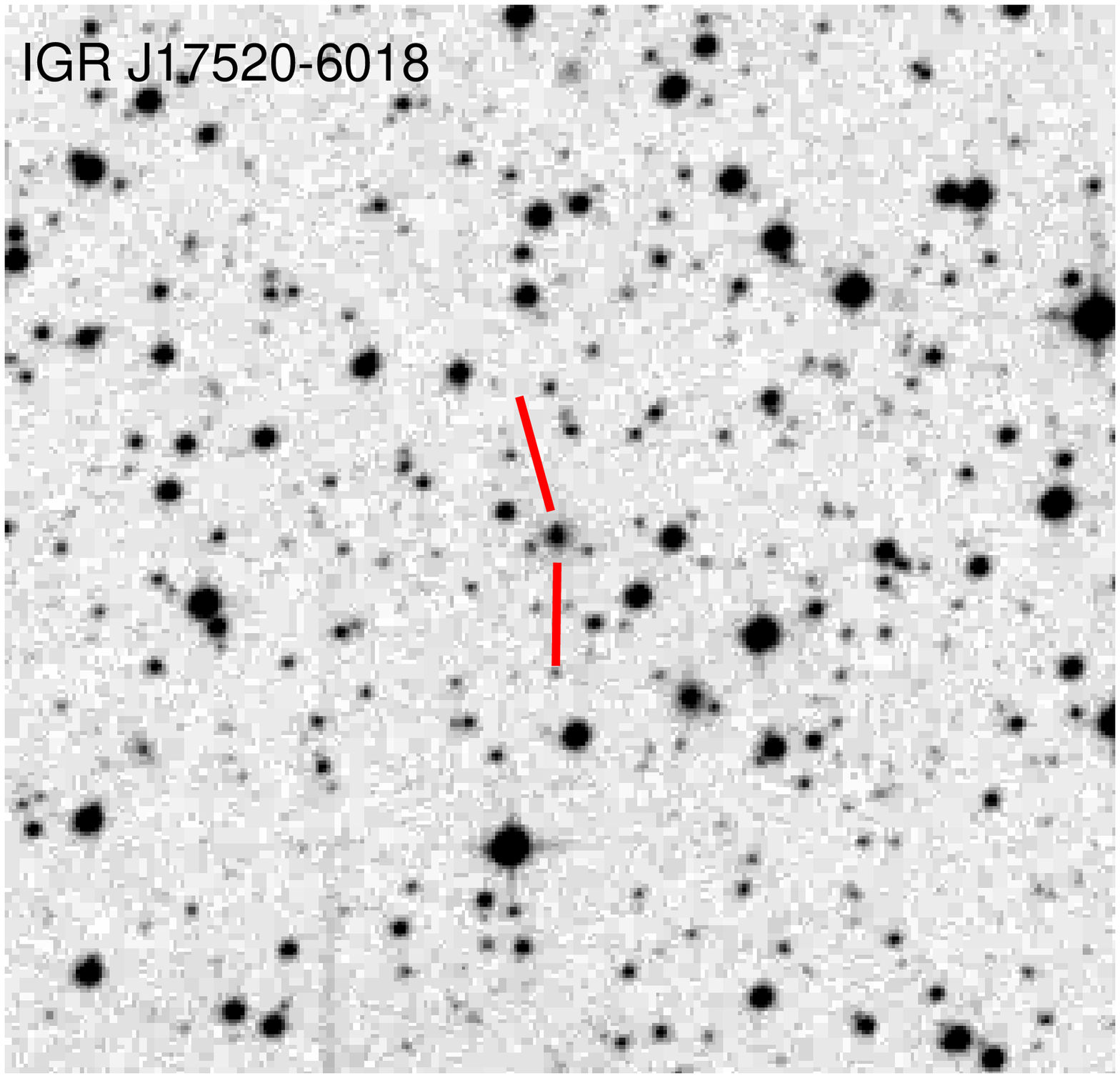,width=5.9cm}}}
\centering{\mbox{\psfig{file=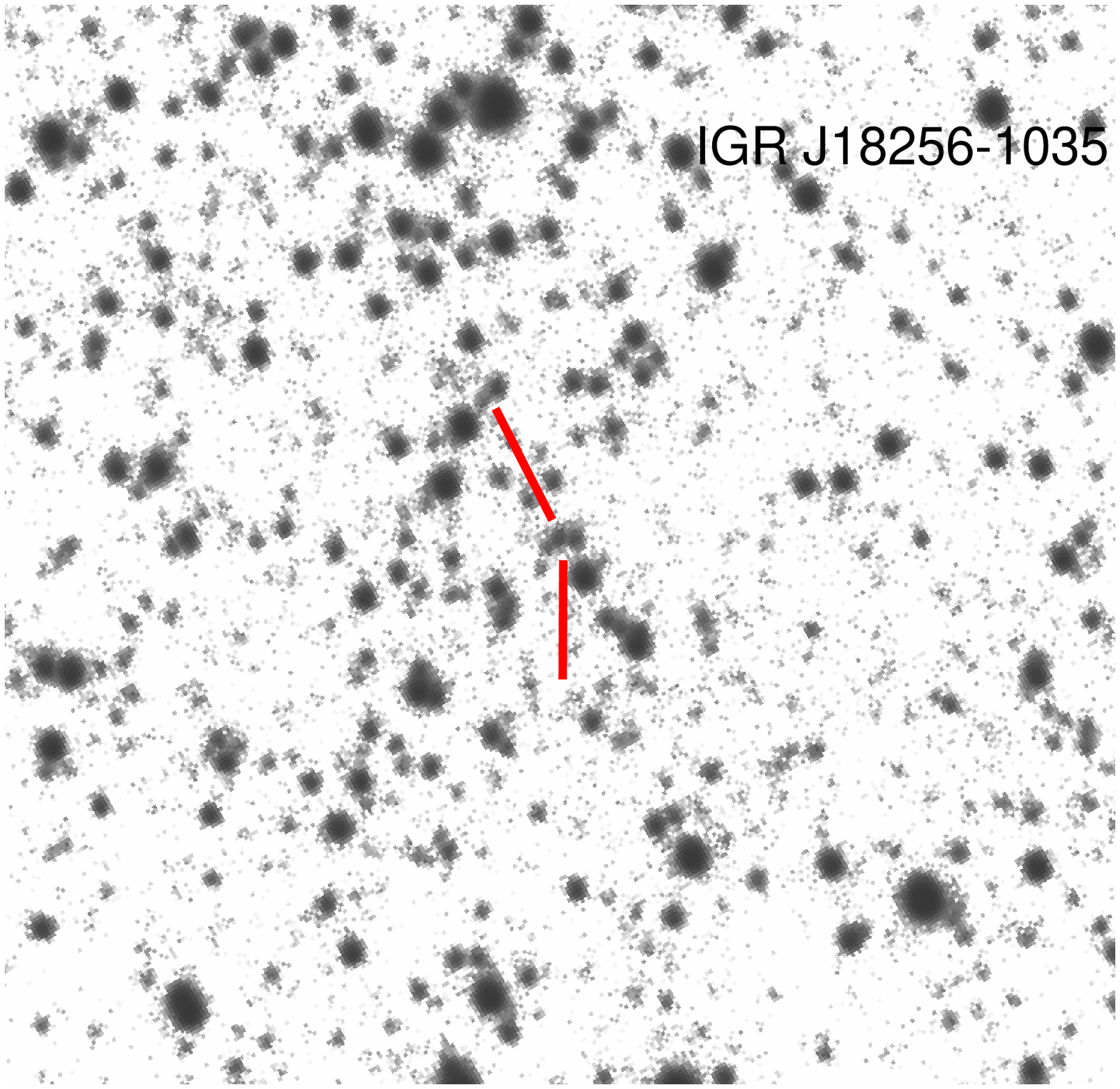,width=5.9cm}}}
\centering{\mbox{\psfig{file=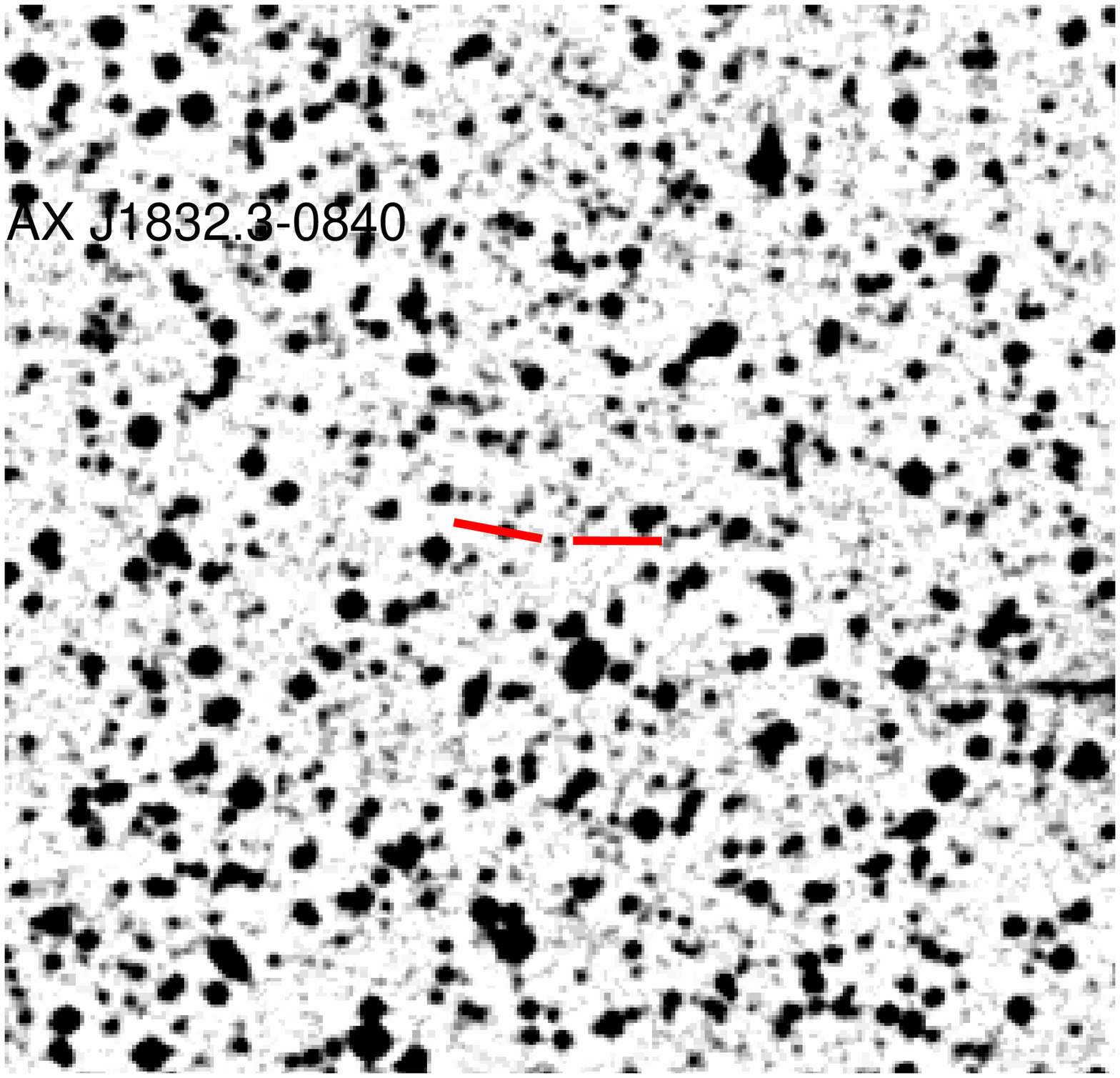,width=5.9cm}}}
\centering{\mbox{\psfig{file=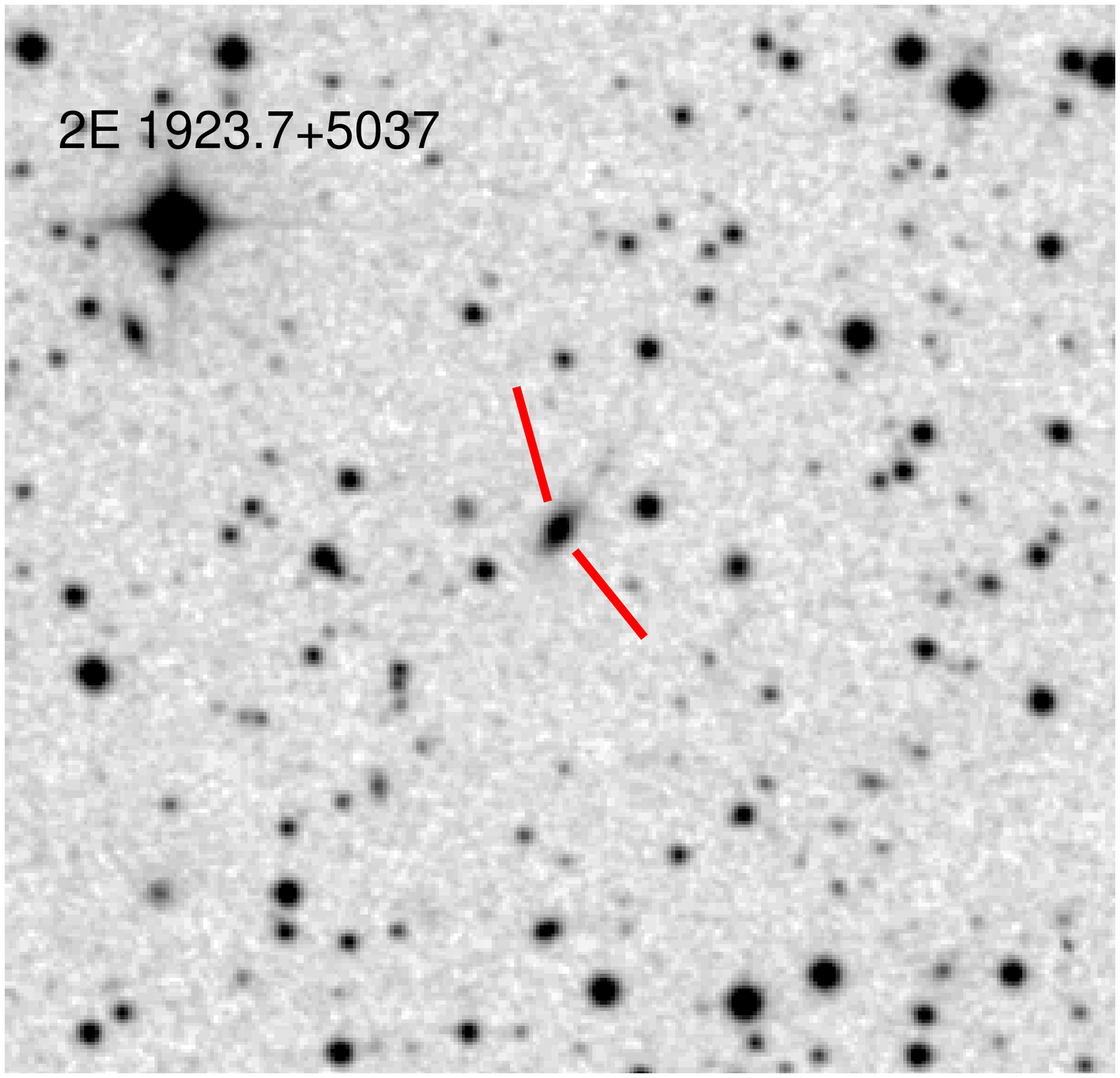,width=5.9cm}}}
\centering{\mbox{\psfig{file=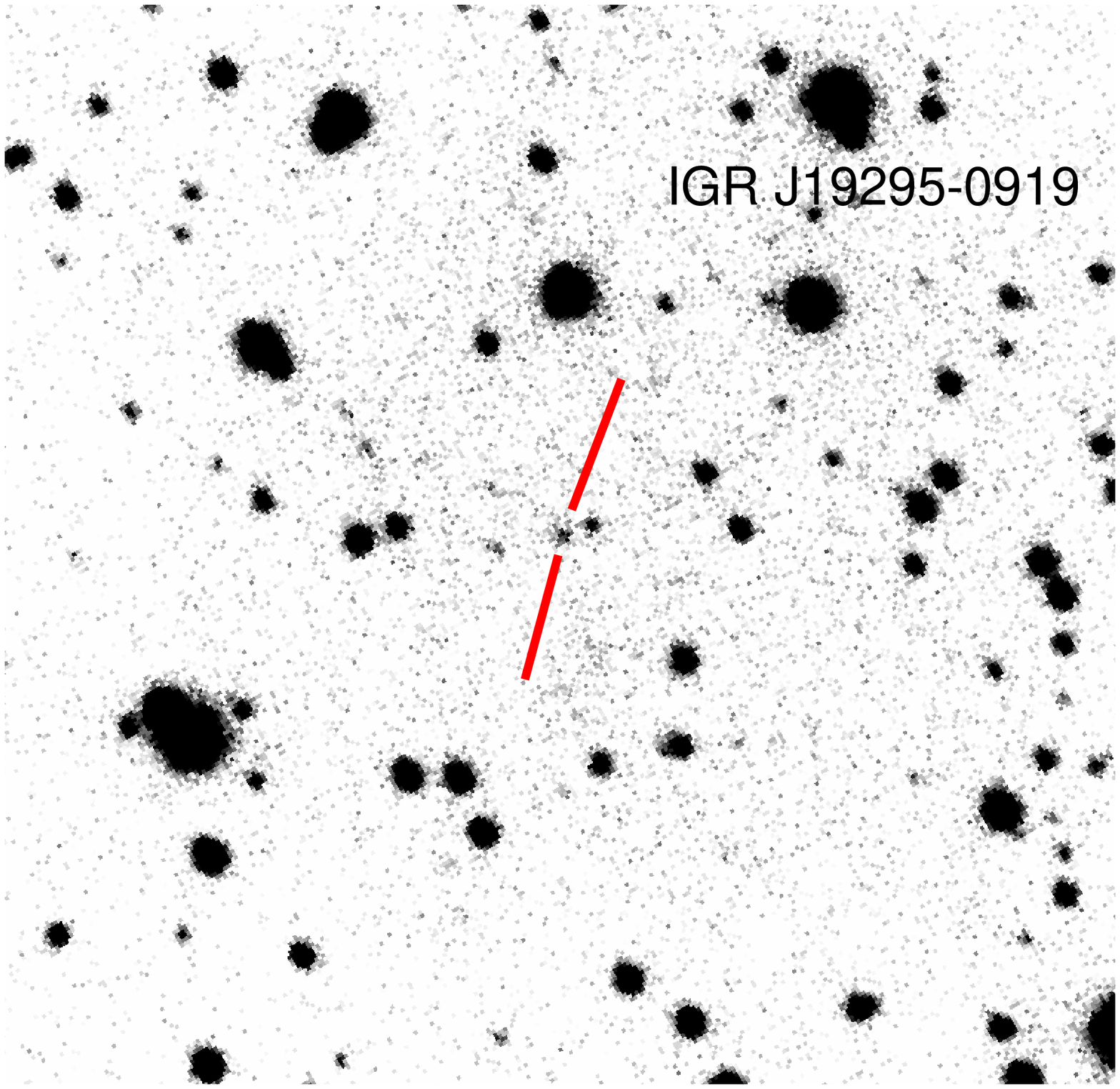,width=5.9cm}}}
\caption{Similar to Fig. 1 but for nine more {\it INTEGRAL} sources of our
sample (see Table 1). The images of the fields of sources IGR 
J17476$-$2253, IGR J17488$-$2338, IGR J18256$-$1035 and IGR J19295$-$0919 
were acquired with TNG+DOLoReS (see Table 2 for details) and cover an area 
of 1$'$$\times$1$'$.}
\end{figure*}

\begin{figure*}
\hspace{-.1cm}
\centering{\mbox{\psfig{file=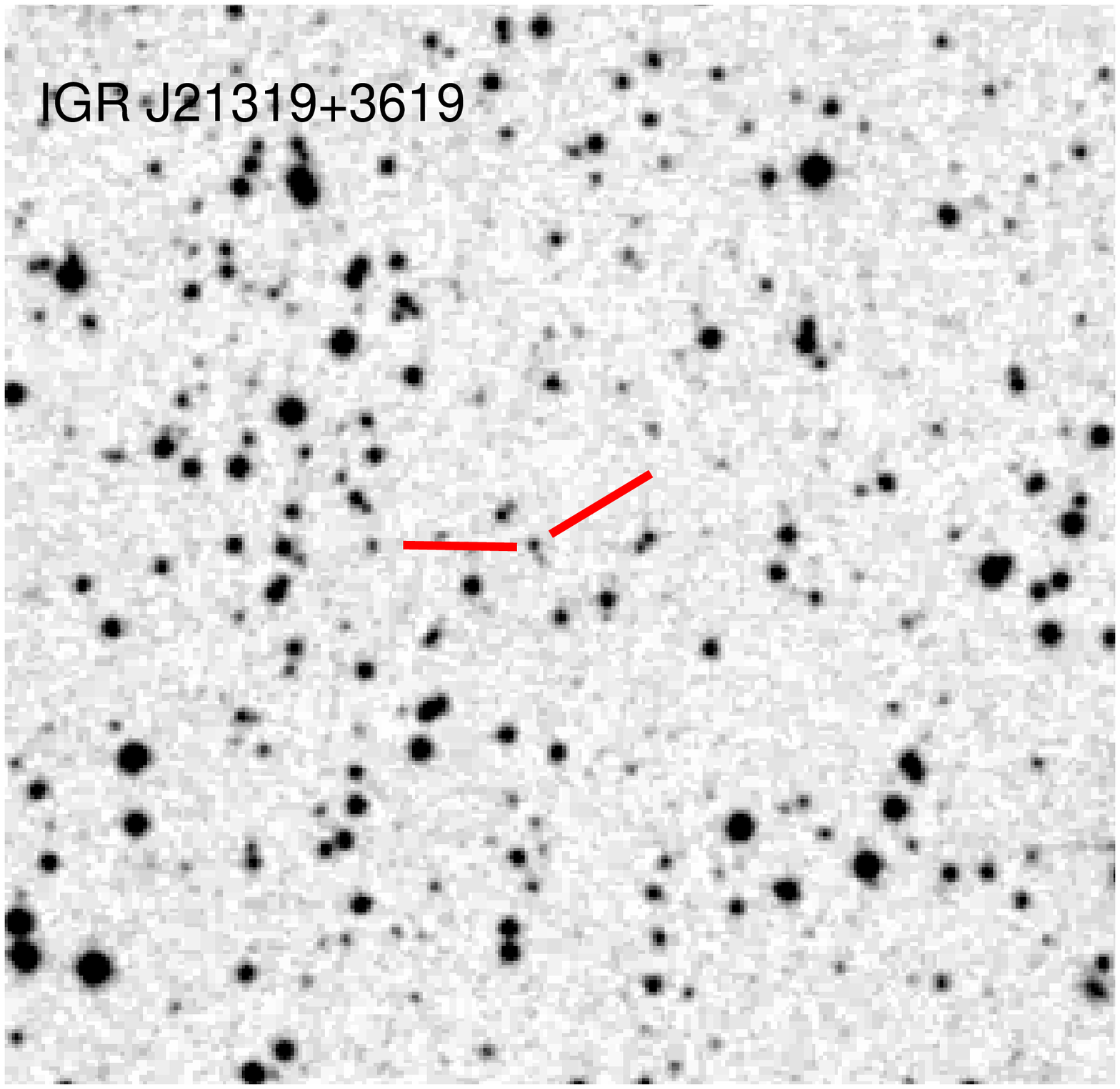,width=5.9cm}}}
\centering{\mbox{\psfig{file=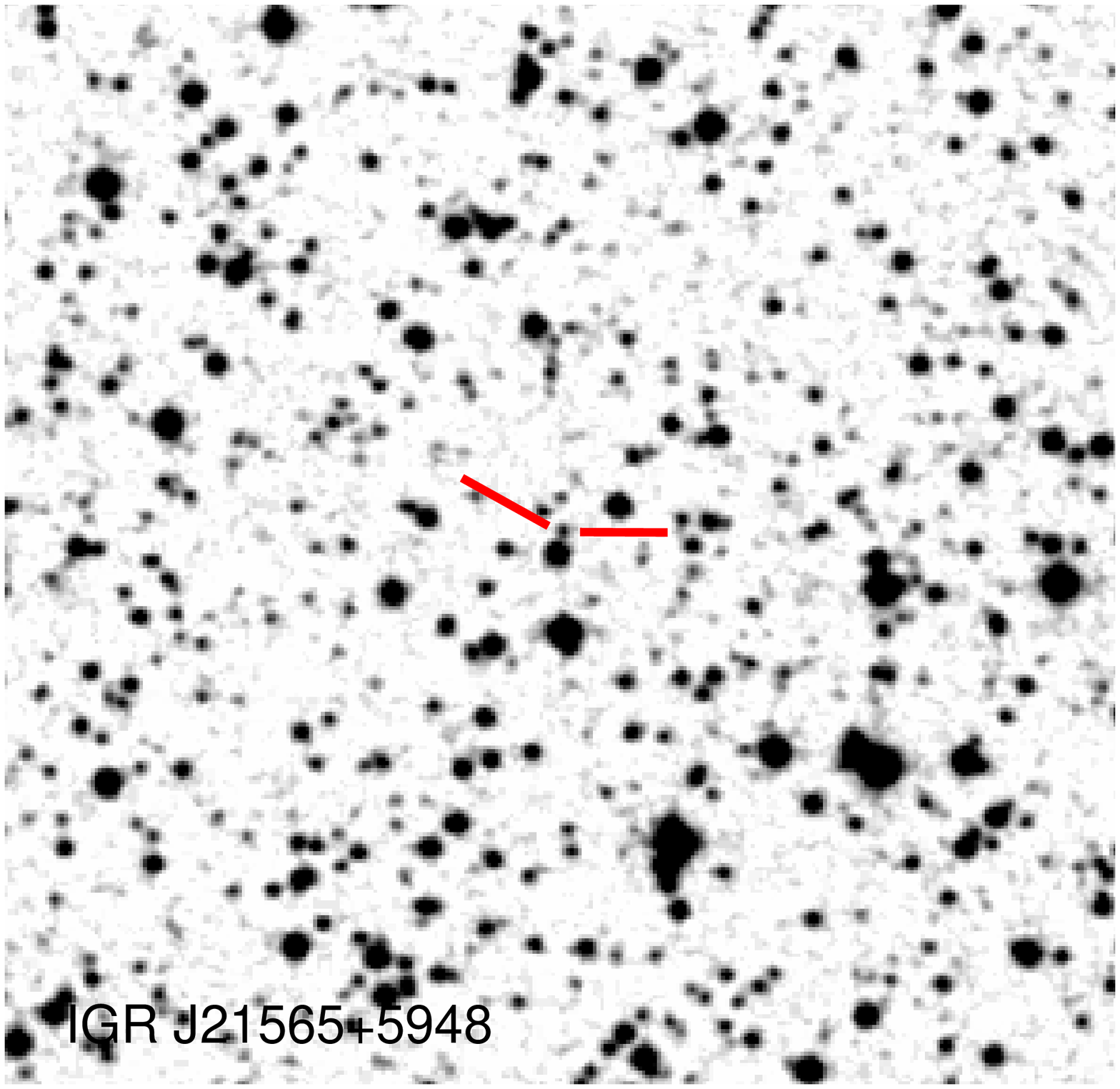,width=5.9cm}}}
\centering{\mbox{\psfig{file=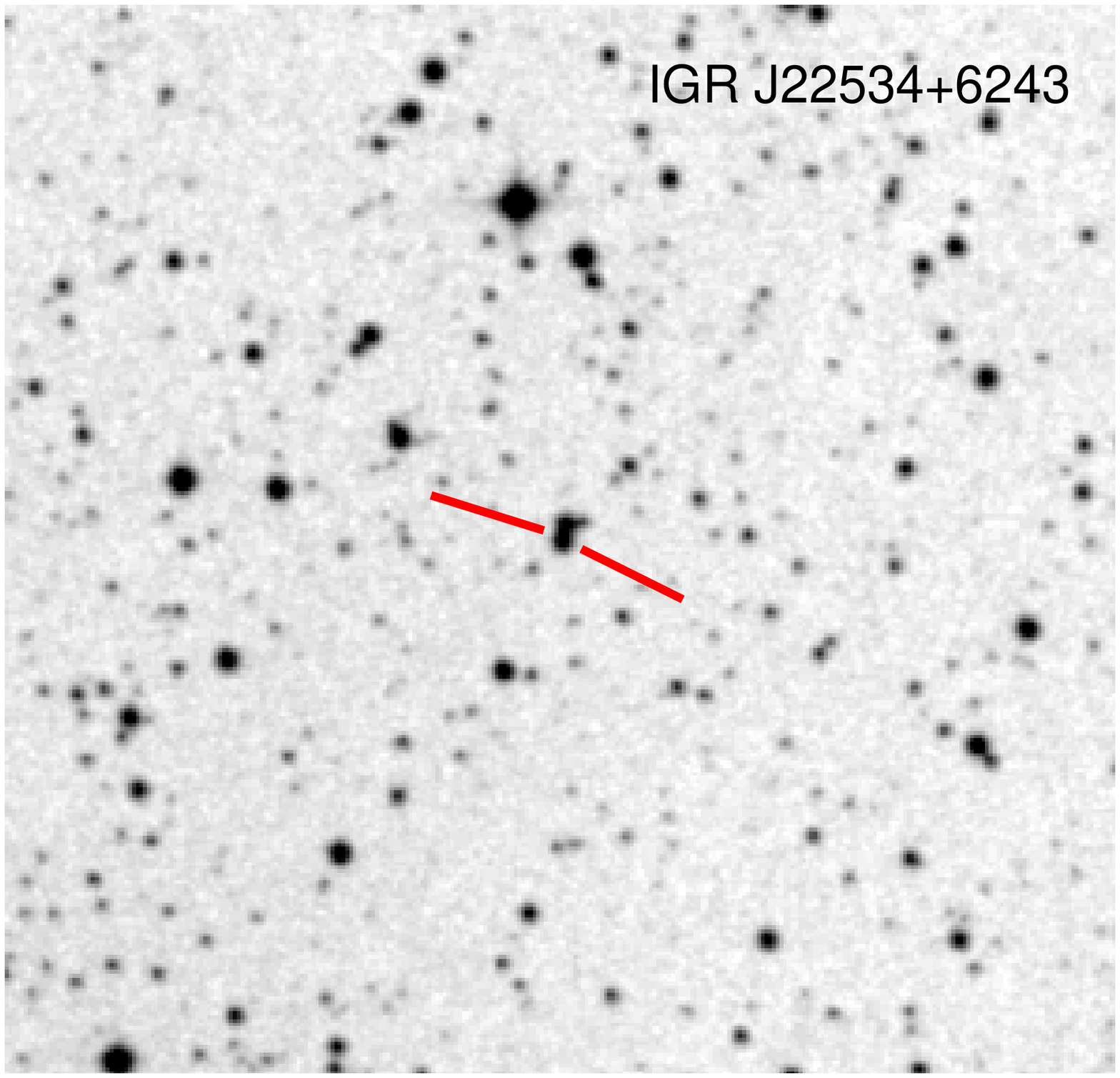,width=5.9cm}}}
\parbox{6cm}{
\psfig{file=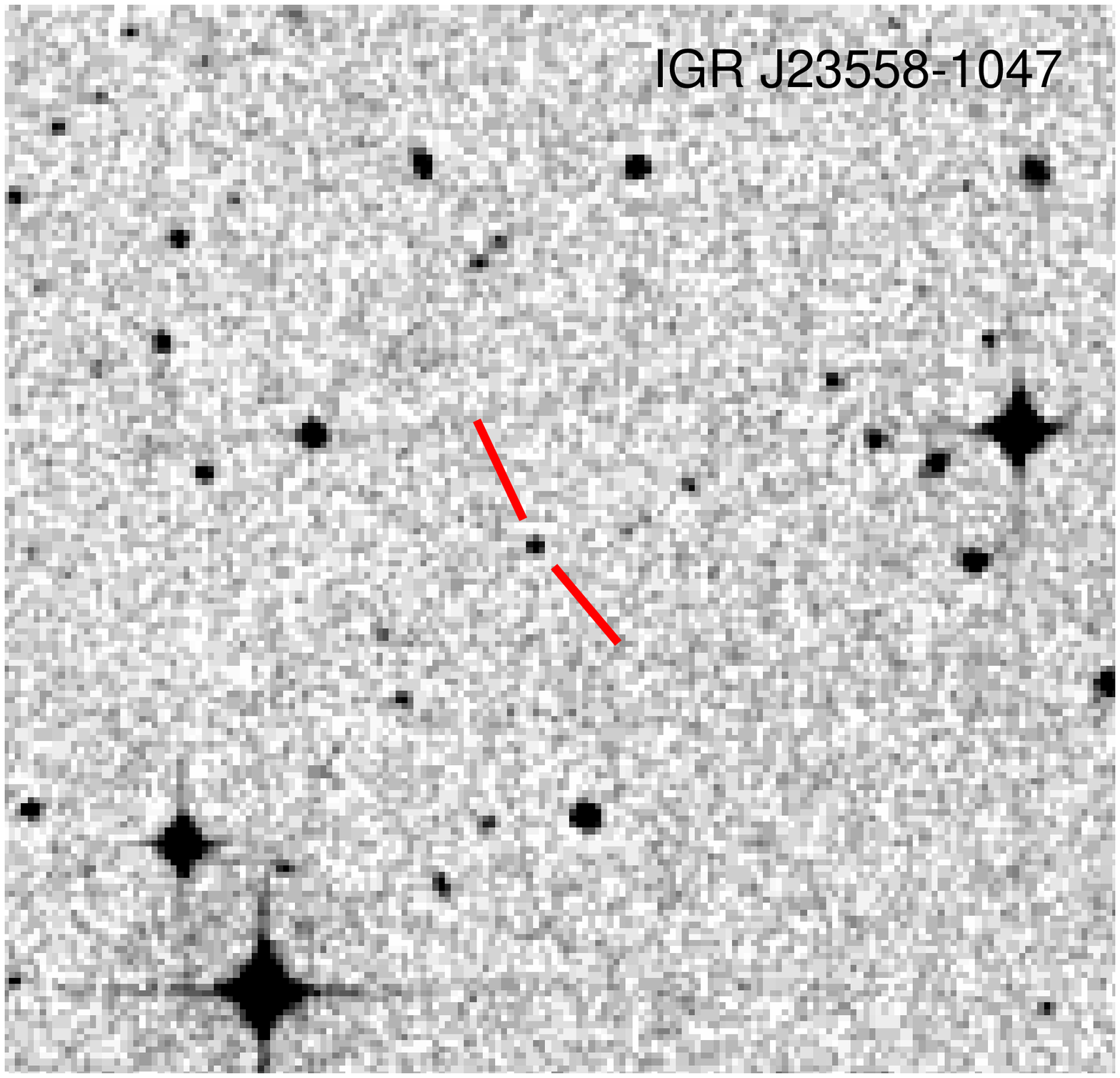,width=5.9cm}
}
\hspace{0.8cm}
\parbox{11cm}{
\vspace{-.8cm}
\hspace{-.3cm}
\caption{Similar to Fig. 1 but for four more {\it INTEGRAL} sources of our
sample (see Table 1). The image of the field of IGR J23558$-$1047 
(bottom left panel) has been extracted from the DSS-I-Blue survey.}}
\end{figure*}

One principal objective of the {\it INTEGRAL} mission (Winkler et al. 
2003) is the systematic survey of the whole sky in the hard X--ray band 
above 20 keV. This is made possible thanks to the unique imaging 
capability of the IBIS instrument (Ubertini et al. 2003) in this spectral 
range, which allows the detection of sources at the mCrab\footnote{1 mCrab 
$\cong$ 2$\times$10$^{-11}$ erg cm$^{-2}$ s$^{-1}$ with a slight 
dependence on the reference energy range.} level with a typical 
localization accuracy of a few arcmin. The fourth and latest IBIS 
catalogue (Bird et al. 2010) contains more than 700 hard X--ray sources 
detected in the 20--100 keV band down to an average flux level of about 1 
mCrab with a positional accuracy that is better than $\sim$5 arcmin. 
Similarly, the 17--60 keV surveys published by Krivonos et al. (2010, 
2012) contain about 500 and 400 sources, respectively (partly overlapping 
between them and with the catalogue of Bird et al. 2010), with sensitivity 
and localization precision comparable to that of Bird et al. (2010).

A non-negligible fraction (for instance, $\sim$30\% in the case of the 
survey of Bird et al. 2010) of these objects had no obvious counterpart at 
other wavelengths and therefore could not be associated with any known 
class of high-energy emitting objects. Multiwavelength observational 
campaigns on these unidentified sources are therefore mandatory to 
pinpoint their nature.

Several approaches using X--ray data analysis have shown their 
effectiveness in identifying the nature of some of these new {\it 
INTEGRAL} sources, such as X--ray timing (e.g., Walter et al. 2006; Sguera 
et al. 2007; Del Santo et al. 2007; La Parola et al. 2010) or spectroscopy 
and imaging (see for instance Rodriguez et al. 2010, Malizia et al. 2010, 
Tomsick et al. 2012, and references therein).

Alternatively and also effectively, cross-correlation with soft X--ray 
catalogues and consequent optical spectroscopy on thereby selected 
candidates allows the determination of the nature and the main 
multiwavelength characteristics of unidentified or poorly studied hard 
X--ray objects. This is especially important, given that very interesting 
classes of objects can be found among hard X--ray emitting sources due to 
the high power of penetration of this radiation through dust and hydrogen 
clouds.

For instance, this approach is vital for the search for nearby 
Compton-thick active galactic nuclei (AGNs) with a line-of-sight hydrogen 
column N$_{\rm H} >$ 10$^{24}$ cm$^{-2}$. Because of the strong absorption 
in their host galaxies, their activity is depressed in the optical 
waveband; as a consequence, heavily absorbed AGNs and Compton thick 
objects are often found among the faintest optical counterparts of {\it 
INTEGRAL} sources. Finding even a few of these highly absorbed active 
galaxies, especially in the local Universe, is of great importance to 
estimate their fraction among the hard X--ray AGN population (e.g., 
Malizia et al. 2009).

The same effect is seen in Galactic objects since intrinsic absorption
characterizes some of the new types of hard X--ray binaries detected by
{\it INTEGRAL} (see e.g. Sguera et al. 2006).

All the above means that a fraction of these objects can substantially be 
absorbed which thus appear relatively faint in the optical. Therefore, 
follow-up with medium-sized telescopes (of 4m-class, for instance) is 
mandatory to properly identify and study the longer-wavelength 
counterparts of these hard X--ray sources.

The use of this class of telescopes also allows us to study the faint end 
of the distribution of high-energy, high-$z$ AGNs lying at a mean distance 
$\sim$5 times larger than that of the average of this type of objects 
known up to now (see Masetti et al. 2012a). It is even possible that these 
objects are indeed different from those at lower redshifts in the sense 
that they belong to a subclass of extreme blazars that emit the bulk of 
their power in the hard X--ray band (Ghisellini et al. 2011).

Here, we carry on the work of identifying {\it INTEGRAL} sources which we 
started in 2004 and which permitted us to identify more than 170 sources 
up to now by means of optical spectroscopy (see Masetti et al. 2004, 
2006abcd, 2008a, 2009, 2010, 2012a [hereafter Papers I-IX, respectively]; 
Masetti et al. 2007, 2008b; Maiorano et al. 2011). The optical spectra of 
the firm or likely counterparts of 30 unidentified, unclassified, or 
poorly studied sources belonging to one or more IBIS surveys (Bird et al. 
2010; Coe et al. 2010; Krivonos et al. 2010, 2012; Bottacini et al. 2012; 
Grebenev et al. 2013) are shown in this paper. We moreover added to our 
sample the spectra of two {\it INTEGRAL} sources reported in Ricci et al. 
(2012) and one in Townsend et al. (2011), thus reaching a total of 33 
objects hereby explored. In this work, we also take the opportunity to 
correct the redshift of source IGR J16388+3557 reported in Paper IX. 
Optical spectroscopy for this sample of sources was acquired using six 
different telescopes and one public spectroscopic archive; among these 
facilities, the use of medium-sized telescopes allows us to pursue the 
exploration of the issues outlined above.

The paper is structured as follows. In Sect. 2, we outline the approach we 
used to choose the sample of {\it INTEGRAL} and optical objects considered 
in this work. In Sect. 3, a description of the observations is given. 
Section 4 shows and discusses the results with an update of the statistics 
on the identifications of {\it INTEGRAL} sources available until now. 
Conclusions are reported in Sect. 5.

We recall that the main results presented here and the information 
concerning the {\it INTEGRAL} sources that have been identified up to now 
(by us or by other groups) using optical or near-infrared (NIR) 
observations are listed in a web page\footnote{{\tt 
http://www.iasfbo.inaf.it/extras/IGR/main.html}} that we maintain as a 
service to the scientific community (Masetti \& Schiavone 2008). Unless 
otherwise stated, errors and limits in this paper are reported at 
1$\sigma$ and 3$\sigma$ confidence levels, respectively. Moreover, this 
work supersedes the optical results presented in the preliminary analyses 
of Masetti et al. (2012b) and Parisi et al. (2012).

\section{Sample selection}

\begin{table*}[th!]
\caption[]{Log of the spectroscopic observations presented in this paper
(see text for details).}
\scriptsize
\begin{center}
\begin{tabular}{llllcccr}
\noalign{\smallskip}
\hline
\hline
\noalign{\smallskip}
\multicolumn{1}{c}{{\it (1)}} & \multicolumn{1}{c}{{\it (2)}} & \multicolumn{1}{c}{{\it (3)}} & \multicolumn{1}{c}{{\it (4)}} & 
{\it (5)} & {\it (6)} & {\it (7)} & \multicolumn{1}{c}{{\it (8)}} \\
\multicolumn{1}{c}{Object} & \multicolumn{1}{c}{RA} & \multicolumn{1}{c}{Dec} & 
\multicolumn{1}{c}{Telescope+instrument} & $\lambda$ range & Disp. & \multicolumn{1}{c}{UT Date \& Time}  & \multicolumn{1}{c}{Exposure} \\
 & \multicolumn{1}{c}{(J2000)} & \multicolumn{1}{c}{(J2000)} & & (\AA) & (\AA/pix) & 
\multicolumn{1}{c}{at mid-exposure} & \multicolumn{1}{c}{time (s)}  \\

\noalign{\smallskip}
\hline
\noalign{\smallskip}

IGR J00515$-$7328              & 00:52:00.59 & $-$73:29:25.5 & Radcliffe+Gr. Spec. & 3750-7750  & 2.3 & 18 Nov 2011, 19:17 & 2$\times$1800 \\ 
RX J0101.8$-$7223              & 01:01:52.28 & $-$72:23:34.1 & CTIO 1.5m+RC Spec.  & 3300-10500 & 5.7 & 24 Sep 2011, 06:38 & 2$\times$1200 \\ 
IGR J02045$-$1156              & 02:04:36.75 & $-$11:59:43.4 & AAT+6dF             & 3900-7600  & 1.6 & 05 Dec 2002, 10:52 & 1200+600 \\ 
IGR J02115$-$4407              & 02:11:43.83$^*$ & $-$44:07:01.0$^*$ & CTIO 1.5m+RC Spec. & 3300-10500 & 5.7 & 29 Dec 2010, 03:48 & 3$\times$1800 \\ 
IGR J02447+7046$\ddagger$      & 02:43:43.05 &   +70:50:38.5 & TNG+DOLoReS         & 3700-8000  & 2.5 & 14 Sep 2012, 02:48 & 1800 \\ 
Swift J0250.2+4650             & 02:50:27.18 &   +46:47:29.4 & SPM 2.1m+B\&C Spec. & 3300-7900  & 4.0 & 04 Dec 2008, 04:37 & 2$\times$1800 \\ 
IGR J02574$-$0303              & 02:57:22.10 & $-$03:06:22.4 & SPM 2.1m+B\&C Spec. & 3300-7900  & 4.0 & 04 Dec 2012, 05:37 & 2$\times$1800 \\ 
IGR J03564+6242                & 03:55:41.29 &   +62:40:56.7 & TNG+DOLoReS         & 3700-8000  & 2.5 & 16 Sep 2012, 02:35 & 1200 \\ 
IGR J05048$-$7340              & 05:05:06.26 & $-$73:39:04.3 & CTIO 1.5m+RC Spec.  & 3300-10500 & 5.7 & 29 Dec 2010, 06:49 & 2$\times$1000 \\ 
IGR J05470+5034                & 05:47:14.88 &   +50:38:25.5 & Copernicus+AFOSC    & 3500-7800  & 4.2 & 01 Mar 2011, 21:50 & 2$\times$1800 \\ 
IGR J06293$-$1359              & 06:29:20.22$^*$ & $-$13:55:23.9$^*$ & TNG+DOLoReS & 3700-8000  & 2.5 & 26 Oct 2011, 05:21 & 3$\times$1800 \\ 
IGR J09034+5329                & 09:03:05.51$^*$ &   +53:30:32.5$^*$ & TNG+DOLoReS & 3700-8000  & 2.5 & 30 Oct 2011, 02:58 & 2$\times$1500 \\ 
Swift J0958.0$-$4208           & 09:57:50.64 & $-$42:08:35.5 & CTIO 1.5m+RC Spec.  & 3300-10500 & 5.7 & 06 Dec 2010, 07:40 & 2$\times$1500 \\ 
IGR J10200$-$1436              & 10:19:37.29$^*$ & $-$14:41:28.3$^*$ & TNG+DOLoReS & 3700-8000  & 2.5 & 05 Dec 2011, 04:08 & 2$\times$1200 \\ 
IGR J14257$-$6117              & 14:25:07.58 & $-$61:18:57.8 & CTIO 1.5m+RC Spec.  & 3300-10500 & 5.7 & 26 Jan 2012, 08:10 & 3$\times$1800 \\ 
IGR J14488$-$4008              & 14:48:50.97 & $-$40:08:45.6 & CTIO 1.5m+RC Spec.  & 3300-10500 & 5.7 & 02 Feb 2012, 07:13 & 2$\times$1000 \\ 
IGR J15415$-$5029              & 15:41:26.43 & $-$50:28:23.3 & CTIO 1.5m+RC Spec.  & 3300-10500 & 5.7 & 21 Mar 2012, 07:40 & 2$\times$1200 \\ 
IGR J16058$-$7253:             & & & & & & & \\
~~~~~~~LEDA 259433             & 16:05:23.24 & $-$72:53:56.2 & CTIO 1.5m+RC Spec.  & 3300-10500 & 5.7 & 15 May 2012, 07:56 & 2$\times$1800 \\ 
~~~~~~~LEDA 259580             & 16:06:06.88 & $-$72:52:41.9 & CTIO 1.5m+RC Spec.  & 3300-10500 & 5.7 & 15 May 2012, 09:00 & 2$\times$1500 \\ 
IGR J17014$-$4306              & 17:01:28.15 & $-$43:06:12.3 & CTIO 1.5m+RC Spec.  & 3300-10500 & 5.7 & 02 Oct 2011, 00:21 & 2$\times$1000 \\ 
IGR J17198$-$3020              & 17:19:48.65 & $-$30:17:26.9 & SPM 2.1m+B\&C Spec. & 3300-7900  & 4.0 & 23 Jun 2012, 06:00 & 1800          \\ 
IGR J17476$-$2253              & 17:47:29.80$^{**}$ & $-$22:52:46.6$^{**}$ & TNG+DOLoReS         & 3700-8000  & 2.5 & 11 Jul 2012, 02:45 & 2$\times$1800 \\ 
IGR J17488$-$2338              & 17:48:39.05$^{**}$ & $-$23:35:21.2$^{**}$ & TNG+DOLoReS         & 3700-8000  & 2.5 & 25 Aug 2012, 01:39 & 2$\times$1800 \\ 
IGR J17520$-$6018              & 17:51:55.80 & $-$60:19:43.2 & CTIO 1.5m+RC Spec.  & 3300-10500 & 5.7 & 24 Sep 2011, 01:36 & 2$\times$1000 \\ 
IGR J18151$-$1052              & 18:15:03.8$^\dagger$ & $-$10:51:35$\dagger$ & TNG+DOLoReS & 3700-8000  & 2.5 & 02 Aug 2011, 22:00 & 2$\times$1500 \\ 
IGR J18256$-$1035              & 18:25:43.83$^{**}$ & $-$10:35:01.3$^{**}$ & TNG+DOLoReS   & 3700-8000  & 2.5 & 21 Ago 2011, 23:20 & 2$\times$1800 \\ 
AX J1832.3$-$0840              & 18:32:19.30$^*$ & $-$08:40:29.8$^*$ & SPM 2.1m+B\&C Spec. & 3300-7900  & 4.0 & 24 Jun 2012, 08:19 & 2$\times$1800 \\ 
2E 1923.7+5037                 & 19:25:02.17 &   +50:43:13.8 & Cassini+BFOSC       & 3500-8700  & 4.0 & 21 Aug 2012, 21:10 & 2$\times$1800 \\ 
IGR J19295$-$0919$\ddagger$    & 19:29:04.02$^{**}$ & $-$09:13:42.9$^{**}$ & TNG+DOLoReS   & 3700-8000  & 2.5 & 02 Sep 2011, 01:41 & 1800 \\  
IGR J21319+3619$\ddagger$      & 21:31:27.37$^*$ & +36:16:50.2$^*$ & TNG+DOLoReS   & 3700-8000  & 2.5 & 18 Aug 2012, 03:15 & 2$\times$1800 \\ 
IGR J21565+5948$\ddagger$      & 21:56:04.18 &   +59:56:04.5 & TNG+DOLoReS         & 3700-8000  & 2.5 & 03 Aug 2011, 03:29 & 2$\times$1500 \\ 
IGR J22534+6243                & 22:53:55.12 &   +62:43:36.8 & SPM 2.1m+B\&C Spec. & 3300-7900  & 4.0 & 22 Jun 2012, 11:07 & 2$\times$900  \\ 
IGR J23558$-$1047              & 23:55:58.82$^*$ & $-$10:46:44.8$^*$ & SPM 2.1m+B\&C Spec. & 3300-7900  & 4.0 & 26 Sep 2011, 07:23 & 2$\times$1200 \\ 

\noalign{\smallskip}
\hline
\noalign{\smallskip}
\multicolumn{8}{l}{Note: if not indicated otherwise, source coordinates were extracted from the 
2MASS catalogue and have an accuracy better than 0$\farcs$1.}\\
\multicolumn{8}{l}{$^\ddagger$: sources for which the 99\% IBIS error circle was considered to 
search for a soft X--ray counterpart.}\\
\multicolumn{8}{l}{$^*$: coordinates extracted from the USNO catalogues, having 
an accuracy of about 0$\farcs$2 (Deutsch 1999; Assafin et al. 2001; Monet et al. 2003).}\\
\multicolumn{8}{l}{$^{**}$: coordinates computed through astrometry calibrated using the USNO 
catalogues on a previously acquired image of the source field (see text).}\\
\multicolumn{8}{l}{$^\dagger$: coordinates extracted from the DSS-II-Red frames having 
an accuracy of $\sim$1$''$.}\\

\noalign{\smallskip}
\hline
\hline
\noalign{\smallskip}
\end{tabular}
\end{center}
\end{table*}

Using the same approach as in our previous Papers I-IX, we considered the 
IBIS surveys of Bird et al. (2010), Krivonos et al. (2010, 2012), Coe et 
al. (2010), Grebenev et al. (2013), and the {\it Swift}/{\it INTEGRAL} 
(``SIX") joint catalogue of Bottacini et al. (2012). Among them, we 
selected unidentified or unclassified hard X--ray sources containing a 
single bright soft X--ray object within the IBIS 90\% confidence level 
error box. The latter information was obtained from either the {\it ROSAT} 
all-sky bright source catalogue (Voges et al. 1999), {\it Swift}/XRT 
pointings (from Krivonos et al. 2009, Rodriguez et al. 2009, Maiorano et 
al. 2010, Sturm et al. 2011, Kennea 2011, Malizia et al. 2011, Landi et 
al. 2012a, Luna et al. 2012, Molina et al. 2012a, Parisi et al. 2012, and 
from the XRT archive\footnote{XRT archival data are freely available at \\ 
{\tt http://www.asdc.asi.it/}}), or {\it Chandra} observations (Kaur et 
al. 2010; Tomsick et al. 2008). As stressed and demonstrated (see Stephen 
et al. 2006 and Papers I-IX), this procedure is extremely effective in 
substantially reducing the search area and in pinpointing a putative 
optical counterpart within the corresponding (sub)arcsecond-sized soft 
X--ray error box.

To increase the positional accuracy of some of the selected sources 
(especially those lying along the Galactic Plane, where the source 
confusion can be an important issue), we also cross-correlated the above 
selected sources with radio catalogues, such as the NVSS (Condon et al. 
1998), SUMSS (Mauch et al. 2003), and MGPS (Murphy et al. 2007) surveys. 
This method allowed the further reduction of the soft X--ray position 
uncertainty down to less than 2$''$ for a few sources, and hence the 
possibility of determining the actual counterpart with a higher degree of 
accuracy.

Among these objects, we then chose those with a single possible optical 
counterpart with magnitude $R \la$ 20 on the DSS-II-Red 
survey\footnote{Available at {\tt http://archive.eso.org/dss/dss}}, so 
that optical spectroscopy could be obtained with a reasonable 
signal-to-noise ratio (S/N) with telescopes having an aperture of at least 
1.5 metres. This allowed us to select 21 IBIS sources.

In four more cases, two soft X--ray sources were found within the 
corresponding IBIS error circle (Landi et al. 2012b, 2013, Molina et al. 
2012b, Tomsick et al. 2012). In these occurrences, we focused on the 
optical object associated with the brighter soft X--ray source, but we 
attempted to get optical spectroscopy on the other putative counterpart as 
well when feasible.

To this sample, we added the object RX J0101.8$-$7223 (Townsend et al. 
2011 and references therein) and the newly-discovered sources IGR 
J02045$-$1156 and IGR J02574$-$0303 (Ricci et al. 2012) because an 
arcsecond-sized soft X--ray position and a likely optical counterpart are 
made available by those authors for them.

We next relaxed the search criterion described at the beginning of 
this Section by considering the IBIS 99\% error circles. This allowed us 
to select four more cases (indicated in Table 1) to be followed up with 
optical spectroscopy (e.g., Landi et al. 2010 and Tomsick et al. 2012). 

Finally, another consistency check concerning the objects from Bird et al. 
(2010) was performed using more recent IBIS maps to see if the selected 
hard X--ray sources are still detected with {\it INTEGRAL}. Indeed, all of 
them were recovered with the possible exceptions of IGR J17198$-$3020 and 
IGR J21319+3619, which however appear to display a high value for the 
associated `bursticity' parametre (see details in Bird et al. 2010), thus 
indicating substantial hard X--ray variability.

In this way, we collected a sample of 32 {\it INTEGRAL} objects with 
possible optical counterparts, which we explored by means of optical 
spectroscopy. Their names and accurate coordinates (to 1$''$ or better; 
see next section) are reported in Table 1, while their optical finding 
charts are shown in Figs. 1-4 with the corresponding putative counterparts 
indicated with tick marks. We did not report the chart that corresponds to 
IGR J18151$-$1052, as it already appears in Lutovinov et al. (2012).

It is stressed that sources IGR J17476$-$2253 (Mescheryakov et al. 2009), 
IGR J18151$-$1052 (Lutovinov et al. 2012) and IGR J21565+5948 (Bikmaev et 
al. 2010) are also included in our final sample. Although already 
identified elsewhere, these objects still have fragmentary 
longer-wavelength information. Our observations are thus presented here to 
confirm their nature and to improve their classification and 
characteristics.

Within this selection, we were also able to determine the actual 
counterpart of source IGR J06293$-$1359, which in Paper VIII was 
tentatively (and, in hindsight, erroneously) identified with a Seyfert 2 
AGN on the basis of an association with a radio object inside the hard 
X--ray error box of this source. This occurred as no soft X--ray 
observations of the field were available at that time. We also refer the 
reader to Paper III concerning the caveats and the shortcomings of 
choosing sources that are not straightforwardly connected with an 
arcsec-sized soft X--ray position within an IBIS uncertainty circle.

For the source naming in Table 1, we directly adopted the names as they 
appear in the relevant works (Bird et al. 2010; Bottacini et al. 2012; 
Ricci et al. 2012; Coe et al. 2010; Grebenev et al. 2013; Krivonos et al. 
2010, 2012), and the IGR alias when available. However, we remark that 
we chose not to use the {\it ROSAT} name as reported in the fourth 
IBIS Survey (Bird et al. 2010) for one of the selected objects (1RXS 
J090320.0+533022), because the soft X--ray emission detected with {\it 
Swift}/XRT within the corresponding 90\% IBIS error circle is not 
positionally consistent with the one reported in the {\it ROSAT} Faint 
Catalogue (actually, no detectable emission appears to be associated with 
this {\it ROSAT} object from the XRT images). Thus, we decided to rename 
this source as IGR J09034+5329 following the usual naming convention for 
sources detected with {\it INTEGRAL}.

\begin{figure*}
\mbox{\psfig{file=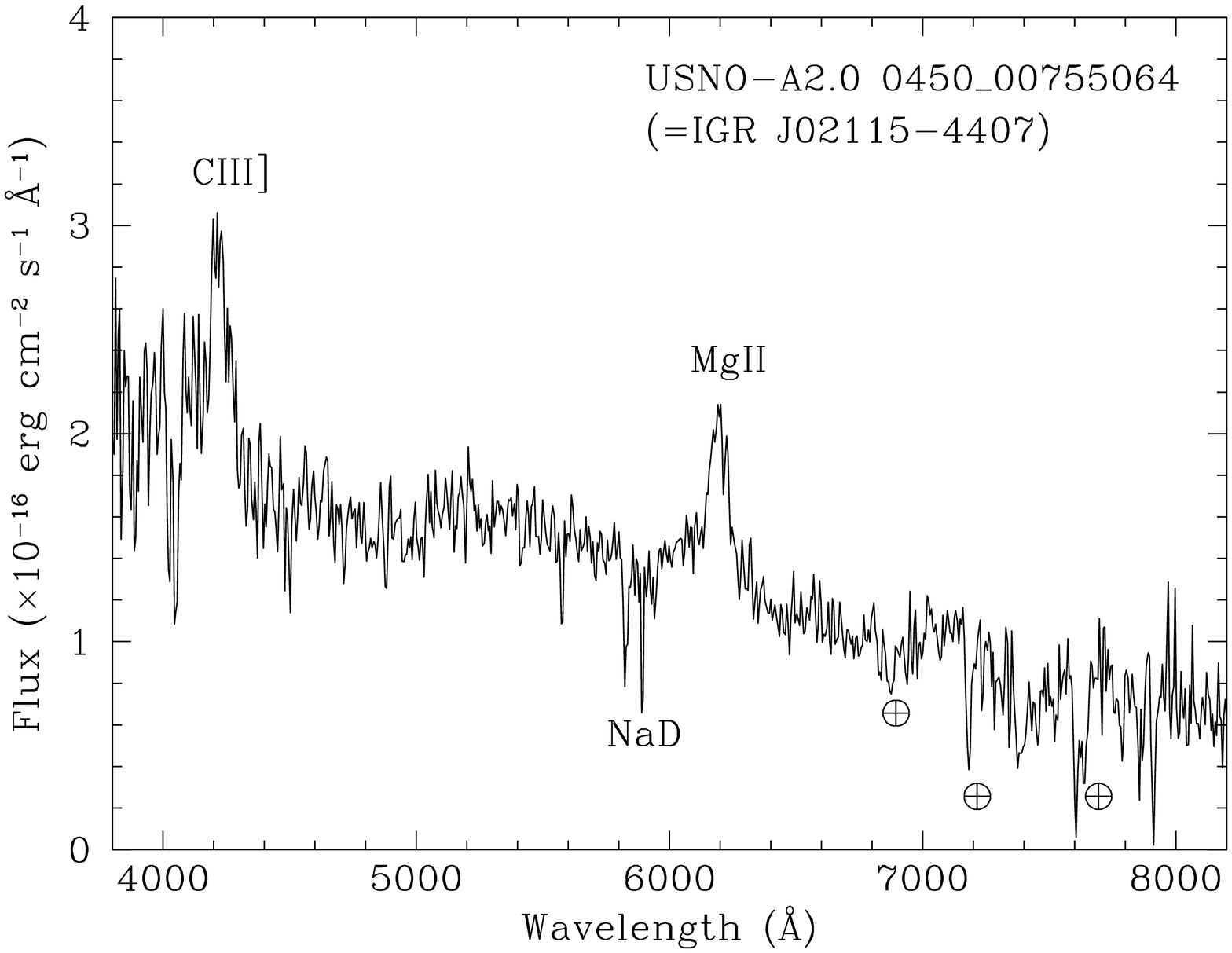,width=9cm,angle=270}}
\mbox{\psfig{file=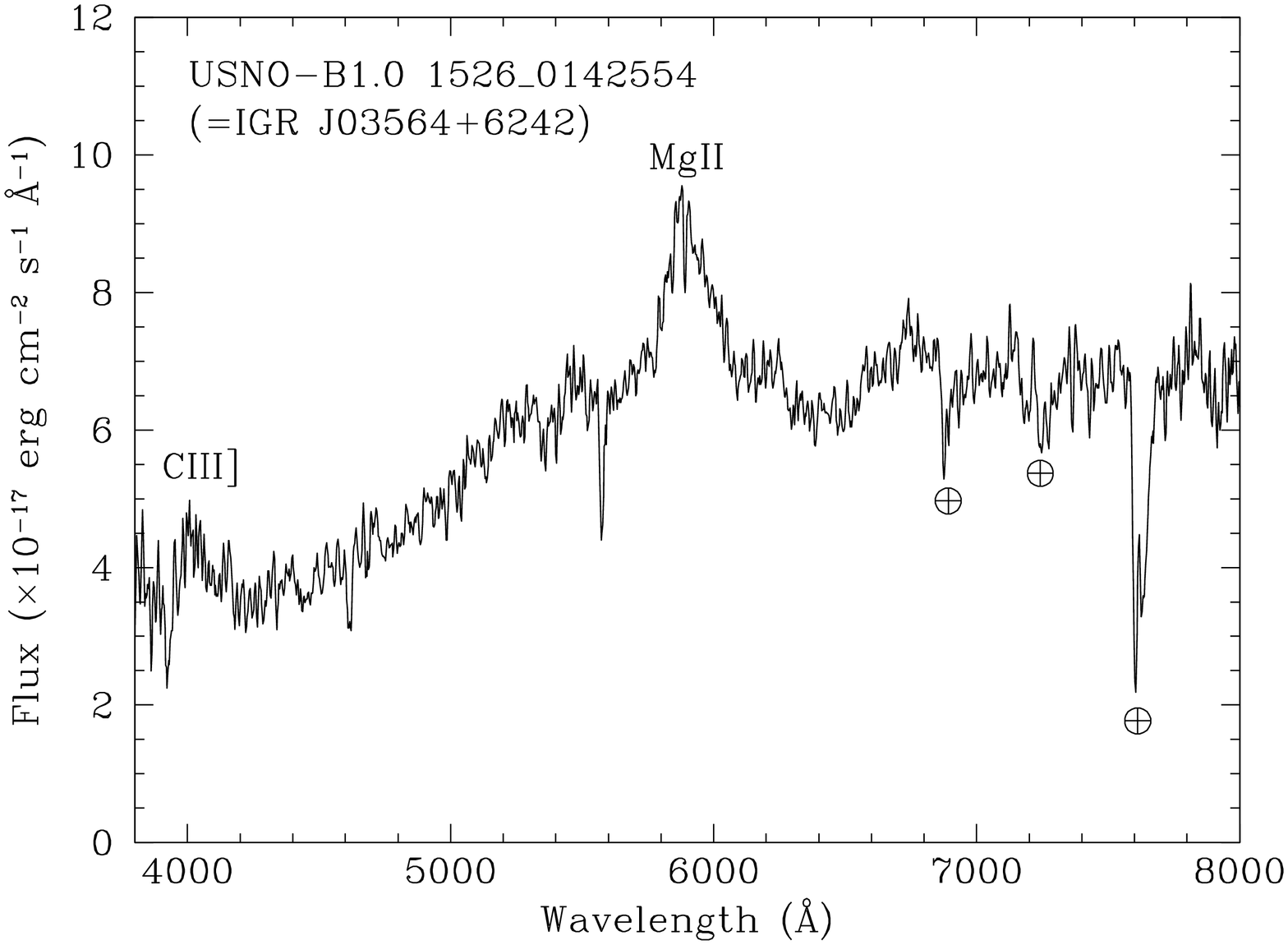,width=9cm,angle=270}}

\vspace{-.9cm}
\mbox{\psfig{file=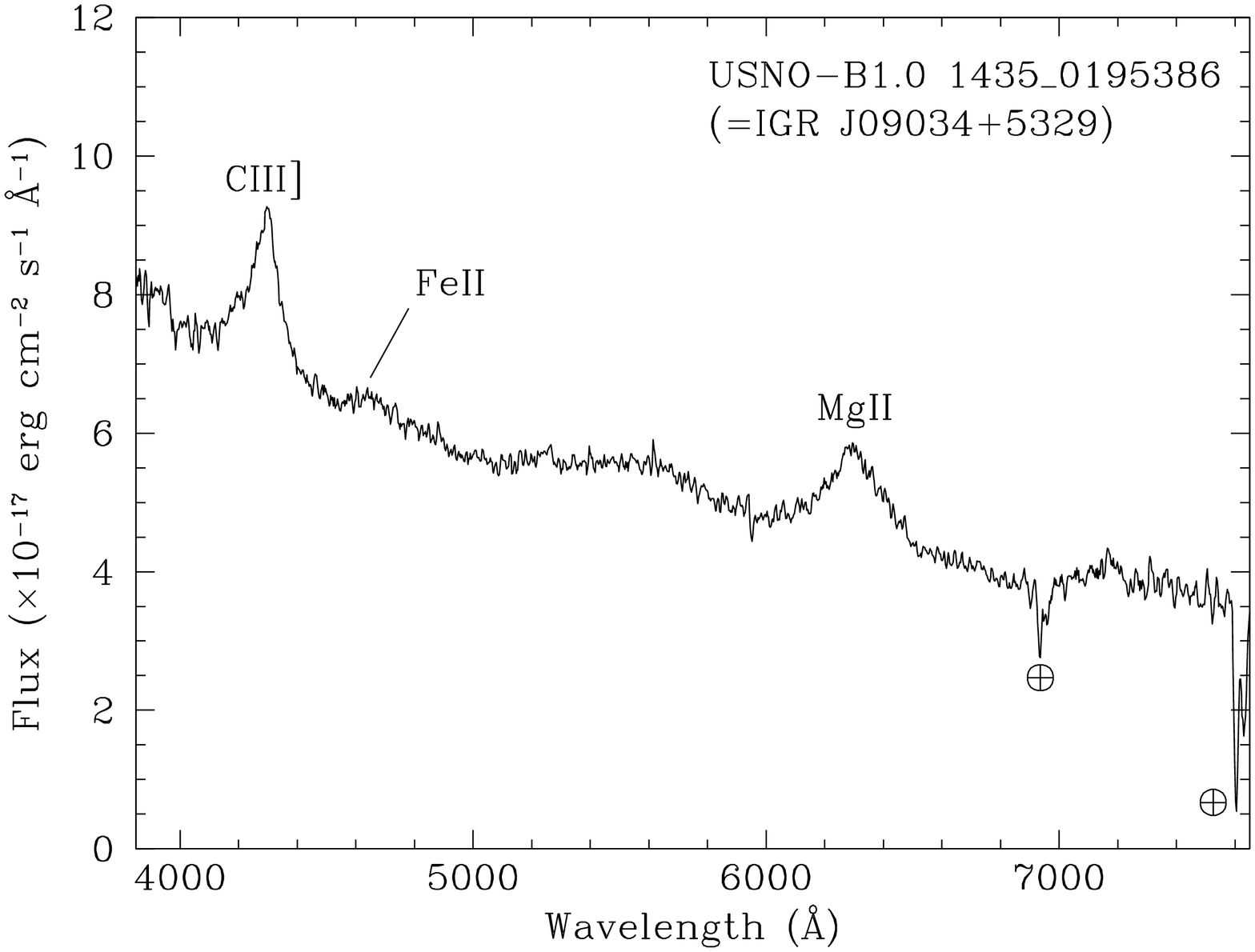,width=9cm,angle=270}}
\mbox{\psfig{file=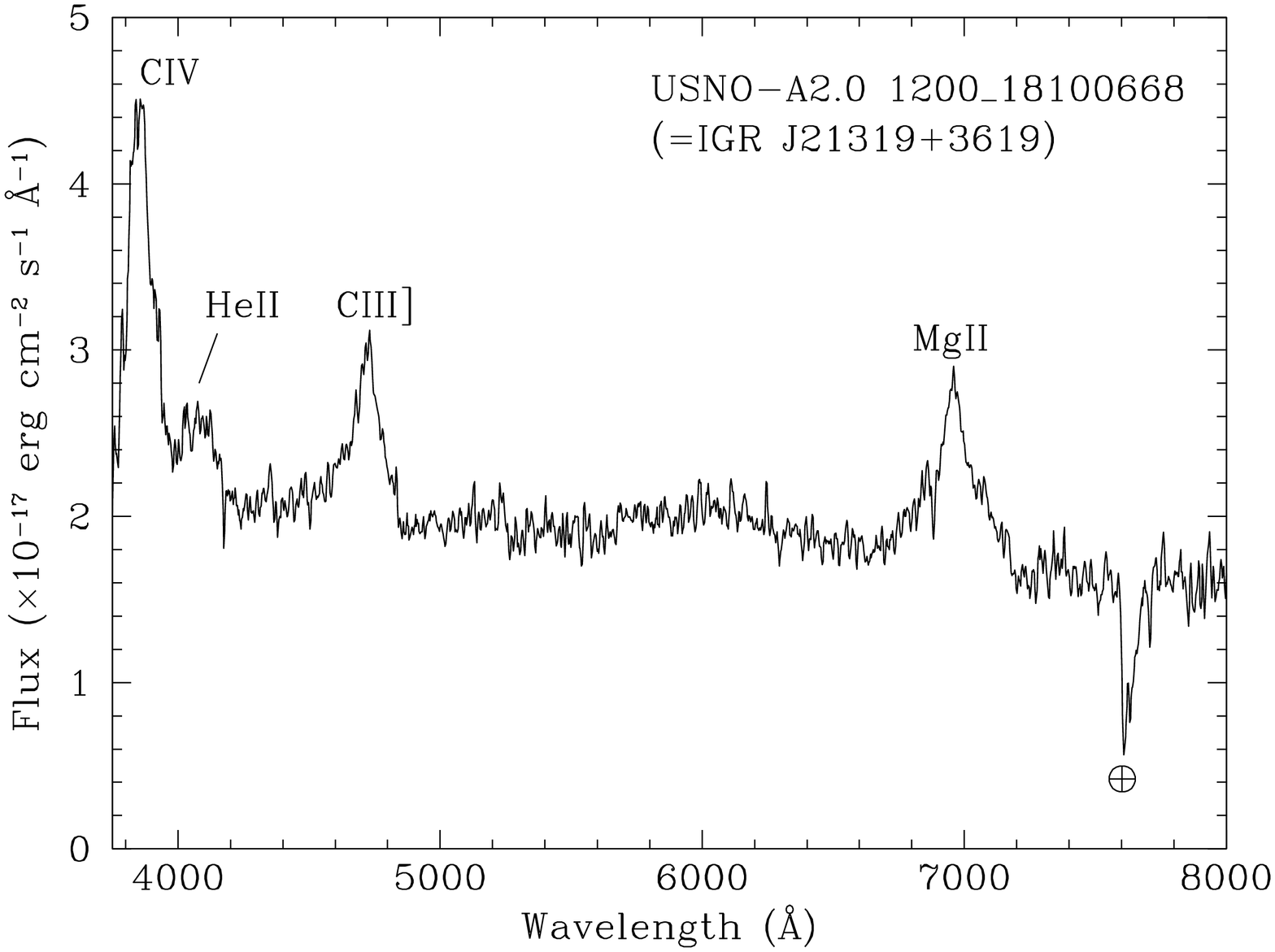,width=9cm,angle=270}}

\vspace{-.9cm}
\mbox{\psfig{file=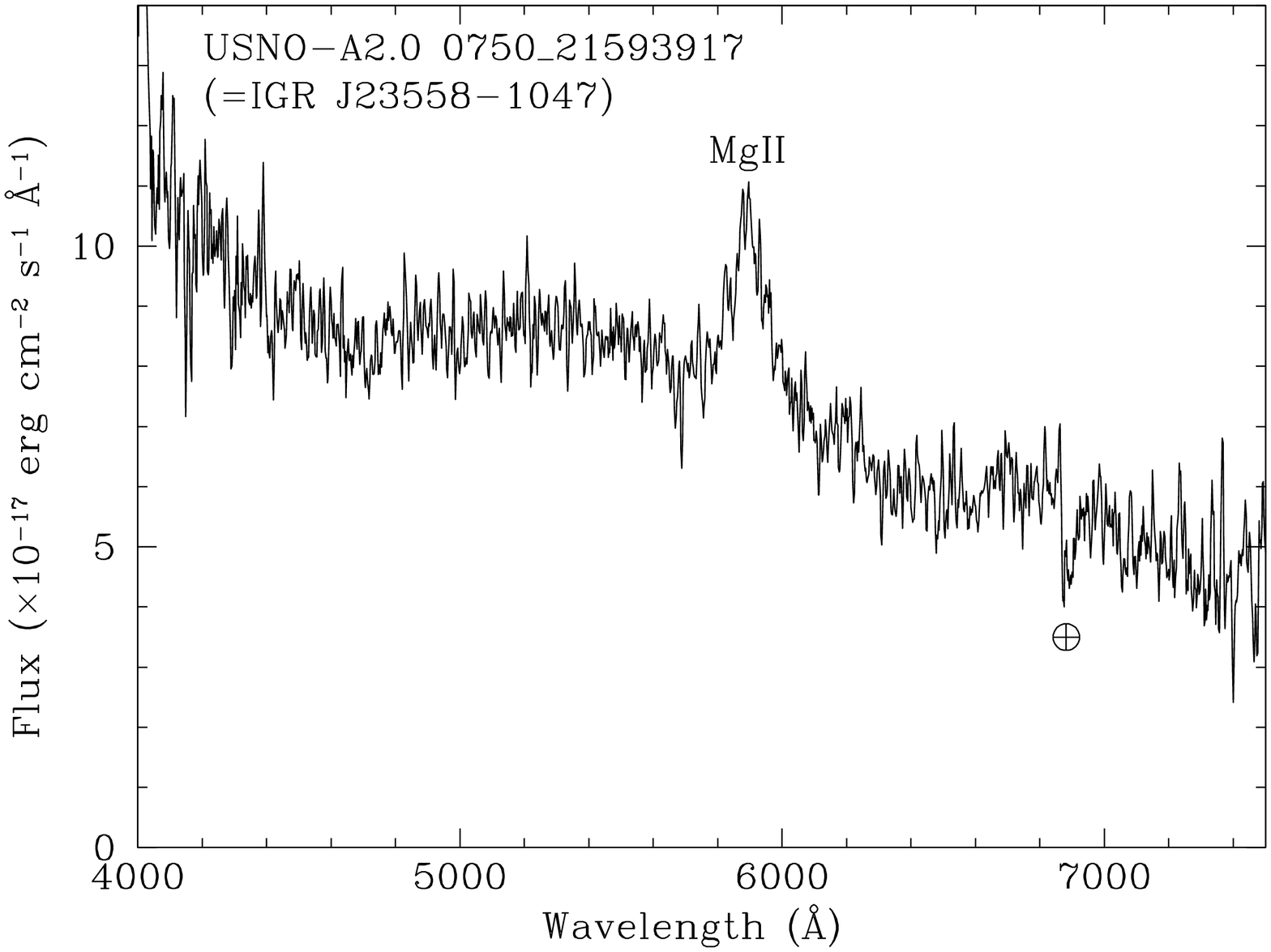,width=9cm,angle=270}}
\mbox{\psfig{file=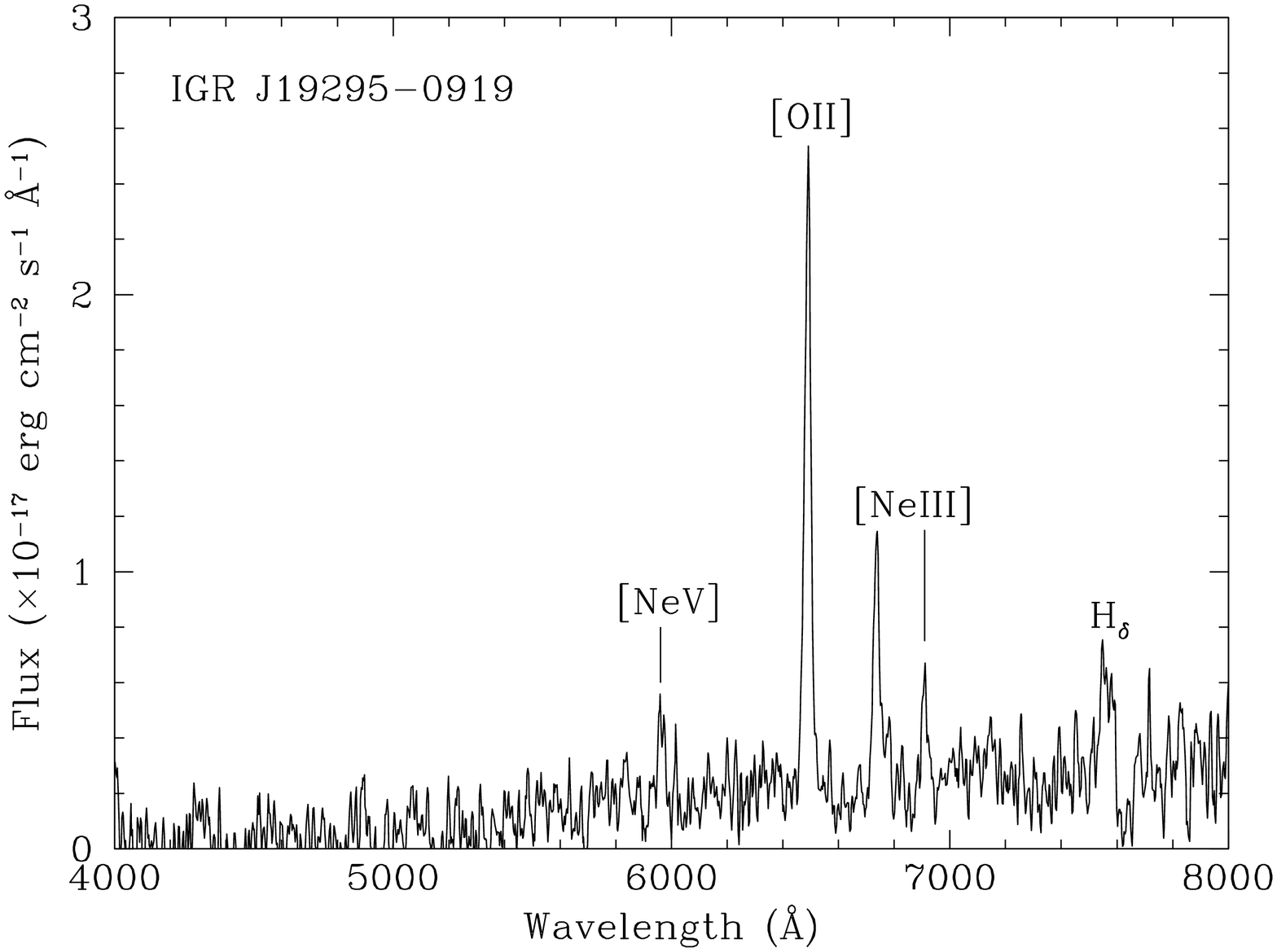,width=9cm,angle=270}}

\caption{Spectra (not corrected for the intervening Galactic absorption) 
of the optical counterparts of the six high-redshift QSOs presented in 
this paper for the first time. For each spectrum, the main spectral 
features are labelled. The symbol $\oplus$ indicates atmospheric telluric 
absorption bands. The TNG and SPM spectra have been smoothed using a 
Gaussian filter with $\sigma$ = 3 \AA. The spectrum of the only Type 
2 QSO present in this sample (IGR J19295$-$0919) is reported in the 
lower right panel of the figure.}
\end{figure*}

\section{Optical observations}

Analogously to Papers VI-IX, almost all of the data presented in this work 
were secured during an observational campaign, which spanned over four 
years (from December 2008 to December 2012) and which involved the use of 
the following telescopes:

\begin{itemize}
\item the 1.5m at the Cerro Tololo Interamerican Observatory (CTIO), Chile;
\item the 1.52m Cassini telescope of the Astronomical Observatory of 
Bologna, in Loiano, Italy; 
\item the 1.82m Copernicus telescope of the Astronomical Observatory of
Asiago, Italy;
\item the 1.9m Radcliffe telescope of the South African Astronomical 
Observatory (Sutherland, South Africa);
\item the 2.1m telescope of the Observatorio Astron\'omico Nacional in San 
Pedro M\'artir (SPM), Mexico;
\item the 3.58m Telescopio Nazionale Galileo (TNG) at the Roque de 
Los Muchachos Observatory in La Palma, Spain.
\end{itemize}

The spectroscopic data acquired at these telescopes have been optimally 
extracted (Horne 1986) and reduced following standard procedures using 
IRAF\footnote{IRAF is the Image Reduction and Analysis Facility made 
available to the astronomical community by the National Optical Astronomy 
Observatories, which are operated by AURA, Inc., under contract with the 
U.S. National Science Foundation. It is available at {\tt 
http://iraf.noao.edu/}}.  Calibration frames (flat fields and bias) were 
taken on the day preceeding or following the observing night. The 
wavelength calibration was performed using lamp data acquired soon after 
each on-target spectroscopic acquisiton; the uncertainty in this 
calibration was $\sim$0.5~\AA~in all cases according to our checks that 
used the positions of background night sky lines. Flux calibration was 
obtained using catalogued spectrophotometric standards. Finally, the data 
of a given object were stacked together to increase the S/N ratio when 
multiple spectra were acquired. 

The spectrum of the optical counterpart of IGR J02045$-$1156 was instead 
retrieved from the Six-degree Field Galaxy Survey\footnote{{\tt 
http://www.aao.gov.au/local/www/6df/}} (6dFGS) archive (Jones et al. 
2004), acquired using the 3.9m Anglo-Australian Telescope of the 
Australian Astronomical Observatory in Siding Spring (Australia). Since 
the 6dFGS archive contains spectra that are not calibrated in flux, we 
considered the optical photometric information in Jones et al. (2005) to 
calibrate the spectrum of this source.

In Table 1, we show a detailed log of all the spectroscopic observations 
presented in this paper. Column 1 indicates the names of the observed {\it 
INTEGRAL} sources. In Cols. 2 and 3, we list the equatorial coordinates of 
the proposed optical counterpart, mostly extracted from the 2MASS (with an 
uncertainty of $\leq$0$\farcs$1: Skrutskie et al. 2006) or the USNO 
catalogues (with accuracies of about 0$\farcs$2: Deutsch 1999; Assafin et 
al. 2001; Monet et al. 2003). The telescope and instrument used for the 
observations are reported in Col. 4, while characteristics of each 
spectrograph are presented in Cols. 5 and 6. Column 7 reports the 
observation date and the UT time at mid-exposure, and Col. 8 provides the 
exposure times and the number of observations for each source.

\begin{table}
\caption{Log of the optical images acquired with TNG+DOLoReS
and used to perform the astrometric calibration of the field 
of selected sources belonging to our sample (see also Fig. 2).}
\scriptsize
\begin{center}
\begin{tabular}{lrrcc}
\noalign{\smallskip}
\hline
\hline
\noalign{\smallskip}
\multicolumn{1}{c}{Object} & \multicolumn{1}{c}{Date and time (UT)} & 
\multicolumn{1}{c}{$T_{\rm exp}$ (s)} & Filter & Seeing \\
\noalign{\smallskip}
\hline
\noalign{\smallskip}

IGR J17476$-$2253 & 21 Jun 2012, 03:15 & 300 & $R$  & 0$\farcs$9 \\
IGR J17488$-$2338 & 01 Sep 2010, 20:55 &   5 & open & 1$\farcs$2 \\
IGR J18256$-$1035 & 21 Aug 2011, 22:46 &   5 & open & 0$\farcs$9 \\
IGR J19295$-$0919 & 22 Aug 2011, 00:00 &  10 & open & 1$\farcs$0 \\

\noalign{\smallskip}
\hline
\hline
\noalign{\smallskip}
\end{tabular}
\end{center}
\end{table}

Before engaging in spectroscopic observations, for four cases (see Table 
2) we acquired a deeper and higher resolution image of the corresponding 
field with respect to that available from the DSS-II-Red survey. All 
images were acquired with TNG+DOLoReS, which covers a field of 
8$\farcm$6$\times$8$\farcm$6 with a scale of 0$\farcs$252 pix$^{-1}$. 
These were then processed to obtain an astrometric solution based on 
USNO-A2.0\footnote{available at: \\ {\tt 
http://archive.eso.org/skycat/servers/usnoa}} reference stars in each 
field. The conservative error in the optical positions can be assumed as 
0$\farcs$252, which was then added in quadrature to the systematic error 
in the USNO-A2.0 catalogue. For these images, the final 1$\sigma$ 
uncertainty in the astrometric solution is thus 0$\farcs$35: this can be 
considered the error in the coordinates of the optical counterparts of the 
{\it INTEGRAL} sources listed in Table 2 (see also Table 1).

With the exception of the field of IGR J17476$-$2253, the above images 
were acquired in white light (that is, with no filter), so no meaningful 
determination of the magnitude of the corresponding optical counterpart 
could be obtained. For the case of IGR J17476$-$2253, we attempted an 
estimate of the optical counterpart $R$-band magnitude by using the 
photometry zero-points available for DOLoReS\footnote{see Table 2 at \\ 
{\tt http://www.tng.iac.es/instruments/lrs/}}. This approach was chosen 
given that no reliable calibrator was available in the image, as all 
USNO-A2.0 stars with a catalogued $R$-band magnitude and close to the 
position of the optical counterpart of this {\it INTEGRAL} source were 
saturated in the TNG acquisition image. Despite the crowdedness of 
the field (see Fig. 3, upper right panel), we used simple aperture 
photometry because the object is apparently extended, so the application 
of the standard point spread function fitting technique was not viable.

\section{Results}

\begin{figure}
\psfig{file=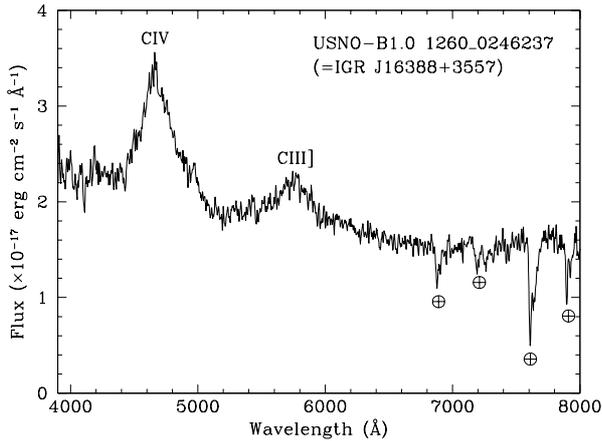,width=9cm,angle=270}
\caption{Spectrum (not corrected for the intervening Galactic 
absorption) of the Type 1 QSO IGR J16388+3557 with the correct emission 
line identifications, which allow us to determine a redshift $z$ = 
2.020$\pm$0.005 for the source. The symbol $\oplus$ indicates atmospheric 
telluric absorption bands. The spectrum has been smoothed using a Gaussian 
filter with $\sigma$ = 3 \AA.}
\end{figure}

A full description of the adopted identification and classification 
criteria for the optical spectra of the selected sources can be found in 
our previous papers (I-IX); the reader is thus referred to them for 
details. The optical magnitudes quoted below, when not stated otherwise, 
are extracted from the USNO-A2.0 catalogue. We briefly recall here that 
the main criteria used for the classification of AGN spectra are from 
Veilleux \& Osterbrock (1987), Winkler (1992), Ho et al. (1993, 1997) and 
Kauffmann et al. (2003).

To estimate the reddening along the line of sight for the Galactic 
sources in our sample when possible and applicable, we assumed an 
intrinsic H$_\alpha$/H$_\beta$ line ratio of 2.86 (Osterbrock 1989) and 
inferred the corresponding color excess by comparing the intrinsic line 
ratio with the measured one by applying the Galactic extinction law of 
Cardelli et al. (1989) and the total-to-selective extinction ratio of 
Rieke \& Lebofsky (1985).

To evaluate the distances of the compact Galactic X--ray sources of our 
sample, we assumed an absolute magnitude M$_V \sim$ +9 and an intrinsic 
color index $(V-R)_0 \sim$ 0 mag (Warner 1995) for cataclysmic variables 
(CVs), whereas we used the intrinsic stellar color indices and absolute 
magnitudes from Lang (1992) and Wegner (1994) for high-mass X--ray 
binaries (HMXBs). Although these approaches basically provide an 
approximate value for the distance of Galactic sources, our past 
experience (Papers I-IX) tells us that these estimates can be considered 
correct to within 50\% of the refined value subsequently determined with 
more precise measurements.

For extragalactic sources, we assume a cosmology with $H_{\rm 0}$ = 65 km 
s$^{-1}$ Mpc$^{-1}$, $\Omega_{\Lambda}$ = 0.7, and $\Omega_{\rm m}$ = 0.3; 
the luminosity distances of the extragalactic objects presented in this 
paper were determined for these parametres using the Cosmology Calculator 
of Wright (2006). When not explicitly stated otherwise, we assume a 
Crab-like spectrum in our X--ray flux estimates except for the catalogued 
{\it XMM-Newton} sources, for which we considered the fluxes reported in 
Saxton et al. (2008) or in Watson et al. (2009). The X--ray luminosities 
reported in Tables 3, 4, 5, 7, 8, and 9 are associated with a letter 
indicating the satellite and/or the instrument, namely {\it ASCA} ({\it 
A}), {\it Swift}/BAT ({\it B}), {\it Chandra} ({\it C}), {\it Einstein} 
({\it E}), {\it INTEGRAL} ({\it I}), {\it XMM-Newton} ({\it N}), {\it 
ROSAT} ({\it R}), and {\it Swift}/XRT ({\it X}), with which the 
measurement of the corresponding X--ray flux was obtained. 

In the following, we present the object identifications by dividing them 
into four broad classes (AGNs, X--ray binaries, CVs, and active stars) as 
we did in some of our previous papers.

\subsection{AGNs}

\begin{table*}
\caption[]{Synoptic table containing the main results for the seven
high-redshift QSOs ($z >$ 0.5; Figs. 5 and 6) identified in the 
present sample of {\it INTEGRAL} sources. The upper part deals with 
the Type 1 QSOs of the sample, while the lower part with the single Type 2 
QSO identified in this paper.}
\scriptsize
\begin{center}
\begin{tabular}{lcccccrcr}
\noalign{\smallskip}
\hline
\hline
\noalign{\smallskip}
\multicolumn{1}{c}{Object} & $F_{\rm [OII]}$ & $F_{\rm MgII}$ &
$F_{\rm C III]}$ & $F_{\rm C IV}$ & $z$ &
\multicolumn{1}{c}{$D_L$ (Mpc)} & $E(B-V)_{\rm Gal}$ & \multicolumn{1}{c}{$L_{\rm X}$} \\
\noalign{\smallskip}
\hline
\noalign{\smallskip}

IGR J02115$-$4407 & --- & 9.5$\pm$1.0 & 6.6$\pm$0.6 & --- & 1.212 & 9025.9 & 0.017 & 150 (20--40; {\it I}) \\
 & --- & [10.0$\pm$1.0] & [12.4$\pm$01.2] & --- & & & & $<$220 (40--100; {\it I}) \\

 & & & & & & & & \\

IGR J03564+6242 & --- & 5.7$\pm$0.5 & 1.4$\pm$0.4 & --- & 1.109 & 8086.6 & 0.910 & 2.0 (2--10; {\it X}) \\
 & --- & [54$\pm$5] & [6$\pm$2] & --- & & & & 190 (20--100; {\it I}) \\

 & & & & & & & & \\

IGR J09034+5329 & --- & 2.9$\pm$0.3 & 2.9$\pm$0.3 & --- & 1.253 & 9405.2 & 0.015 & 180 (20--40; {\it I}) \\
 & --- & [3.0$\pm$0.3] & [3.0$\pm$0.3] & --- & & & & $<$270 (40--100; {\it I}) \\

 & & & & & & & & \\

IGR J16388+3557 & --- & --- & 4.5$\pm$0.5 & 1.9$\pm$0.3 & 2.020 & 16936.9 & 0.021 & 750 (20--100; {\it I}) \\
 & --- & --- & [5.0$\pm$0.5] & [2.0$\pm$0.3] & & & & \\

 & & & & & & & & \\

IGR J21319+3619 & --- & 2.4$\pm$0.2 & 1.40$\pm$0.15 & 2.3$\pm$0.2 & 1.488 & 11631.4 & 0.212 & $<$50 (20--40; {\it I}) \\
 & --- & [3.2$\pm$0.3] & [2.8$\pm$0.3] & [5.7$\pm$0.6] & & & & $<$90 (40--100; {\it I}) \\

 & & & & & & & & \\

IGR J23558$-$1047 & --- & 4.9$\pm$0.5 & --- & --- & 1.108 & 8077.6 & 0.034 & 0.66 (2--10; {\it X}) \\
 & --- & [5.0$\pm$0.5] & --- & --- & & & & 480 (20--100; {\it I}) \\

\noalign{\smallskip}
\hline
\noalign{\smallskip}

IGR J19295$-$0919 & 0.65$\pm$0.03 & --- & --- & --- & 0.741 & 4918.3 & 0.292 & 50 (20--100; {\it I}) \\
 & [1.32$\pm$0.07] & --- & --- & --- & & & & \\

\noalign{\smallskip} 
\hline
\noalign{\smallskip} 
\multicolumn{9}{l}{Note: emission-line fluxes are reported both as 
observed and (between square brackets) corrected for the intervening Galactic} \\ 
\multicolumn{9}{l}{absorption $E(B-V)_{\rm Gal}$ along the object line of sight 
(from Schlegel et al. 1998). Line fluxes are in units of 10$^{-15}$ erg cm$^{-2}$ s$^{-1}$,} \\
\multicolumn{9}{l}{X--ray luminosities are in units of 10$^{45}$ erg s$^{-1}$,
and the reference band (between round brackets) is expressed in keV.} \\
\multicolumn{9}{l}{In the last column, the upper case letter indicates the satellite and/or the 
instrument with which the corresponding X--ray flux} \\
\multicolumn{9}{l}{measurement was obtained (see text).} \\
\multicolumn{9}{l}{The typical error of the redshift measurement is $\pm$0.001 
except for the spectrum of IGR J16388+3557, for which an uncertainty} \\
\multicolumn{9}{l}{of $\pm$0.005 is assumed.} \\
\noalign{\smallskip} 
\hline
\hline
\end{tabular} 
\end{center} 
\end{table*}

\begin{figure*}
\mbox{\psfig{file=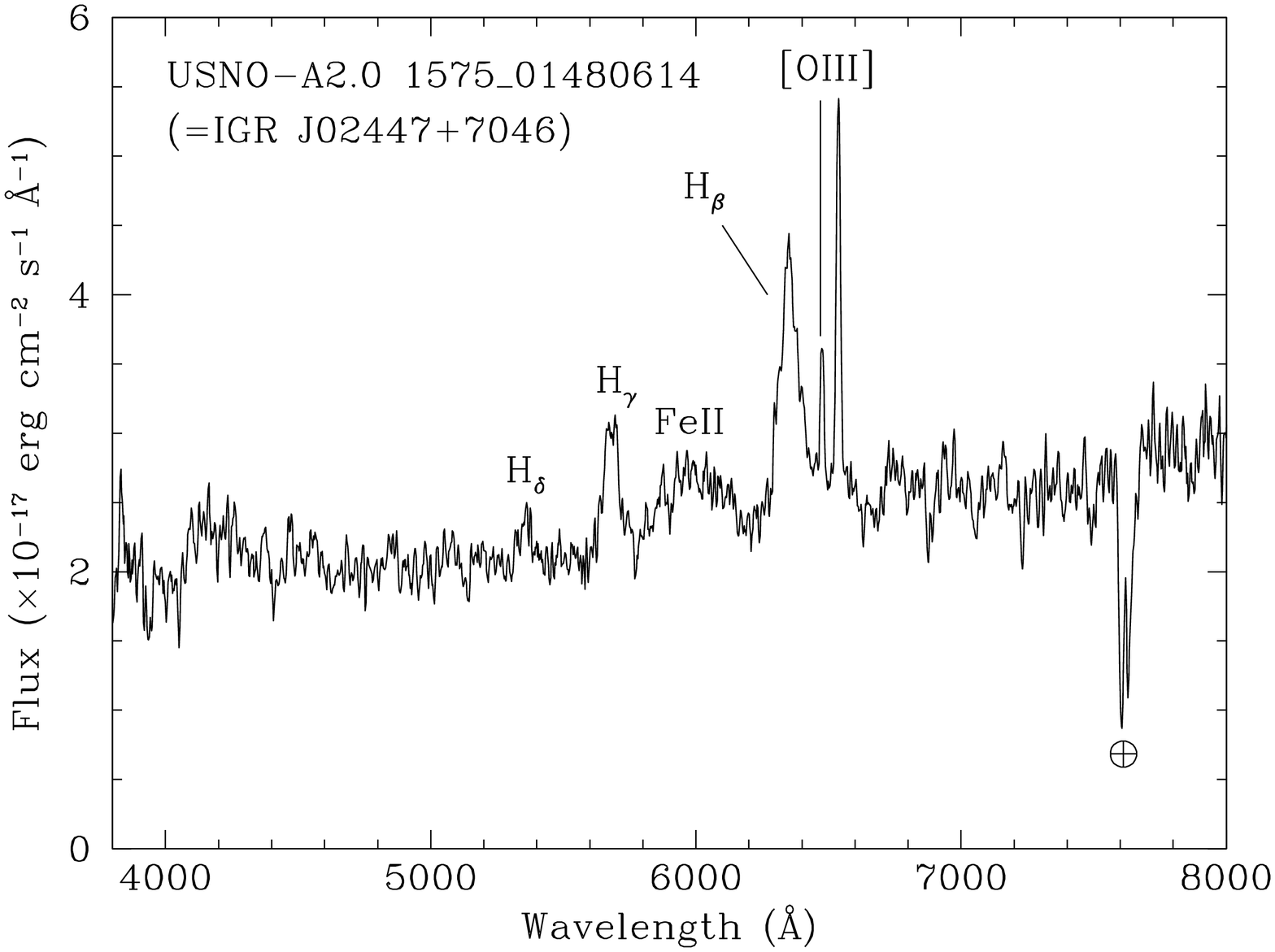,width=9cm,angle=270}}
\mbox{\psfig{file=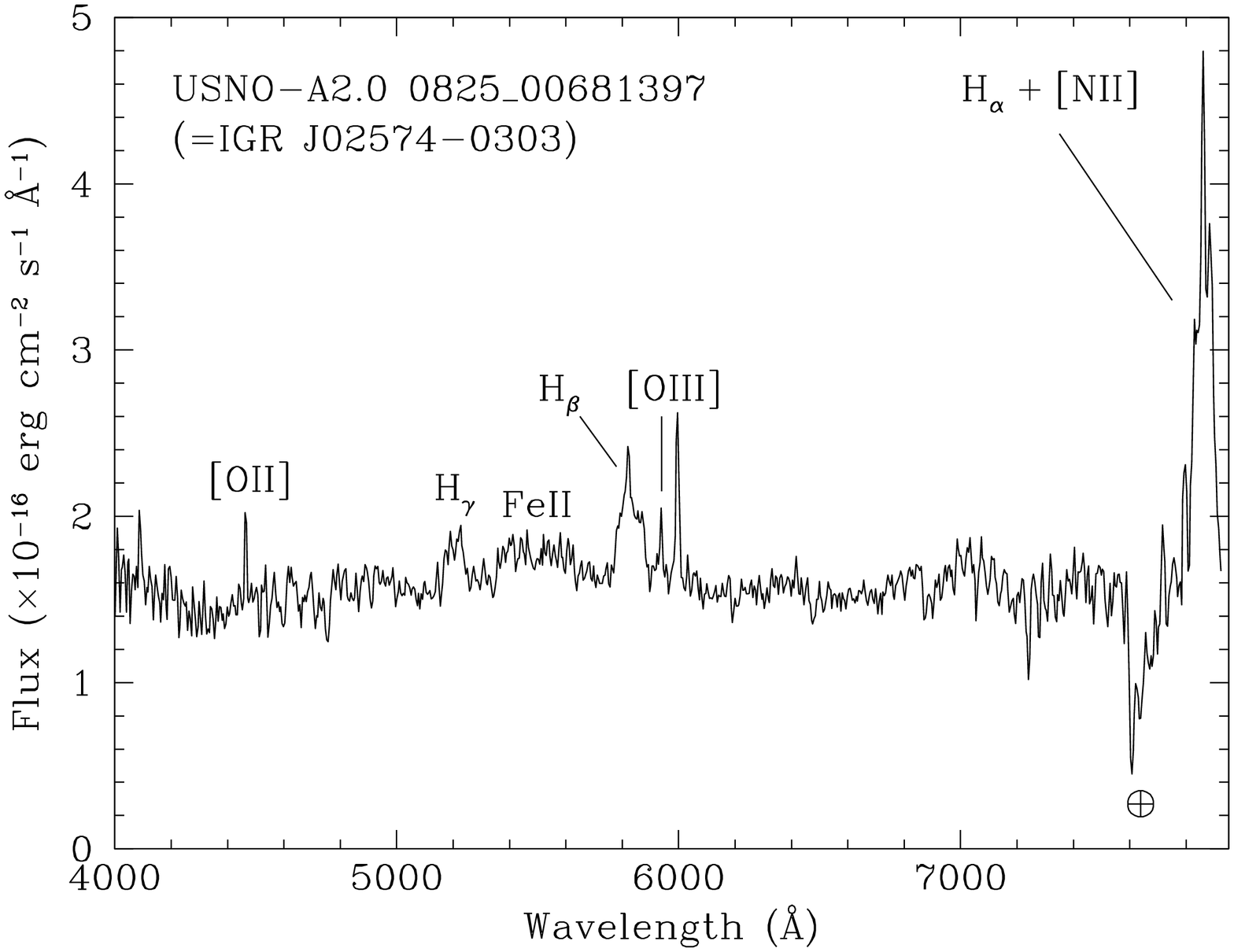,width=9cm,angle=270}}

\vspace{-.9cm}
\mbox{\psfig{file=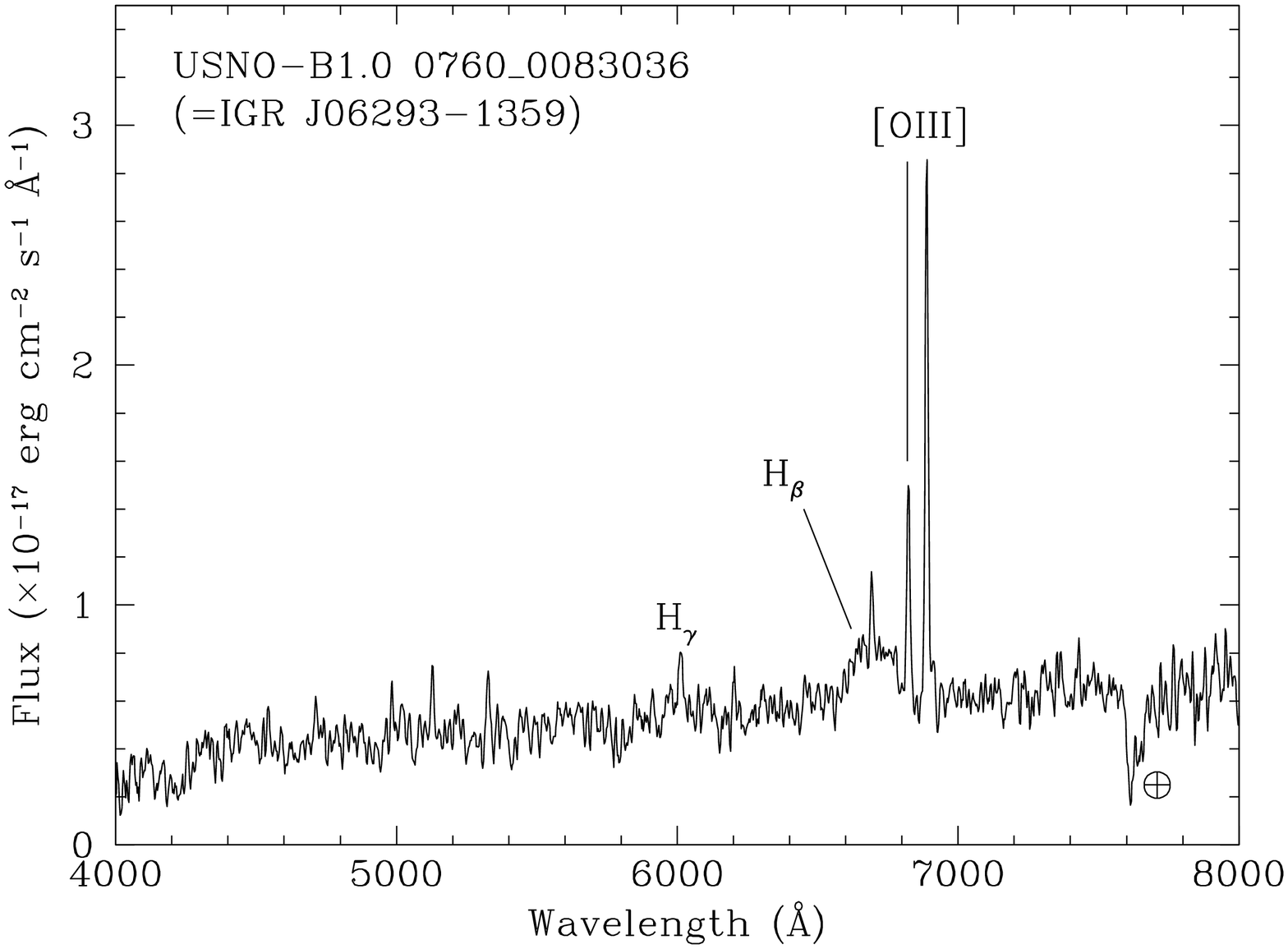,width=9cm,angle=270}}
\mbox{\psfig{file=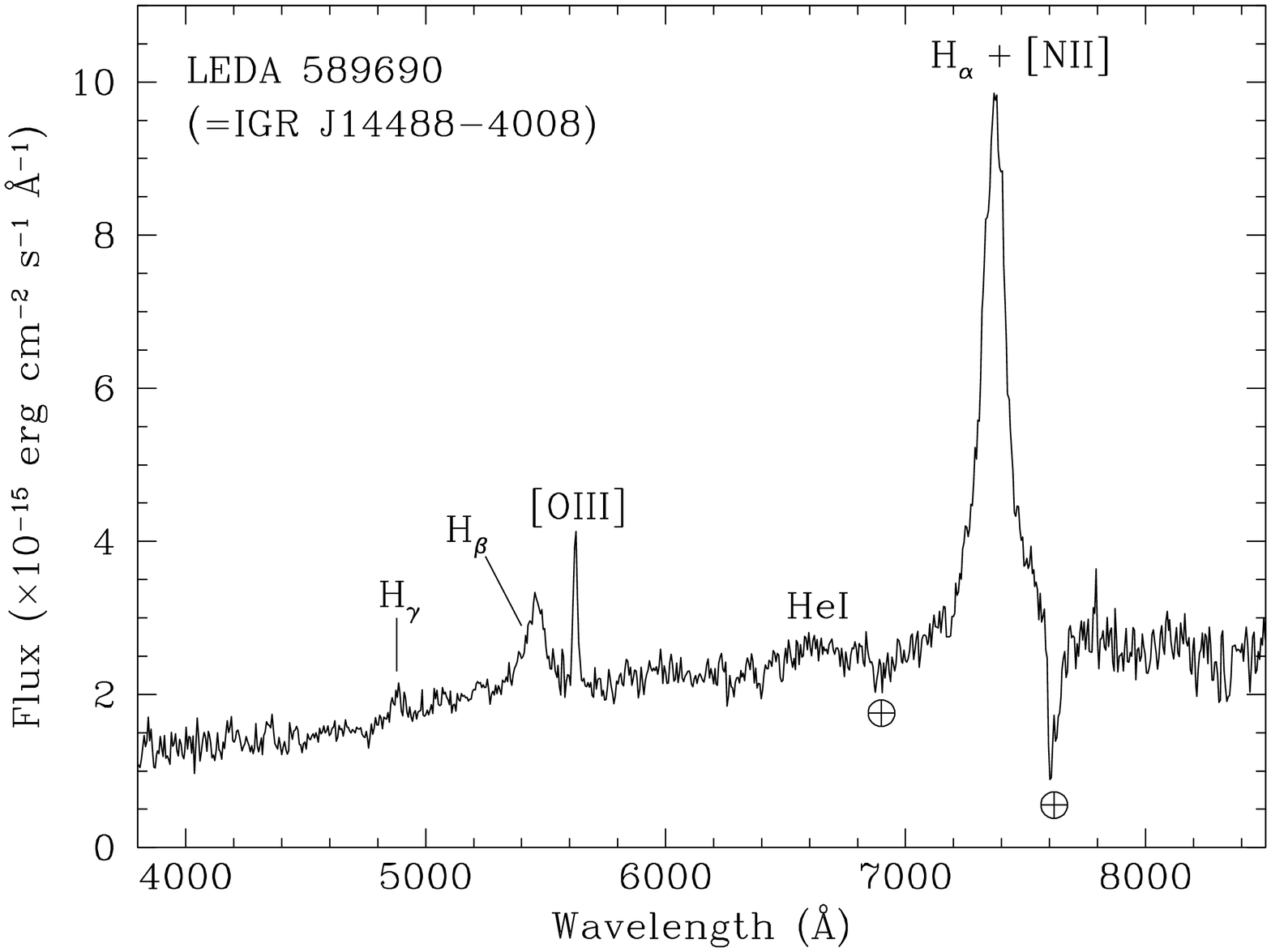,width=9cm,angle=270}}

\vspace{-.9cm}
\mbox{\psfig{file=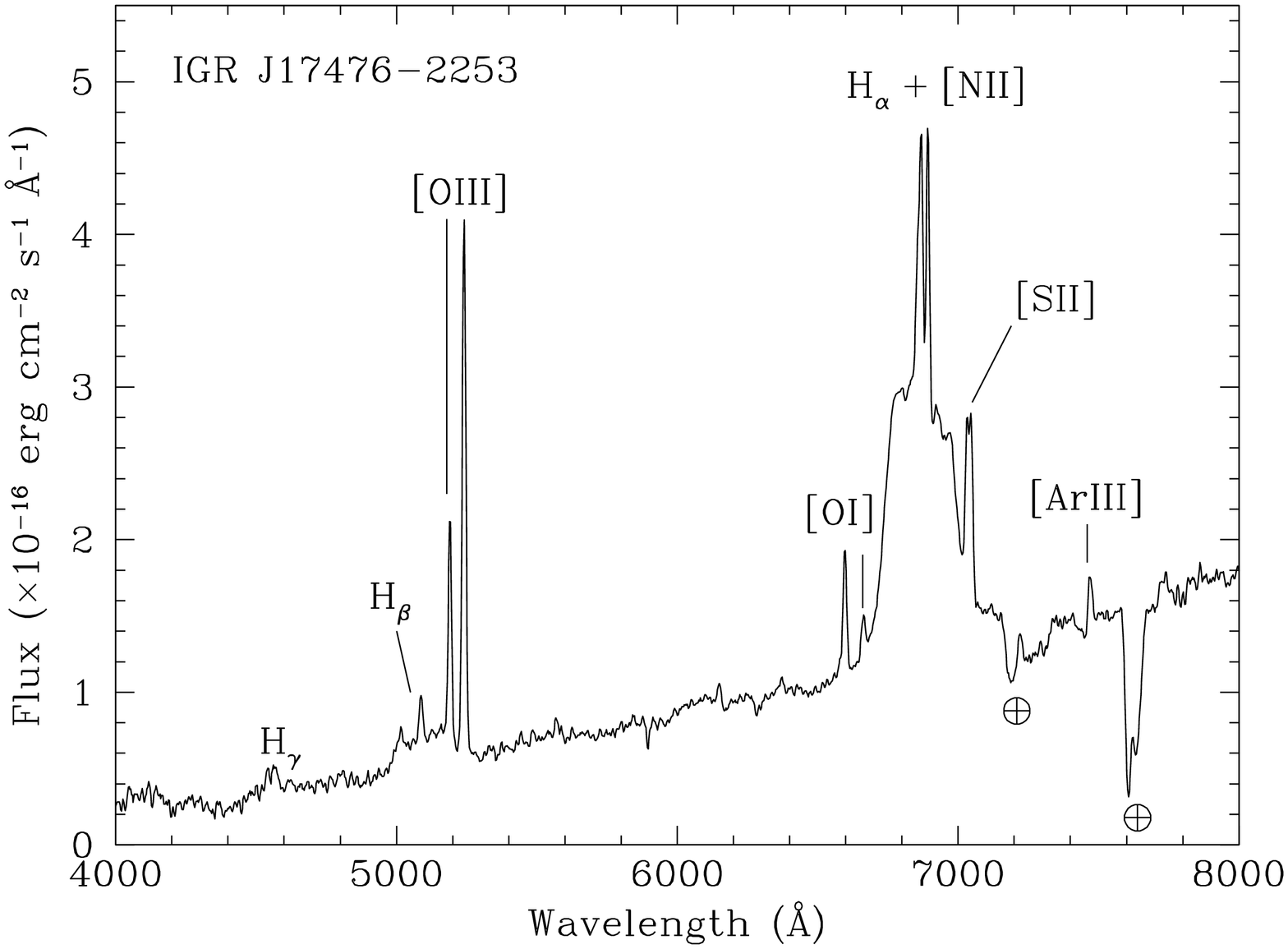,width=9cm,angle=270}}
\mbox{\psfig{file=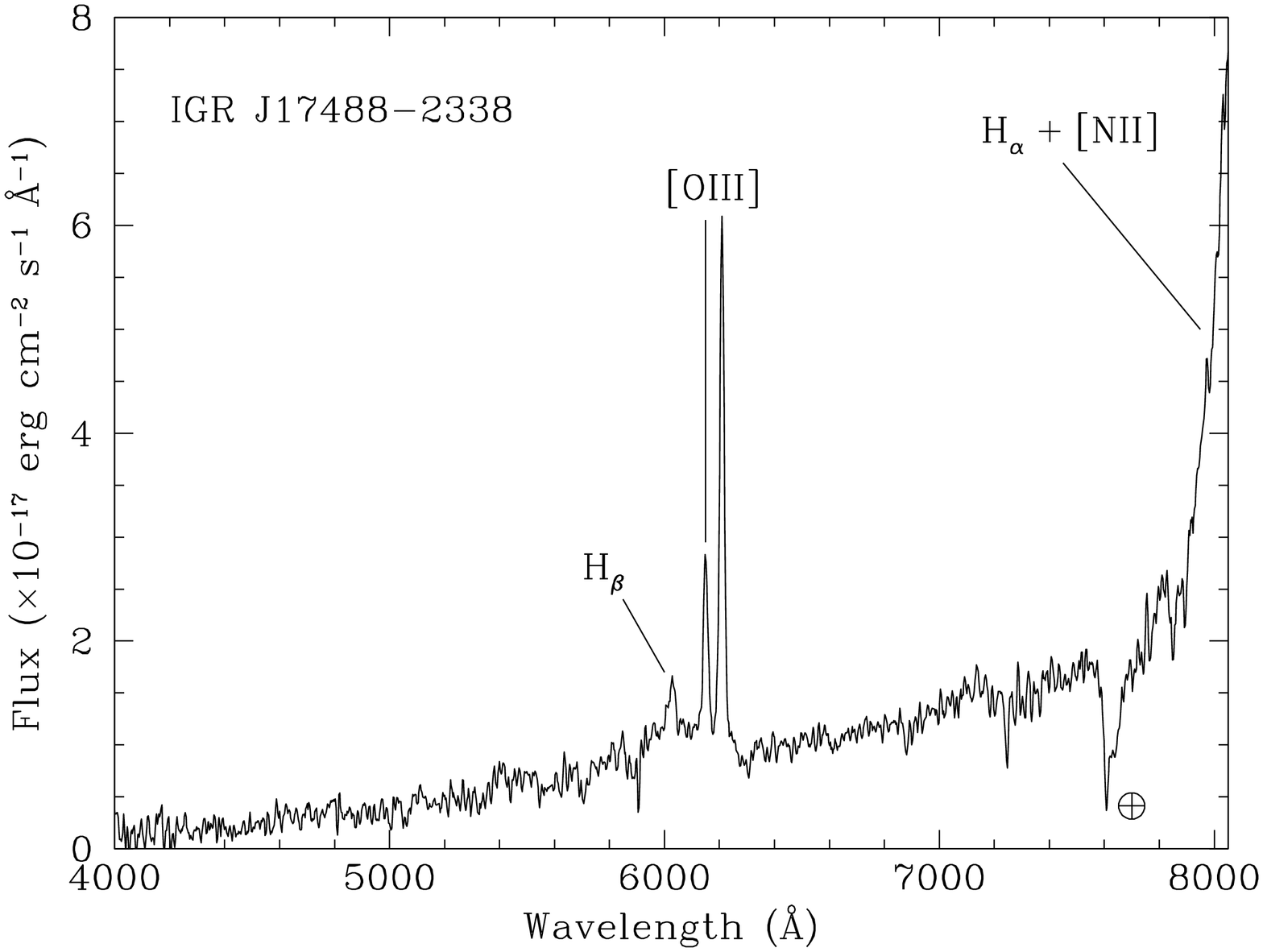,width=9cm,angle=270}}

\vspace{-.9cm}
\mbox{\psfig{file=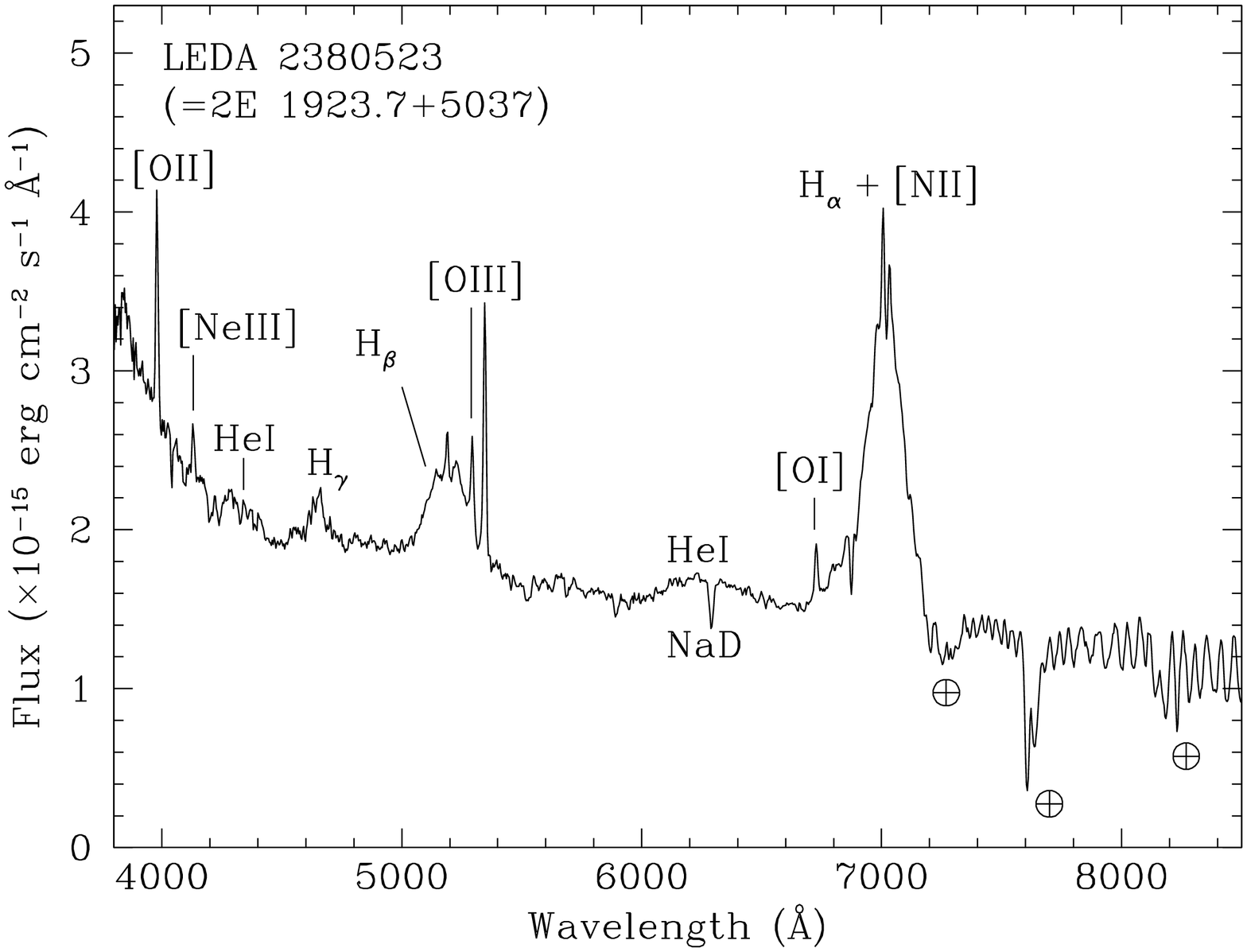,width=9cm,angle=270}}
\mbox{\psfig{file=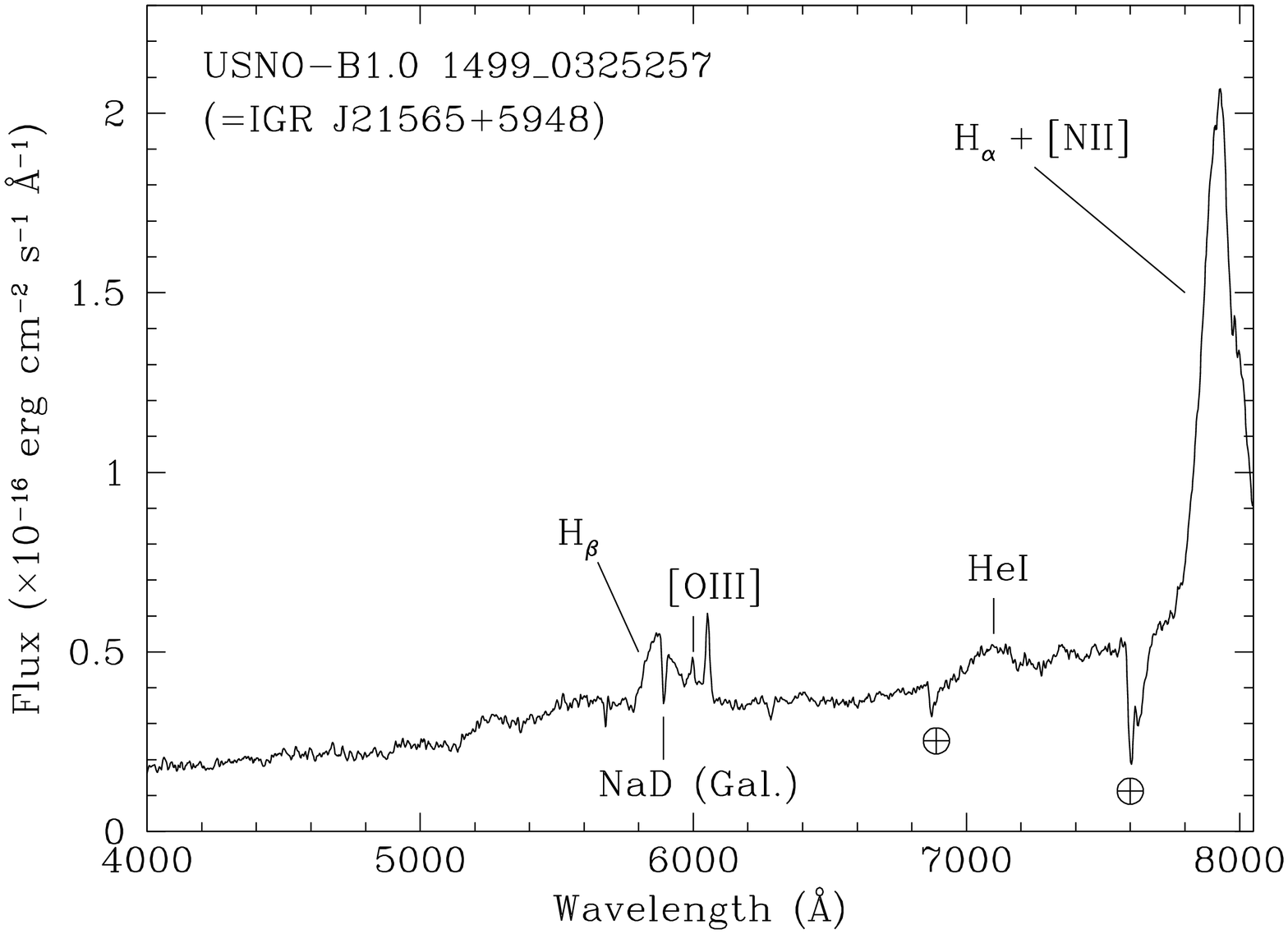,width=9cm,angle=270}}

\caption{Spectra (not corrected for the intervening Galactic absorption) 
of the optical counterparts of the eight low-redshift, broad emission-line 
AGNs belonging to the sample of {\it INTEGRAL} sources presented in 
this paper. For each spectrum, the main spectral features are labelled. 
The symbol $\oplus$ indicates atmospheric telluric absorption bands.
The TNG spectra have been smoothed using a Gaussian filter with 
$\sigma$ = 3 \AA.}
\end{figure*}

\begin{table*}
\caption[]{Synoptic table containing the main results for the eight
low-$z$ broad emission-line AGNs (Fig. 7) identified or observed in the 
present sample of {\it INTEGRAL} sources.}
\scriptsize
\begin{center}
\begin{tabular}{lccccrcr}
\noalign{\smallskip}
\hline
\hline
\noalign{\smallskip}
\multicolumn{1}{c}{Object} & $F_{\rm H_\beta}$ & $F_{\rm [OIII]}$ & Class & $z$ &
\multicolumn{1}{c}{$D_L$ (Mpc)} & $E(B-V)_{\rm Gal}$ & \multicolumn{1}{c}{$L_{\rm X}$} \\
\noalign{\smallskip}
\hline
\noalign{\smallskip}

IGR J02447+7046 & 2.0$\pm$0.2 & 0.66$\pm$0.06 & Sy1.2 & 0.306 & 1710.8 & 0.770 & 3.7 (20--100; {\it I}) \\
 & [11.6$\pm$1.2] & [3.9$\pm$0.4] & & & & & \\

 & & & & & & & \\

IGR J02574$-$0303 & 5.5$\pm$0.6 & 1.51$\pm$0.15 & Sy1.2 & 0.197 & 1037.8 & 0.047 & 0.13 (0.5--10; {\it X}) \\
 & [5.6$\pm$0.6] & [1.63$\pm$0.16] & & & & & 2.6 (17--80; {\it I}) \\

 & & & & & & & \\

IGR J06293$-$1359 & 0.47$\pm$0.07 & 0.31$\pm$0.03 & Sy1.5 & 0.376 & 2174.8 & 0.317 & 9.6 (20--40; {\it I}) \\
 & [1.2$\pm$0.2] & [0.62$\pm$0.06] & & & & & $<$7.9 (40--100; {\it I}) \\

 & & & & & & & \\

IGR J14488$-$4008 & 120$\pm$20 & 45$\pm$4 & Sy1.2 & 0.123 & 619.0 & 0.109 & 0.11 (2--10; {\it X}) \\
 & [160$\pm$30] & [58$\pm$6] & & & & & 0.17 (20--40; {\it I}) \\
 & & & & & & & $<$0.17 (40--100; {\it I}) \\
 & & & & & & & 0.33 (17--60; {\it I}) \\
 & & & & & & & 0.60 (14--195; {\it B}) \\
 & & & & & & & 0.38 (15--150; {\it B}) \\

 & & & & & & & \\

IGR J17476$-$2253 & 6.0$\pm$0.6 & 5.73$\pm$0.17 & Sy1.5 & 0.047 & 224.5 & 1.009 & 0.016--0.030 (2--10; {\it X}) \\
 & [127$\pm$13] & [121$\pm$4] & & & & & 0.11 (20--100; {\it I}) \\
 & & & & & & & 0.060 (17--60; {\it I}) \\
 & & & & & & & 0.11 (14--195; {\it B}) \\
 & & & & & & & 0.090 (15--150; {\it B}) \\

 & & & & & & & \\

IGR J17488$-$2338 & 1.08$\pm$0.16 & 1.13$\pm$0.03 & Sy1.5 & 0.240 & 1295.7 & 1.570 & 0.70 (0.2--12; {\it N}) \\
 & [62$\pm$9] & [54.8$\pm$1.6] & & & & & 0.46 (2--10; {\it X}) \\
 & & & & & & & 2.8 (20--100; {\it I}) \\

 & & & & & & & \\

2E 1923.7+5037 & 137$\pm$10 & 27.9$\pm$1.4 & Sy1.2 & 0.068 & 329.8 & 0.093 & 0.013 (0.1--2.4; {\it X}) \\
 & [173$\pm$12] & [36.1$\pm$1.8] & & & & & 0.014 (0.16--3.5; {\it E}) \\
 & & & & & & & 0.13 (0.2--12; {\it N}) \\
 & & & & & & & 0.10 (18--55; {\it I+B}) \\
 & & & & & & & 0.13 (14--195; {\it B}) \\

 & & & & & & & \\

IGR J21565+5948 & 0.55$\pm$0.06 & 2.8$\pm$0.4 & Sy1 & 0.209 & 1108.7 & 1.267 & 0.010 (0.1--2.4; {\it R}) \\
 & [68$\pm$10] & [13.7$\pm$1.4] & & & & & 0.21 (0.3--10; {\it C}) \\
 & & & & & & & 0.18 (2--10; {\it X}) \\
 & & & & & & & 1.4 (20--100; {\it I}) \\

\noalign{\smallskip} 
\hline
\noalign{\smallskip} 
\multicolumn{8}{l}{Note: emission-line fluxes are reported both as 
observed and (between square brackets) corrected for the intervening Galactic} \\ 
\multicolumn{8}{l}{absorption $E(B-V)_{\rm Gal}$ along the object line of sight 
(from Schlegel et al. 1998). Line fluxes are in units of 10$^{-15}$ erg cm$^{-2}$ s$^{-1}$,} \\
\multicolumn{8}{l}{X--ray luminosities are in units of 10$^{45}$ erg s$^{-1}$, 
and the reference band (between round brackets) is expressed in keV.} \\ 
\multicolumn{8}{l}{In the last column, the upper case letter indicates the satellite 
and/or the instrument with which the corresponding X--ray flux} \\
\multicolumn{8}{l}{measurement was obtained (see text). The typical error of the redshift measurement is $\pm$0.001.} \\
\noalign{\smallskip} 
\hline
\hline
\end{tabular} 
\end{center} 
\end{table*}

\begin{figure*}
\mbox{\psfig{file=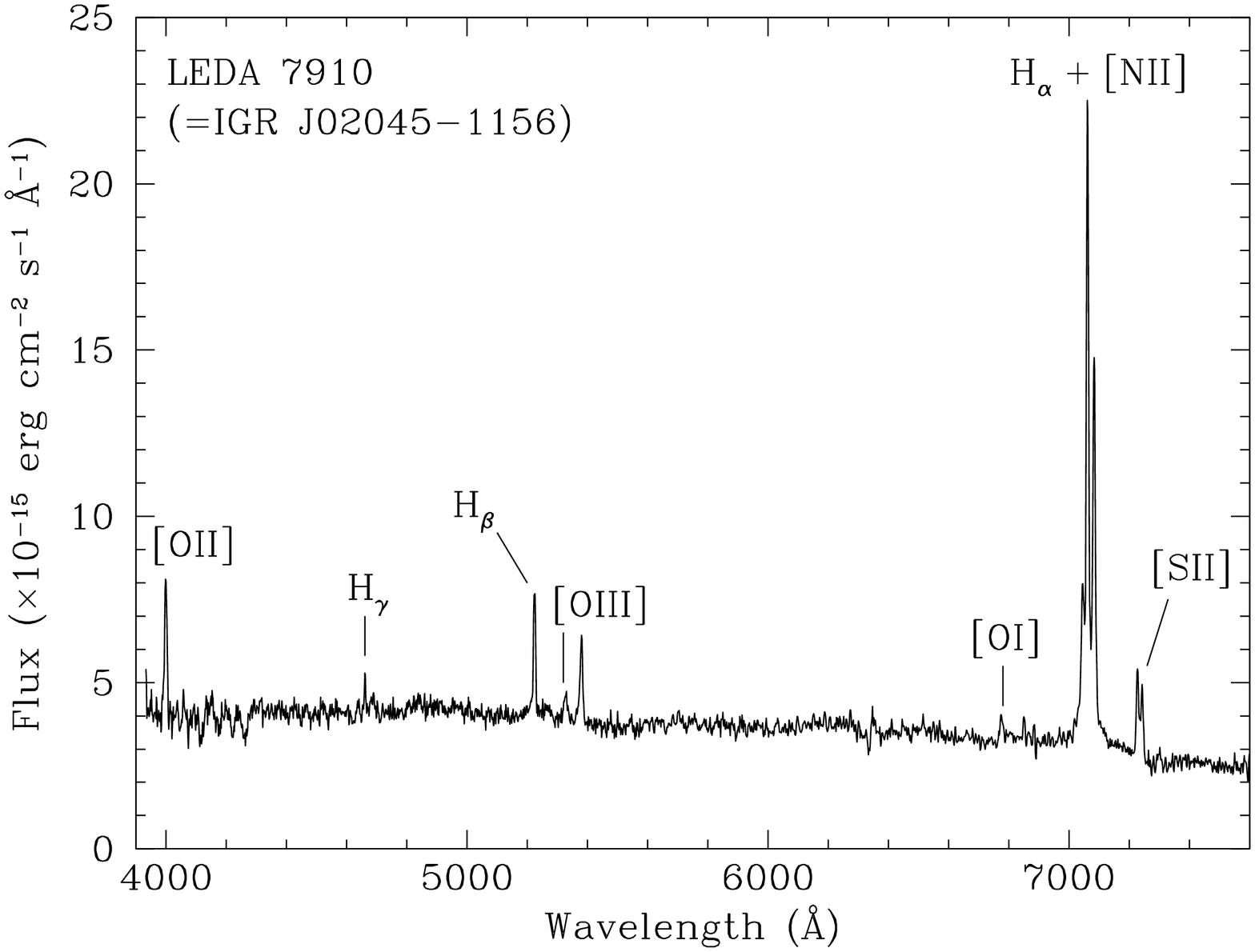,width=9cm,angle=270}}
\mbox{\psfig{file=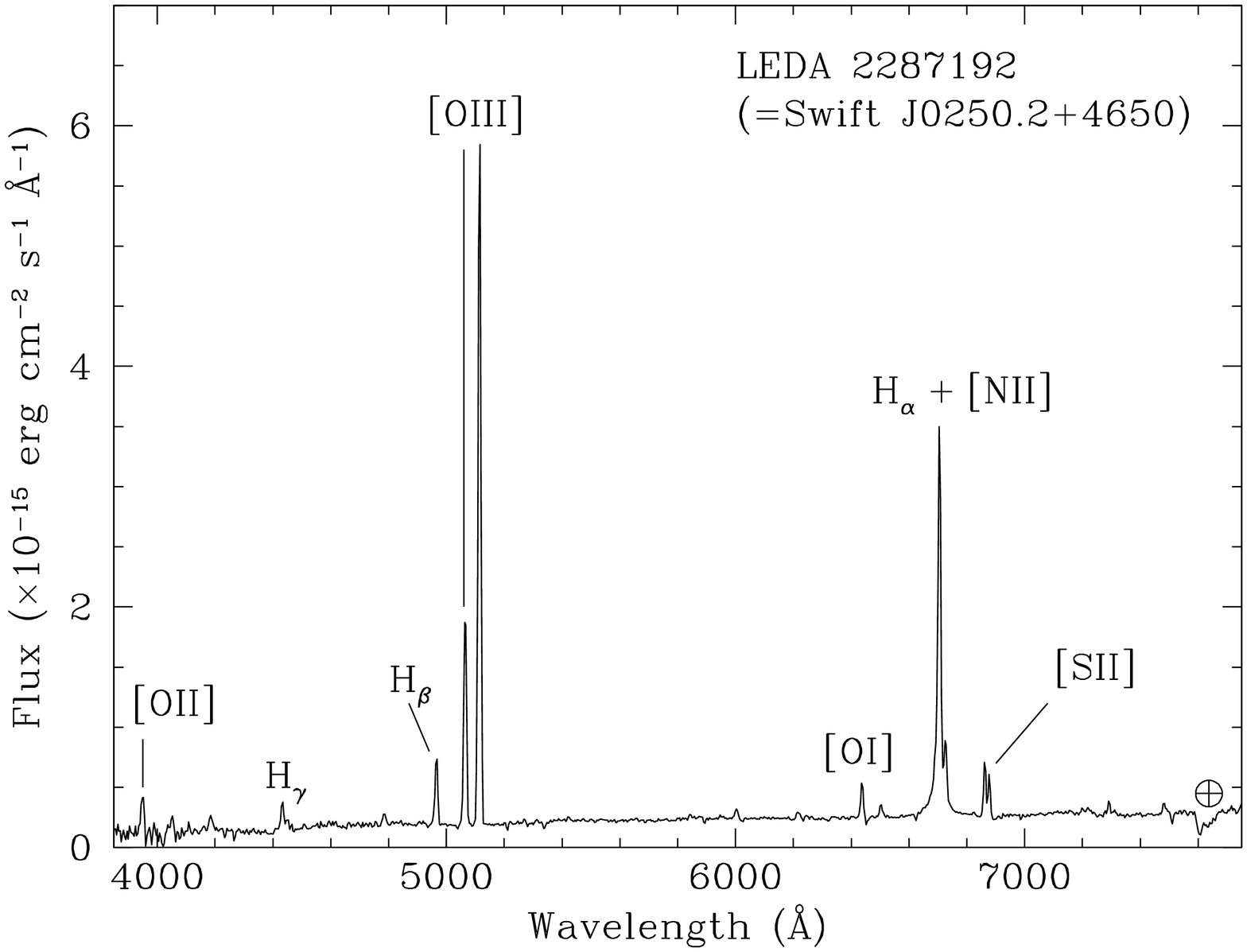,width=9cm,angle=270}}

\vspace{-.9cm}
\mbox{\psfig{file=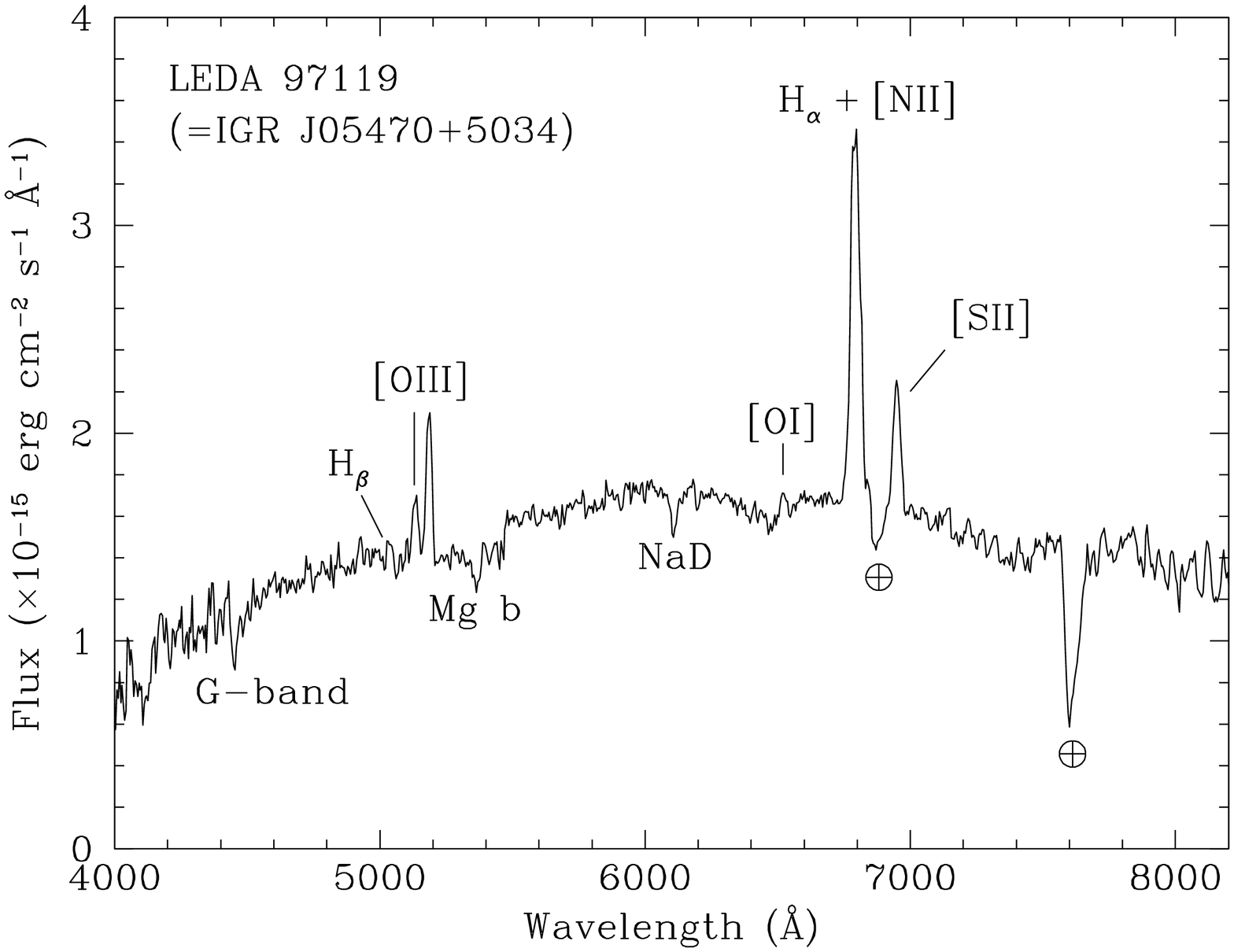,width=9cm,angle=270}}
\mbox{\psfig{file=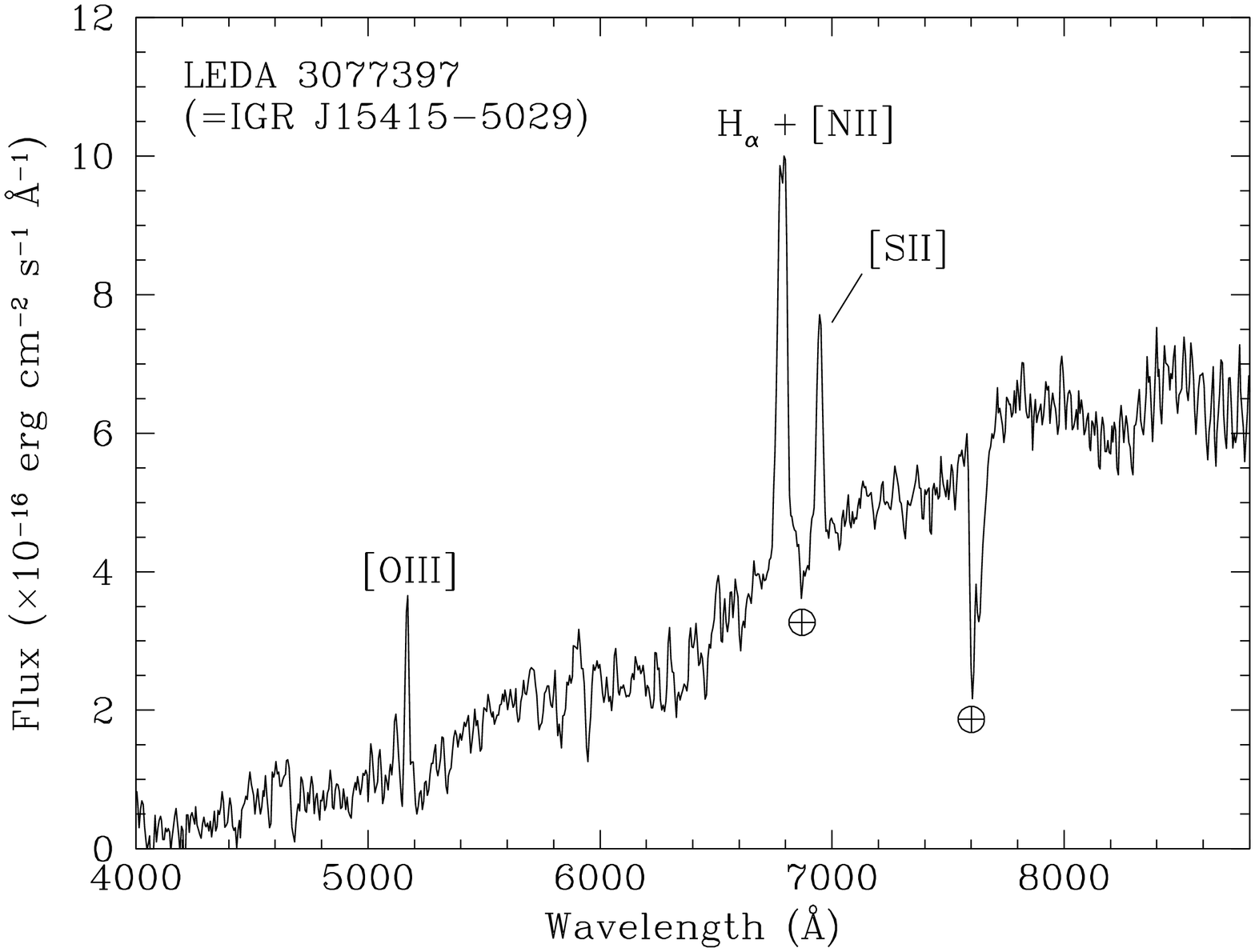,width=9cm,angle=270}}

\vspace{-.9cm}
\mbox{\psfig{file=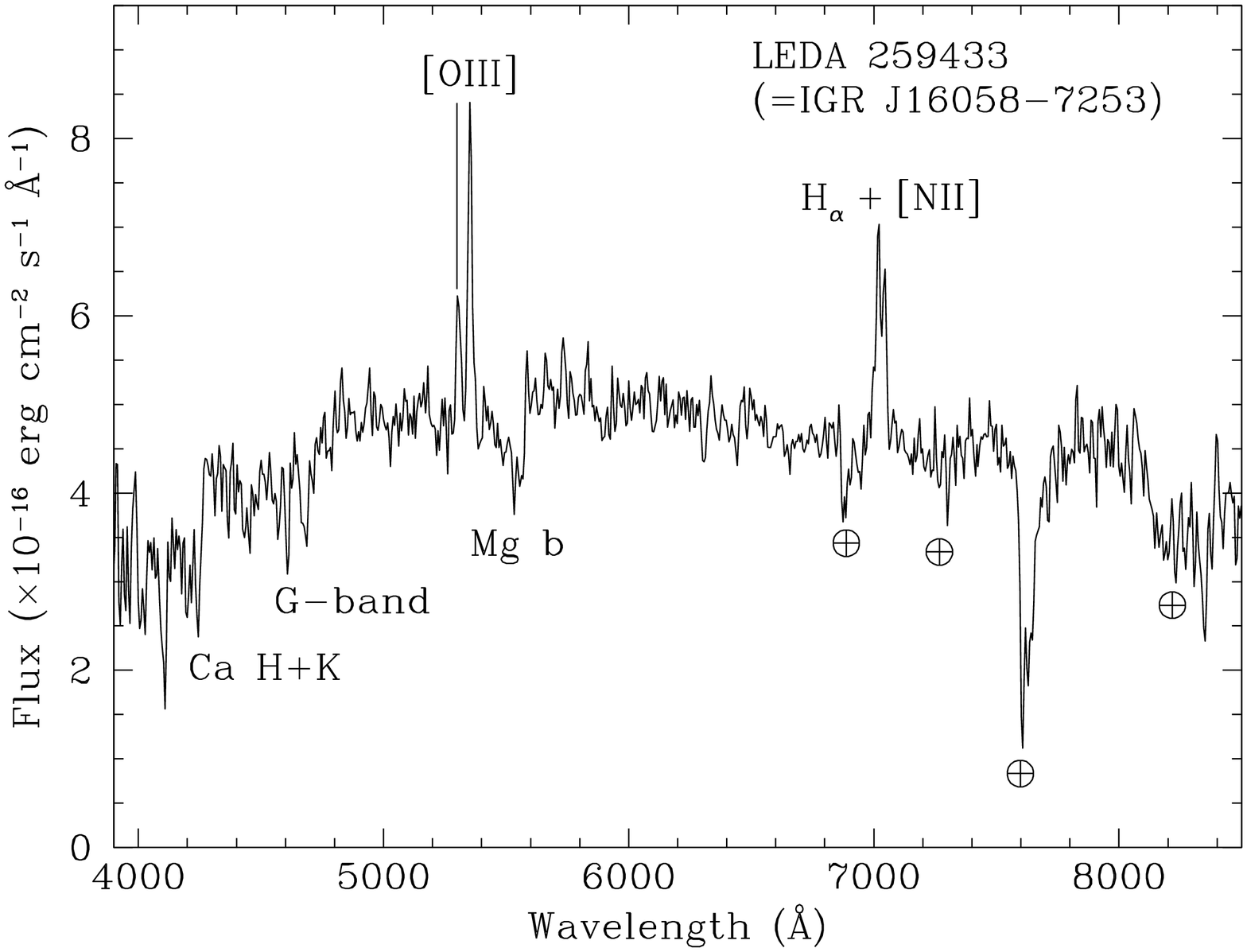,width=9cm,angle=270}}
\mbox{\psfig{file=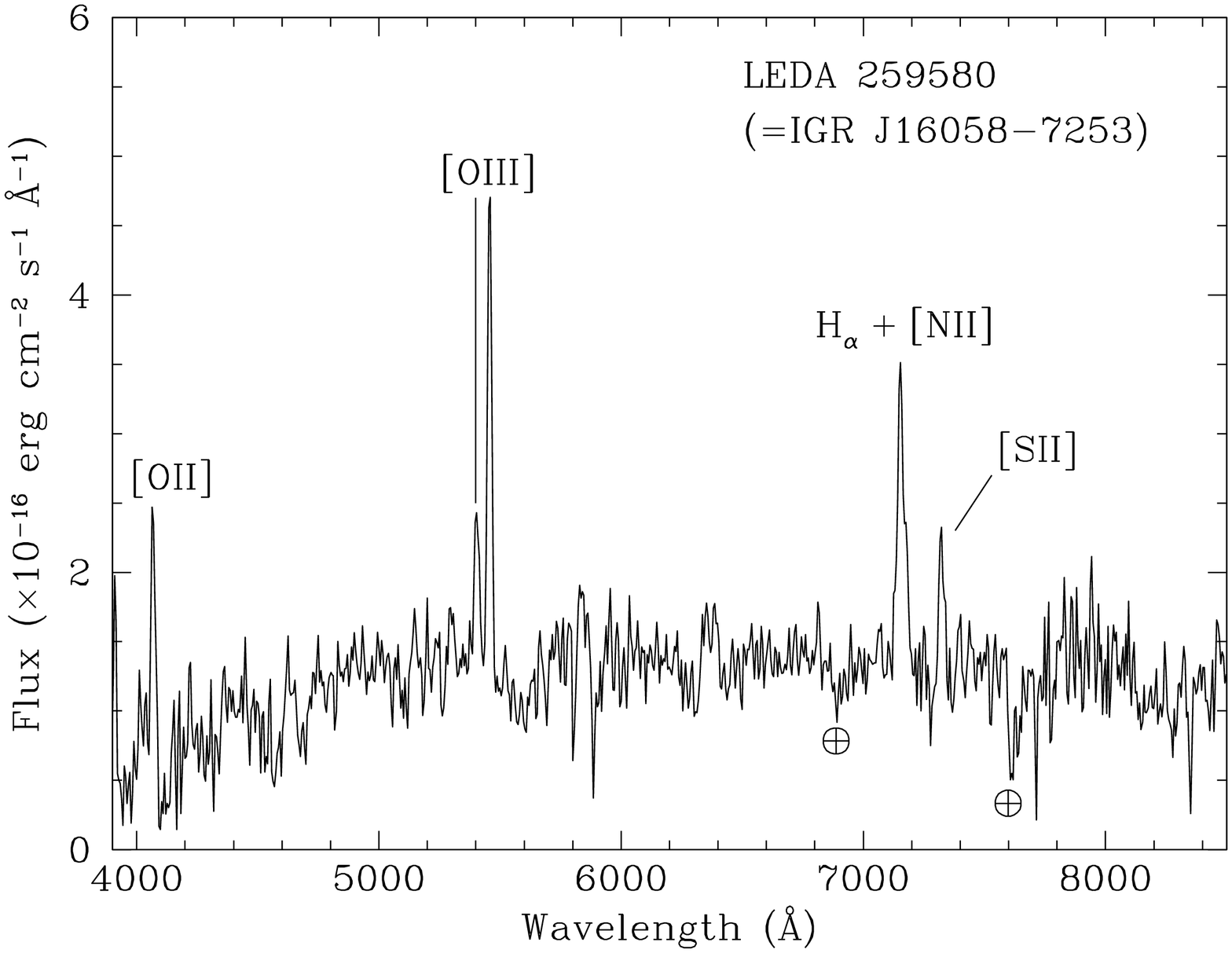,width=9cm,angle=270}}

\vspace{-.9cm}
\parbox{9.5cm}{
\psfig{file=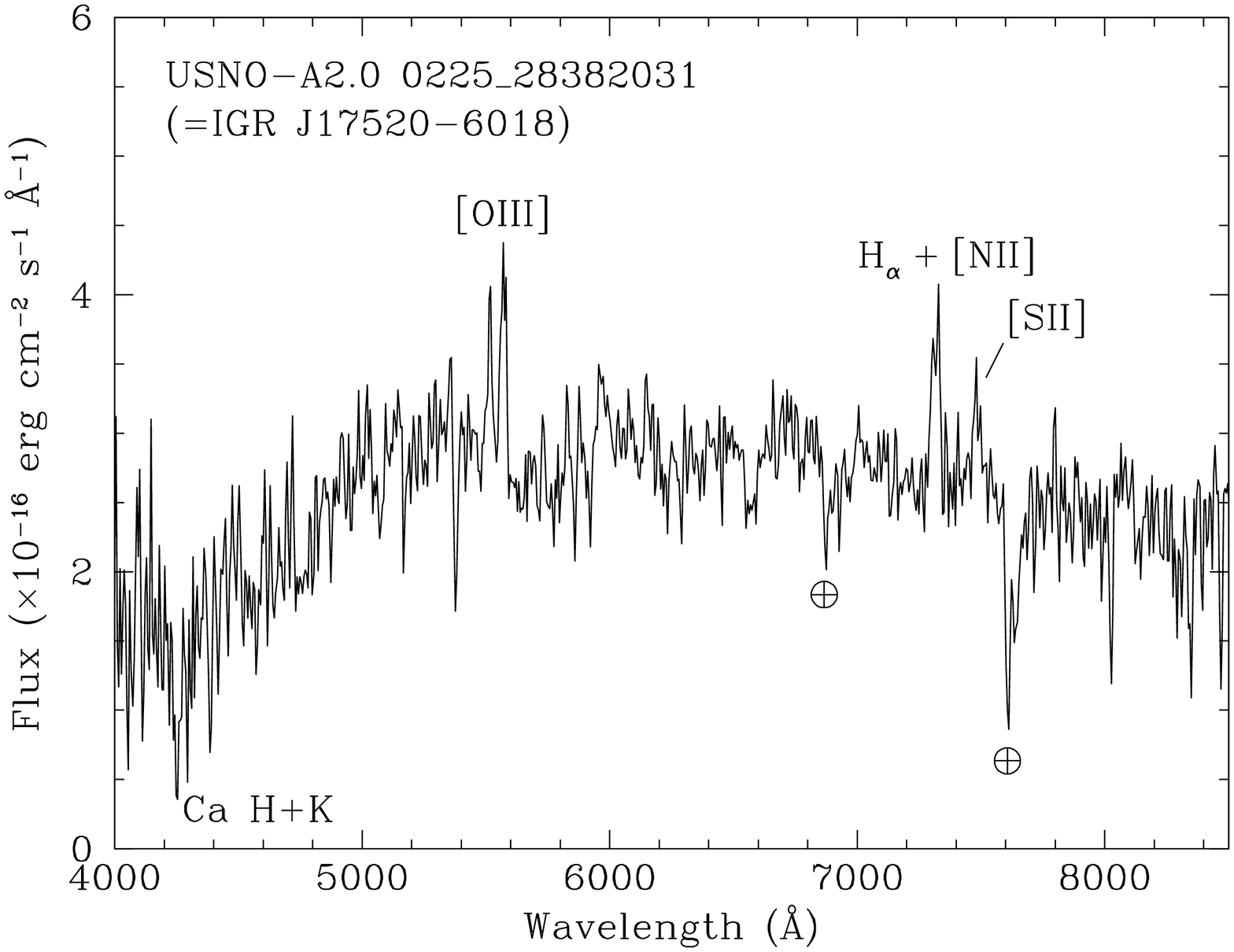,width=9cm,angle=270}
}
\hspace{0.8cm}
\parbox{8cm}{
\vspace{-.5cm}
\caption{Spectra (not corrected for the intervening Galactic absorption) 
of the optical counterparts of the seven low-redshift, narrow emission-line 
AGNs belonging to the sample of {\it INTEGRAL} sources presented in 
this paper. For each spectrum, the main spectral features are labelled. 
The symbol $\oplus$ indicates atmospheric telluric absorption bands.}}
\end{figure*}

\begin{table*}
\caption[]{Synoptic table containing the main results for the seven 
low-redshift narrow emission-line AGNs (Fig. 8) and for the two XBONGs
(Fig. 9) identified or observed in the present sample of {\it INTEGRAL} 
sources.}
\scriptsize
\begin{center}
\begin{tabular}{lcccccrccr}
\noalign{\smallskip}
\hline
\hline
\noalign{\smallskip}
\multicolumn{1}{c}{Object} & $F_{\rm H_\alpha}$ & $F_{\rm H_\beta}$ & $F_{\rm [OIII]}$ & Class & $z$ &
\multicolumn{1}{c}{$D_L$} & \multicolumn{2}{c}{$E(B-V)$} &
\multicolumn{1}{c}{$L_{\rm X}$} \\
\cline{8-9}
\noalign{\smallskip}
 & & & & & & (Mpc) & Gal. & AGN & \\
\noalign{\smallskip}
\hline
\noalign{\smallskip}

IGR J02045$-$1156 & 196$\pm$10 & 32$\pm$3 & 31$\pm$3 & LINER & 0.0727 & 353.7 & 0.023 & 0.68 & 0.61 (0.1--2.4; {\it R}) \\
 & [196$\pm$10] & [35$\pm$4] & [33$\pm$3] & & & & & & 4.3 (0.2--12; {\it N}) \\
 & & & & & & & & & 1.4 (0.3--10; {\it X}) \\
 & & & & & & & & & 24 (17--80; {\it I}) \\

 & & & & & & & & & \\

Swift J0250.2+4650 & 31.4$\pm$1.6 & 6.5$\pm$0.7 & 67$\pm$2 & Sy2 & 0.021 & 98.4 & 0.194 & 0.20 & 0.20 (0.2--12; {\it N}) \\
 & [48$\pm$3] & [13.7$\pm$1.4] & [124$\pm$4] & & & & & & 1.7 (17--60; {\it I}) \\
 & & & & & & & & & 2.3 (14--195; {\it B}) \\
 & & & & & & & & & 2.0 (15--150; {\it B}) \\

 & & & & & & & & & \\

IGR J05048$-$7340 & in abs. & in abs. & $<$8 & XBONG & 0.014 & 69.0 & 0.127 & --- & 0.31 (20--60; {\it I}) \\
 & [in abs.] & [in abs.] & [$<$12] & & & & & & 0.48 (14--195; {\it B}) \\
 & & & & & & & & & 0.43 (15--150; {\it B}) \\

 & & & & & & & & & \\

IGR J05470+5034 & 29$\pm$3 & 3.4$\pm$1.0 & 20$\pm$2 & Sy2 & 0.036 & 170.6 & 0.233 & 0.83 & 0.006 (0.1--2.4; {\it R}) \\
 & [48$\pm$5] & [7$\pm$2] & [40$\pm$4] & & & & & & 5.9 (17--60; {\it I}) \\
 & & & & & & & & & 8.3 (14--195; {\it B}) \\
 & & & & & & & & & 5.6 (15--150; {\it B}) \\

 & & & & & & & & & \\

IGR J10200$-$1436 & --- & $<$0.03 & $<$0.07 & XBONG & 0.391 & 2277.2 & 0.104 & --- & 6.8 (0.1--2.4; {\it X}) \\
 & --- & [$<$0.05] & [$<$0.08] & & & & & & 740 (20--40; {\it I}) \\
 & & & & & & & & & $<$810 (40--100; {\it I}) \\

 & & & & & & & & & \\

IGR J15415$-$5029 & 12.1$\pm$1.8 & $<$0.17 & 4.7$\pm$0.7 & likely Sy2 & 0.032 & 151.2 & 0.862 & $>$0.85 & 0.15 (0.3--10; {\it C}) \\
 & [86$\pm$13] & [$<$13] & [66$\pm$10] & & & & & & 1.7 (17--60; {\it I}) \\
 & & & & & & & & & 3.0 (20--100; {\it I}) \\
 & & & & & & & & & 2.5 (15--150; {\it B}) \\

 & & & & & & & & & \\

IGR J16058$-$7253$^*$ & & & & & & & & & \\

(LEDA 259433) & 4.6$\pm$0.5 & in abs. & 7.6$\pm$0.7 & likely Sy2 & 0.069 & 334.8 & 0.094 & --- & 4.4 (2--10; {\it X}) \\
              & [8.5$\pm$0.9] & [in abs.] & [9.4$\pm$0.9] & & & & & & \\

(LEDA 259580) & 5.7$\pm$0.5 & $<$2 & 7.8$\pm$0.8 & Sy2 & 0.090 & 373.0 & 0.094 & $>$0 & 4.8 (2--10; {\it X}) \\
              & [6.4$\pm$0.6] & [$<$3] & [1.05$\pm$0.11] & & & & & & \\

 & & & & & & & & & \\

IGR J17520$-$6018 & 2.7$\pm$0.8 & in abs. & 4.4$\pm$1.1 & likely Sy2 & 0.112 & 559.6 & 0.091 & --- & 9.7 (2--10; {\it X}) \\
 & [3.2$\pm$0.9] & [in abs.] & [5.4$\pm$1.4] & & & & & & 86 (20--100; {\it I}) \\
 & & & & & & & & & 56 (14--195; {\it B}) \\

\noalign{\smallskip} 
\hline
\noalign{\smallskip} 
\multicolumn{10}{l}{Note: emission-line fluxes are reported both as 
observed and (between square brackets) corrected for the intervening Galactic} \\ 
\multicolumn{10}{l}{absorption $E(B-V)_{\rm Gal}$ along the object line of sight 
(from Schlegel et al. 1998). Line fluxes are in units of 10$^{-15}$ erg cm$^{-2}$ s$^{-1}$,} \\
\multicolumn{10}{l}{X--ray luminosities are in units of 10$^{43}$ erg s$^{-1}$,
and the reference band (between round brackets) is expressed in keV.} \\ 
\multicolumn{10}{l}{In the last column, the upper case letter indicates the satellite and/or the 
instrument with which the corresponding X--ray flux} \\
\multicolumn{10}{l}{measurement was obtained (see text). The typical error 
of the redshift measurement is $\pm$0.001 except for the 6dFGS spectrum of IGR J02045$-$1156,} \\
\multicolumn{10}{l}{for which an uncertainty of $\pm$0.0003 can be assumed.} \\
\multicolumn{10}{l}{$^*$We did not attempt to estimate the total hard X--ray luminosity of 
IGR J16058$-$7253 and the contributions of the two} \\
\multicolumn{10}{l}{AGNs to this quantity due to the lack of information on their X--ray spectra.} \\
\noalign{\smallskip} 
\hline
\hline
\end{tabular}
\end{center}
\end{table*}

\begin{figure*}
\mbox{\psfig{file=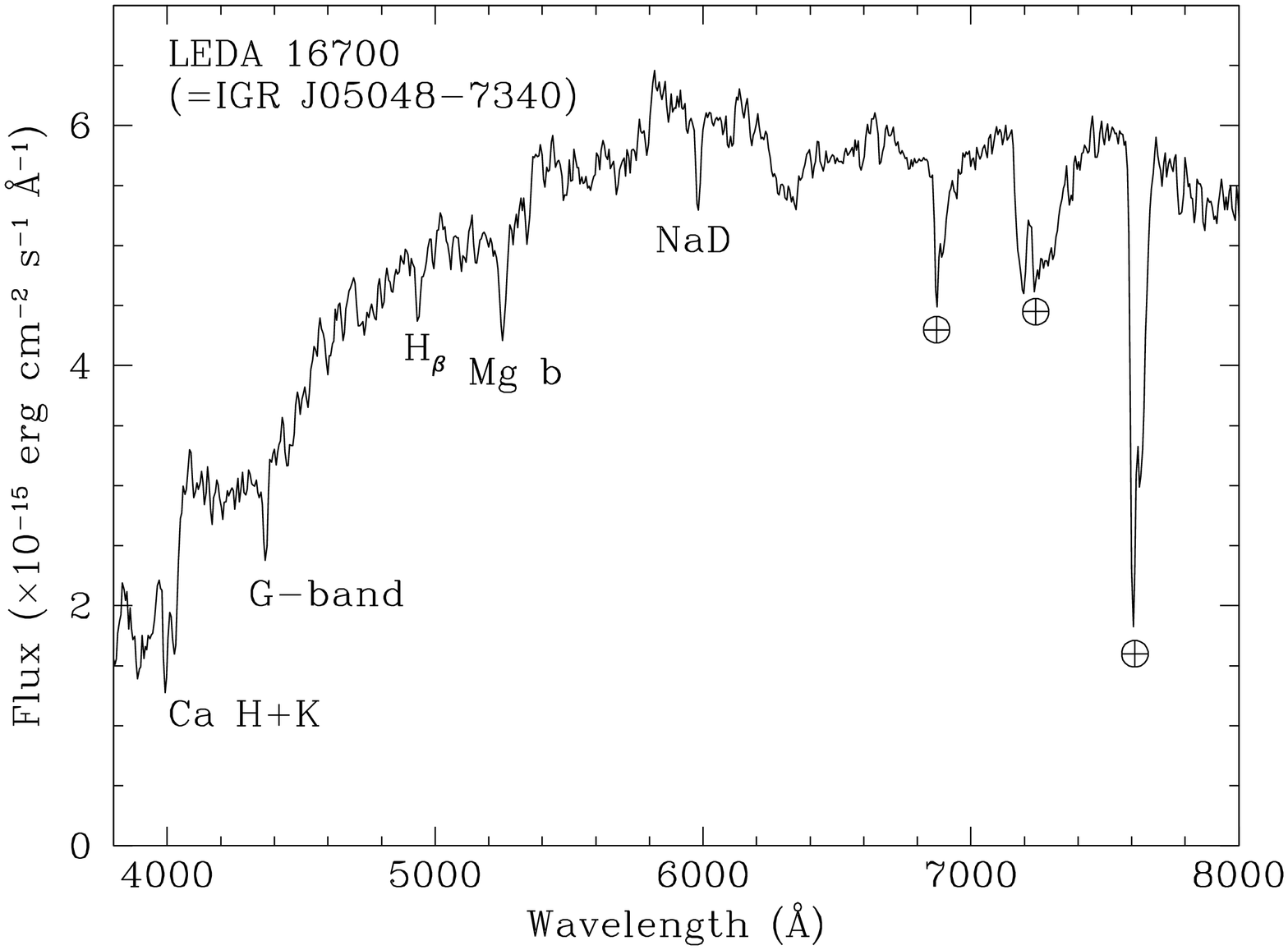,width=9cm,angle=270}}
\mbox{\psfig{file=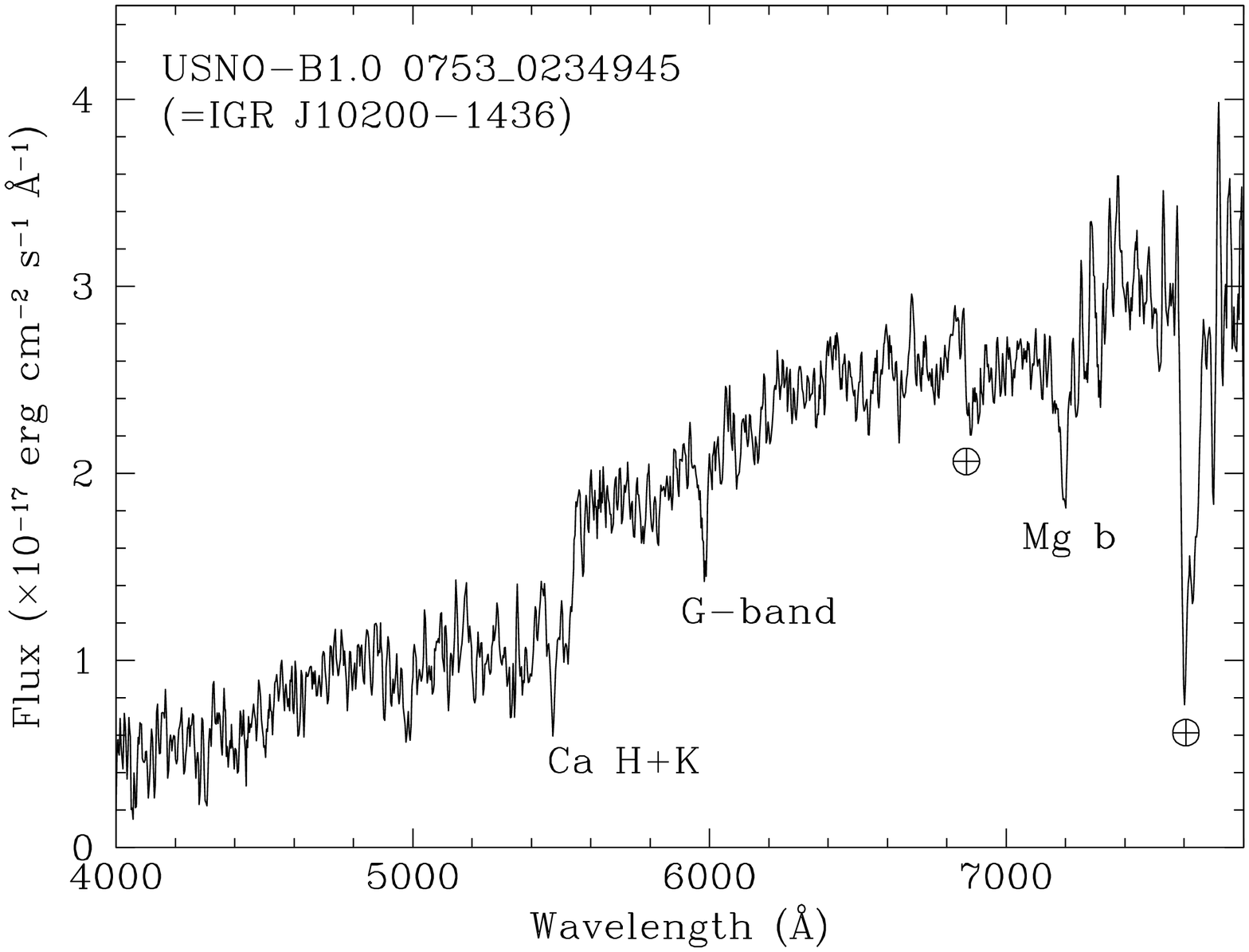,width=9cm,angle=270}}

\caption{Spectra (not corrected for the intervening Galactic absorption) 
of the optical counterparts of the two XBONGs belonging to the sample of 
{\it INTEGRAL} sources presented in this paper. For each spectrum, the 
main spectral features are labelled. The symbol $\oplus$ indicates 
atmospheric telluric absorption bands. The spectrum of IGR J10200$-$1436
has been smoothed using a Gaussian filter with $\sigma$ = 3 \AA.}
\end{figure*}

It is found that 23 objects in our sample have optical spectra that allow 
their classification as AGNs (see Figs. 4-8). The majority of them exhibits
redshifted broad and/or narrow emission lines typical of nuclear galactic 
activity: 13 sources can be classified as Type 1 (broad-line) and 
eight as Type 2 (narrow-line) AGNs. 

In addition, two X--ray objects (IGR J05048$-$7340 and IGR J10200$-$1436) 
are associated with galaxies showing absorption features only (see Fig. 
9). Using the approach of Laurent-Muehleisen et al. (1998), we find that 
these sources do not show any suggestion of AGN activity: we thus classify 
them as X--ray bright, optically normal galaxies (XBONGs; see Comastri et 
al. 2002). We also note that we identify IGR J10200$-$1436 as the farthest 
XBONG known up to now within the hard X--ray surveys made with {\it 
INTEGRAL}.

Moreover, we remark that 6 AGNs (5 broad-line and one narrow-line) 
identified here are found with a high redshift ($z >$ 0.5; see Table 3). 
In particular, all broad-line cases lie at $z >$ 1; we also identify a 
narrow-line hard X--ray emitting AGN at high redshift (IGR J19295$-$0919; 
see below).

The division of low-redshift Type 1 AGNs in terms of subclasses is 
reported in Table 4. In this same table we do not report the fluxes of the 
broad H$_\alpha$ emission because it is heavily blended with other 
emission lines (mainly [N {\sc ii}]) in all cases.

The main observed and inferred parametres for each of these broad classes 
of AGNs are displayed in Tables 3-5. In these tables, X--ray luminosities 
were computed from the fluxes reported in McDowell (1994), Voges et al. 
(1999, 2000), {\it ROSAT} Team (2000), Saxton et al. (2008), Rodriguez et 
al. (2009), Bird et al. (2010), Cusumano et al. (2010), Landi et al. 
(2010, 2012b), Maiorano et al. (2010), Malizia et al. (2011, 2012a), 
Bottacini et al. (2012), Krivonos et al. (2012), Molina et al. (2012a,b), 
Ricci et al. (2012), Tomsick et al. (2012), Baumgartner et al. (2013), and 
Grebenev et al. (2013).

In this paper, we also report the redshift value for most AGNs of our 
sample (18 out of 23) for the first time. Among them, we confirm and 
refine the value for the photometric redshift ($z$ = 1.1) of the optical 
counterpart of IGR J23558$-$1047 proposed by Richards et al. (2009). The 
redshifts of the remaining five cases are compatible with those reported 
in the literature (Paturel et al. 2003; Mescheryakov et al. 2009; Bikmaev 
et al. 2010). We moreover give a more accurate classification for IGR 
J17476$-$2253 in terms of Seyfert 1.5 galaxy (the source was previously 
identified by Mescheryakov et al. 2009 as a Seyfert 1 AGN).

We take this opportunity to correct and/or improve the following points. 
First, we give (see Table 4) the correct identification for the actual 
optical counterpart of IGR J06293$-$1359 (in Fig. 7) as a Seyfert 1.5 
galaxy at $z$ = 0.376; the tentative identification given in Paper VIII 
should thus be discarded for the reasons given in Section 2 above. Next, 
we correct the line identifications (given in Paper IX) in the spectrum of 
the optical counterpart of IGR J16388+3557 as shown in Fig. 6. This leads 
to a corrected redshift $z$ = 2.020 (see Table 3).

We moreover revise the AGN class of galaxy LEDA 7910, proposed as the 
optical counterpart of IGR J02045$-$1156 (Ricci et al. 2012). The presence 
of pure narrow emission lines in its spectrum (Fig. 8, upper left panel) 
is at variance with the (intermediate) Seyfert 1 spectrum and 
classification reported by Augarde et al. (1994). Rather, we classify it 
as a LINER (Heckman 1980) on the basis of its emission line ratios in the 
spectrum presented here. This indicates that this source changed its 
spectral appearance between the late 1980s and the 6dFGS observation of 
December 2002.

Let us now focus on the optical and X--ray properties of other interesting 
AGN sources within our sample.

First, we would like to draw the reader's attention on the spectrum of the 
optical counterpart of IGR J19295$-$0919 (Fig. 5, bottom right panel). 
Although the high-$z$ nature of this source does not allow us to apply the 
emission line ratio diagnostics for the AGN classification (all relevant 
lines lie in the infrared bands in the observer's frame), we can 
nevertheless classify the object as a Type 2 (i.e., narrow-line) QSO due 
to the large X--ray luminosity, which is at least four orders of magnitude 
larger than that of the brightest starbursts (David et al. 1992). Only 
five objects of this kind were detected up to now with {\it INTEGRAL} 
(Malizia et al. 2012b); IGR J19295$-$0919 is the sixth and farthest one, 
being located at redshift $z$=0.741.

Figure 7 reports the spectra of the low-redshift ($z <$ 0.5) broad-line 
AGNs in our sample. It is noteworthy that two of them, IGR 
J17476$-$2253 and 2E 1923.7+5037, show double-peaked Balmer lines in 
emission (this is particularly evident in the case of the H$_\alpha$ line 
for both objects). This characteristic is not frequent in the optical 
spectra of broad-line AGNs and may indicate asymmetries in their 
broad-line region (BLR), such as an elliptical disk, warps, spiral shocks, 
or hot spots (Strateva et al. 2003), or a lower accretion rate with 
respect to AGNs with single-peaked broad emissions (Eracleous \& Halpern 
1994; Ho et al. 2000).

The value for the $R$-band magnitude of the optical counterpart of IGR 
J17476$-$2253 extracted from the field image of this source is reported 
here. Considering all the caveats described in Sect. 3, we obtained a 
magnitude $R$ = 18.4$\pm$0.3 for the optical counterpart of IGR 
J17476$-$2253, where the large error mainly reflects the systematic 
uncertainties of the photometry. This value is broadly consistent with 
that reported by Mescheryakov et al. (2009).

Regarding IGR J16058$-$7253, we confirm the suggestion of Landi et al. 
(2012b) that this hard X--ray emission is the sum of the contributions of 
two narrow-line AGNs within the IBIS error circle of the source but at 
different redshifts (see Table 5). By applying the statistical approach of 
Tomsick et al. (2012) and using the X--ray fluxes of Landi et al. (2012b), 
we see that the probability for each of the two X--ray emitting galaxies, 
LEDA 259433 and LEDA 259580, that can be contained by chance in the hard 
X--ray error circle is less than 3$\times$10$^{-3}$ with a slight 
dependence on the chosen error size (Cusumano et al. 2010; Krivonos et al. 
2012; Baumgartner et al. 2013). This further strengthens the above 
mentioned indication of Landi et al. (2012b) and makes this hard X--ray 
object similar to the case of IGR J20286+2544, which is most likely the 
combined high-energy emission of two X--ray variable AGNs (Paper IV; 
Winter et al. 2008). In Table 5, we did not attempt to evaluate either the 
total hard X--ray luminosity of IGR J16058$-$7253 or the contributions of 
the two AGNs to this quantity due to the present lack of an accurate model 
of their X--ray spectra.

We also point out that Molina et al. (2012b) reports the presence of two 
soft X--ray sources within the IBIS error circle of IGR J14488$-$4008. Our 
identification as a Seyfert 1.2 galaxy (LEDA 589690) corresponds to the 
brighter of the two X--ray objects (the one labelled as `N1' by Molina et 
al. 2012b). Indeed, we also acquired an optical spectrum of source `N2' of 
Molina et al. (2012b) with the 1.5m CTIO telescope on 2 February 2012 and 
found that it is a G-type Galactic star with no peculiarities. Likewise, 
source \#2 of Malizia et al. (2011) in the field of IGR J23558$-$1047 
(though formally outside the 90\% IBIS error circle) was observed by us 
from Loiano on 22 August 2011 and again shows the optical spectrum of a 
normal star of G type. Source \#1 of Landi et al. (2010) in the error 
circle of IGR J21565+5948 was spectroscopically confirmed as a normal 
A-type star by Bikmaev et al. (2010). Therefore, we will not discuss these 
objects further. In a similar vein, we focused our attention on LEDA 
3077397 following the statistical considerations of Tomsick et al. (2012) 
with regard to the putative optical counterpart of IGR J15415$-$5029.

We now consider the Type 2 galaxies for which an estimate of the 
absorption local to the AGN is possible, and we apply the diagnostic $T$ 
of Bassani et al. (1999) to their cases, i.e. the ratio of the measured 
2--10 keV X--ray flux to the unabsorbed flux of the [O {\sc 
iii}]$\lambda$5007 forbidden emission line. This approach allows us to 
infer their Compton nature. We find that the value of $T$ is 3.4, 14.1, 
and 0.57 for LEDA 7910, LEDA 2287192, and LEDA 3077397, respectively, 
after correction of the [O {\sc iii}]$\lambda$5007 emission line flux for 
the absorption local to the corresponding AGN (see Table 5). This points 
to a Compton thick source classification for the latter AGN, whereas the 
two other cases lie in the Compton thin regime. We stress that the above 
values should actually be considered as upper limits for $T$ because the 
soft X--ray fluxes that we used refer to bands which are wider than the 
ones for which this method should be applied (see Table 5).

Actually, this latter result is at variance with the findings of Tomsick 
et al. (2012), who did not detect any absorption local to the AGN for LEDA 
3077397. A possible way out to this discrepancy lies in the evidence 
(e.g., Trippe et al. 2011) that Compton thick AGNs may fictitiously 
display no appreciable X--ray absorption with a flat X--ray spectrum. This 
is due to the large actual $N_{\rm H}$, which suppresses the direct 
continuum below 10 keV, so that only the reflected component is observed 
in this energy range with a spectrum characterized by a photon index 
$\Gamma <$ 0.5 (as indeed observed from LEDA 3077397 by Tomsick et al. 
2012).

Finally, we applied the following prescriptions to infer the mass of the 
central black hole in the broad-line AGNs of our sample. The choice of the 
method varies according to the spectral emission line used for the 
estimate.
For the cases in which the H$_\beta$ emission is within the covered 
optical spectral range, we used the approach of Wu et al. (2004) and Kaspi 
et al. (2000). Otherwise, we apply the formulae from McLure \& Jarvis 
(2002) or Assef et al. (2011), which use the information afforded by the 
Mg {\sc ii} or C {\sc iv} broad emissions, respectively. In this way we 
were able to compute an estimate of the mass of the central black hole for 
all 13 broad-line AGNs presented in this paper for the first time plus the 
correct value for IGR J16388+3557 (see Table 6). In all cases, we assumed 
a null local absorption.

As already remarked in Paper IX, the main sources of error in these mass 
estimates generally come from the determination of the flux of the 
employed emission lines, which is around 15\% in the present sample (see 
Tables 3 and 4), and from the scatter in the scaling relation between the 
size of the BLR and the diagnostic line luminosity (Vestergaard 2004). As 
a whole, we expect the typical error to be about 50\% of the black hole 
mass value.

In Table 6, we also list the apparent Eddington ratios for these AGNs.
These were computed using the observed X--ray fluxes and/or upper limits 
in the 20--100 keV band. When needed, we rescaled the observed 
luminosities to the above spectral range assuming a photon index $\Gamma$ 
= 1.8.

Even considering all the above caveats and making the same considerations 
as in Paper IX, we see that the values of the Eddington ratios in Table 6 
suggest a very energetic nature for several Type 1 AGNs of our sample, 
especially those at high redshift, which is in line with cases already 
found in hard X--ray surveys (see e.g. Lanzuisi et al. 2012; Bassani et 
al. 2012). This supports the scenario of Ghisellini et al. (2011),
according to which many powerful blazars emit most of their energy in the 
MeV band or below, and implies that these objects are more efficiently 
searched in hard X--rays rather than in the GeV range.

\begin{table}
\caption{BLR gas velocities (in km s$^{-1}$), central black 
hole masses (in units of 10$^8$ $M_\odot$), and apparent Eddington ratios 
for the 14 broad line AGNs discussed in this paper.}
\begin{center}
\begin{tabular}{lrcc}
\noalign{\smallskip}
\hline
\hline
\noalign{\smallskip}
\multicolumn{1}{c}{Object} & \multicolumn{1}{c}{$v_{\rm BLR}$} & $M_{\rm BH}$ & $L_X/L_{\rm Edd}$\\
\noalign{\smallskip}
\hline
\noalign{\smallskip}

IGR J02447+7046   &  4000 &  1.4  &       0.2   \\
IGR J02574$-$0303 &  3900 &  0.42 & $\sim$0.5   \\
IGR J06293$-$1359 &  5700 &  0.87 &    $<$1.6   \\
IGR J14488$-$4008 &  5300 &  3.8  &    $<$0.007 \\
IGR J17476$-$2253 & 11100 &  3.6  &       0.002 \\
IGR J17488$-$2338 &  8200 & 13    &    $<$0.006 \\
2E 1923.7+5037    &  9700 &  5.8  & $\sim$0.002 \\
IGR J21565+5948   &  5700 &  5.6  &       0.02  \\

\noalign{\smallskip}
\hline
\noalign{\smallskip}

IGR J02115$-$4407 &  6100 &  8.9 & $<$3.3 \\
IGR J03564+6242   & 11300 & 63   &    0.2 \\
IGR J09034+5329   & 11000 & 20   & $<$1.8 \\
IGR J21319+3619   & 10700 & 19   & $<$0.6 \\
IGR J23558$-$1047 &  8200 & 11   &    3.5 \\

\noalign{\smallskip}
\hline
\noalign{\smallskip}

IGR J16388+3557   & 23000 & 160 & 0.4 \\

\noalign{\smallskip}
\hline
\noalign{\smallskip}
\multicolumn{4}{l}{Note: the final uncertainties on the black hole mass} \\
\multicolumn{4}{l}{estimates are about 50\% of their values. The} \\
\multicolumn{4}{l}{velocities were determined using H$_\beta$, Mg {\sc ii},} \\
\multicolumn{4}{l}{or C {\sc iv} emissions (upper, central and lower part} \\
\multicolumn{4}{l}{of the table, respectively), whereas the apparent} \\
\multicolumn{4}{l}{Eddington ratios were computed using the (observed} \\
\multicolumn{4}{l}{or rescaled, see text) 20--100 keV luminosities.} \\
\noalign{\smallskip}
\hline
\hline
\noalign{\smallskip}
\end{tabular}
\end{center}
\end{table}

\subsection{X--ray binaries}

\begin{figure*}
\mbox{\psfig{file=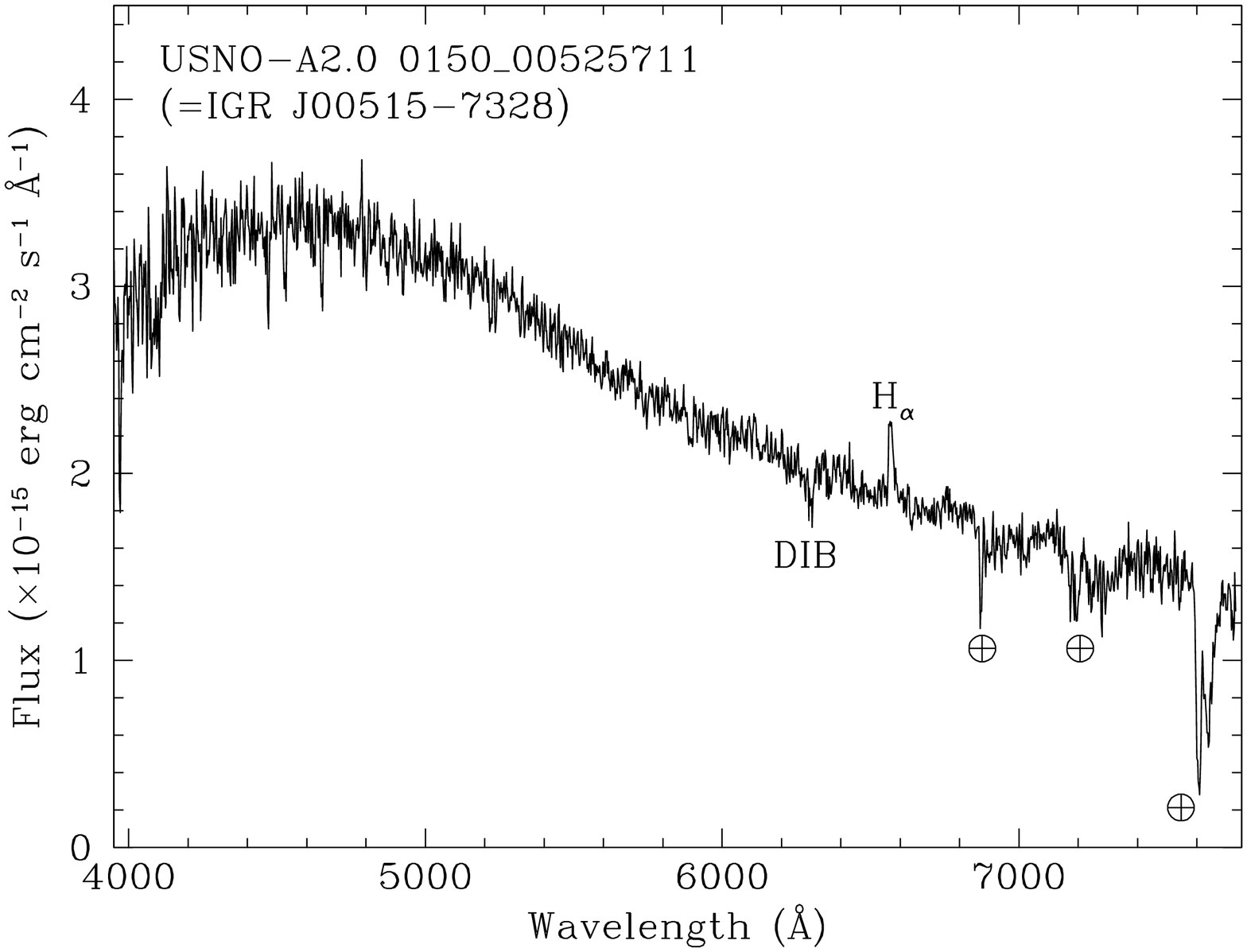,width=9cm,angle=270}}
\mbox{\psfig{file=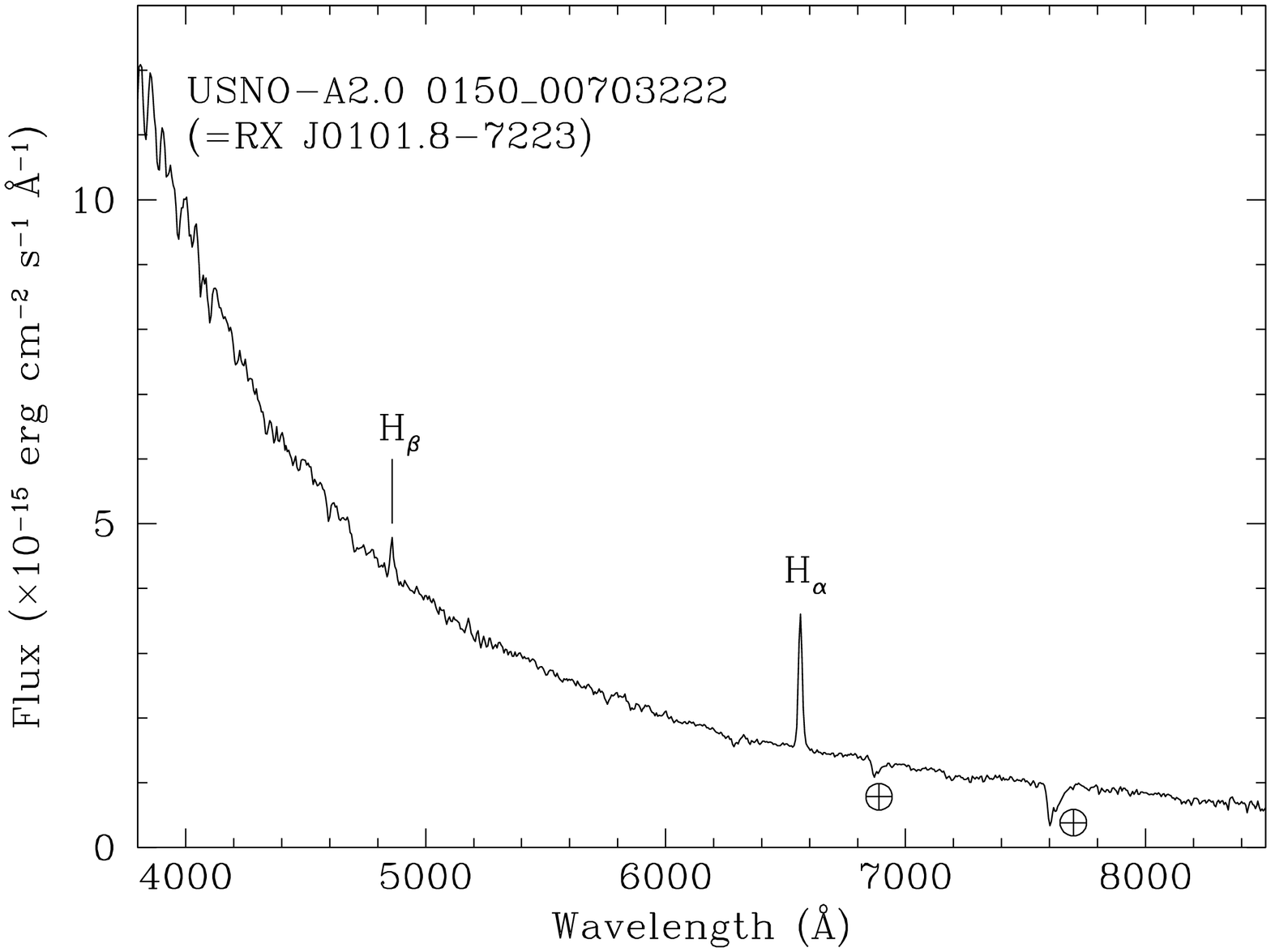,width=9cm,angle=270}}

\vspace{-.9cm}
\mbox{\psfig{file=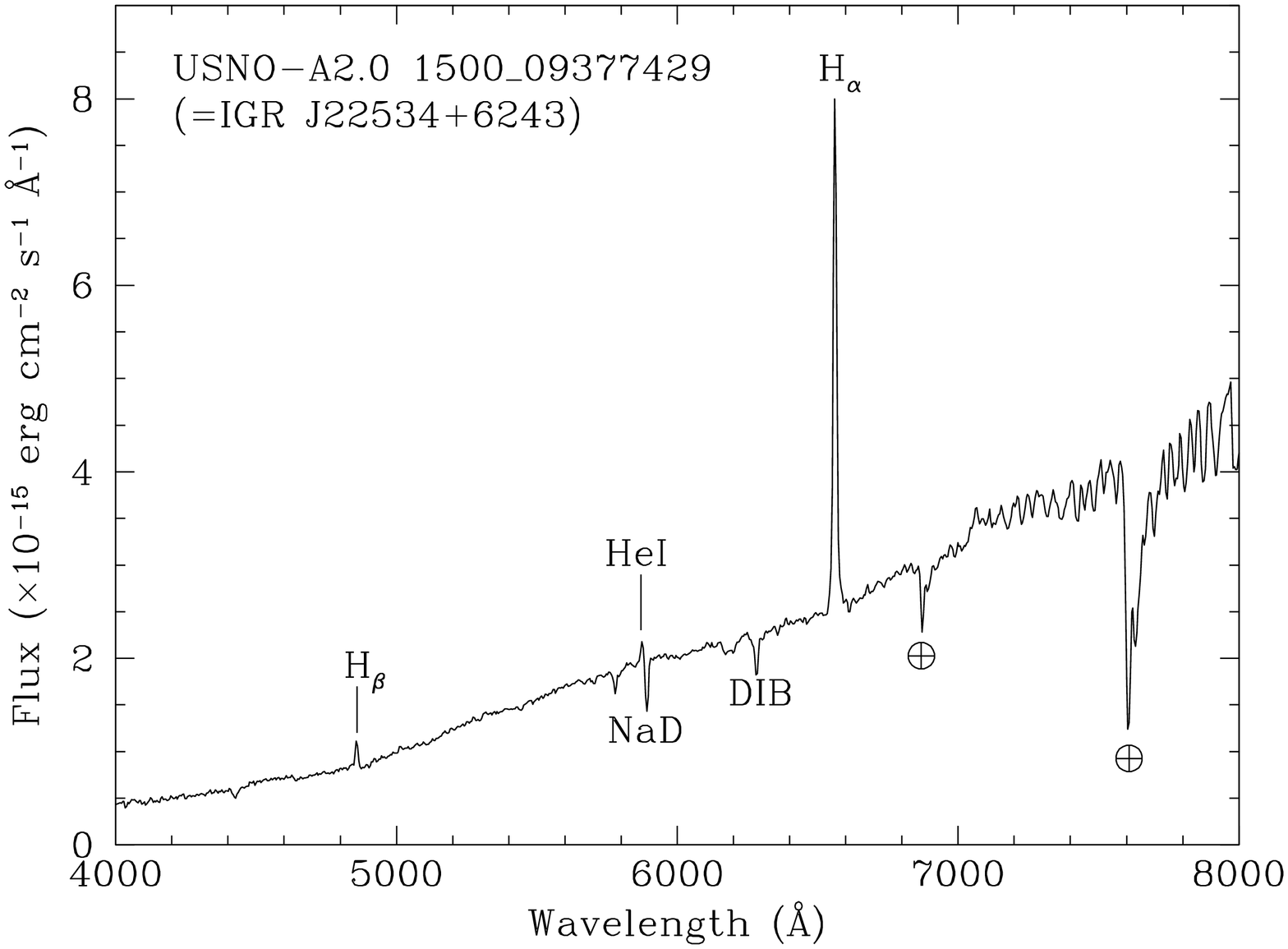,width=9cm,angle=270}}
\mbox{\psfig{file=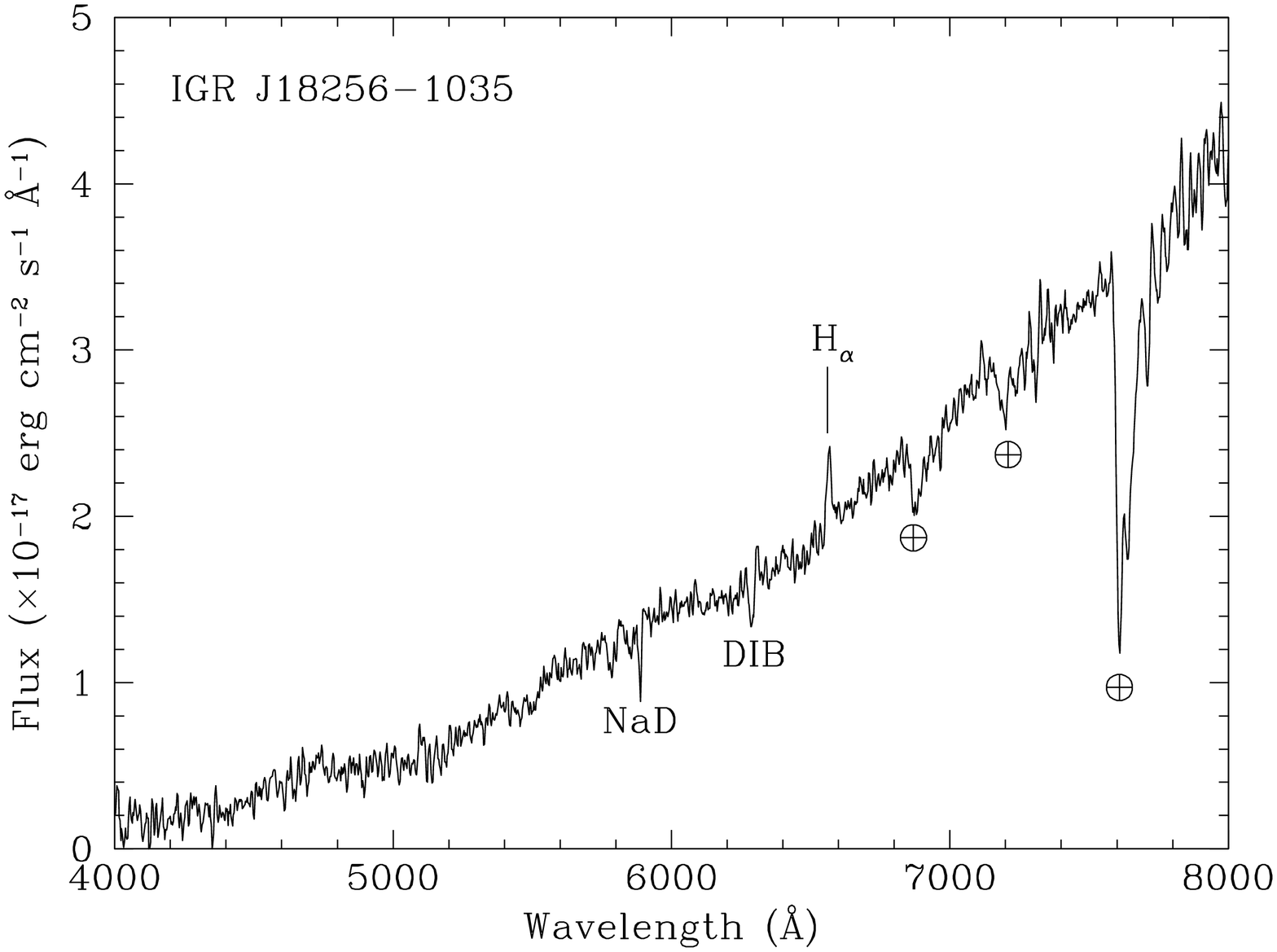,width=9cm,angle=270}}

\caption{Spectra (not corrected for the intervening Galactic absorption) 
of the optical counterparts of the four X--ray binaries belonging to the 
sample of {\it INTEGRAL} sources presented in this paper. For each 
spectrum, the main spectral features are labelled. The symbol $\oplus$ 
indicates atmospheric telluric absorption bands.}
\end{figure*}

Among the other objects identified in our sample of {\it INTEGRAL} sources 
in this paper we found four cases, which could be classified as X--ray 
binaries. Three of them show the characteristics of HMXBs --- two are
located in the Small Magellanic Cloud (SMC) and one in the Galaxy --- 
while one is classified as a generic X--ray binary but is probably 
a Low-Mass one (LMXB; see below).

The objects IGR J00515$-$7328, RX J0101.8$-$7223, and IGR J22534+6243 are 
identified as HMXBs due to the presence of an H$_\alpha$ emission at a 
redshift consistent with 0 superimposed on an intrinsically blue spectral 
continuum; in the latter two cases, He {\sc i} and/or H$_\beta$ emission 
lines are also detected (see Fig. 10 and Table 7). We however note that 
the spectral shape of IGR J22534+6243 appears to be substantially modified 
by intervening reddening. This points to the presence of interstellar dust 
along the source line of sight. This result is indeed usual for Galactic 
HMXBs detected with {\it INTEGRAL} (e.g., Papers III-IX) and indicates 
that the object lies far from the Earth.

The HMXB identification for RX J0101.8$-$7223 and IGR J22534+6243 is also
independently supported by the detection of X--ray pulsations (Townsend et 
al. 2013; Esposito et al. 2013), strongly pointing to the presence of an 
accreting neutron star hosted in these systems.

Concerning IGR J18256$-$1035, an H$_\alpha$ emission found again at 
redshift 0 is detected on a red spectral continuum. The faintness of the 
optical counterpart along with the optical spectral shape and the NIR 
non-detection in the 2MASS catalogue (Skrutskie et al. 2006) does not 
favor the presence of an early type companion and, consequently, a 
classification as HMXB. Given the relative faintness of the H$_\alpha$ 
emission line with respect to that detected in similarly reddened CVs (see 
Paper IX and Sect. 4.3), we suggest to classify this source as a LMXB.

For all these objects, Table 7 lists the relevant optical spectral 
information along with the main parametres determined from the available 
X--ray and optical data. When not indicated otherwise, X--ray luminosities 
in Table 7 were calculated using the fluxes extracted from Wang \& Wu 
(1992), Voges et al. (2000), Landi et al. (2007), Tomsick et al. (2008), 
Bird et al. (2010), Kennea (2011), Sturm et al. (2011), Krivonos et al. 
(2012), Landi et al. (2013), and Esposito et al. (2013). Constraints on 
distance, reddening, spectral type, and X--ray luminosity shown in Table 7 
for the identified HMXBs were derived by considering the absolute 
magnitudes of early-type stars and by applying the method described in 
Papers III and IV for the classification of this type of X--ray sources.

Regarding the determination of the distance for the X--ray binaries 
discussed in this section, we assumed that IGR J00515$-$7328 and RX 
J0101.8$-$7223 are located in the SMC and thus at a distance of 60 kpc 
(Harries et al. 2003), whereas a distance of 8 kpc was assumed for IGR 
J18256$-$1035; no inferences on this value were otherwise possible due to 
the lack of sufficient optical/NIR information on its counterpart. For IGR 
J22534+6243, we exclude the possibility that its companion is of 
luminosity class I because it would lie outside the Galactic disk ($>$8 
kpc from Earth) according to the map of Leicht \& Vasisht (1998). 
Likewise, the presence of substantial reddening towards this source would 
place the source in or beyond the Perseus Arm of the Galaxy, suggesting a 
distance $>$3 kpc. This figure is also too large to support a luminosity 
class V for the optical companion. This analysis indicates that all HMXBs 
identified here most likely belong to the subclass of Be/X binaries.

Considering the equivalent widths (EWs) of the H$_\alpha$ emission of the 
three HMXBs in our sample and following the empirical relation of Antoniou 
et al. (2009) that connects this observable with the orbital period of 
Be/X binaries, we infer that RX J0101.8$-$7223 and IGR J22534+6243 may 
have $P_{\rm orb} \approx$ 100 d, whereas this value is $\la$20 d for IGR 
J00515$-$7328. For RX J0101.8$-$7223, this suggestion is confirmed by the 
periodicity analysis of the optical light curve by Townsend et al. (2013).

It should be stressed that our optical results are broadly consistent with 
those of Townsend et al. (2013) and Esposito et al. (2013), which 
presented optical spectra of the counterparts of RX J0101.8$-$7223 and IGR 
J22534+6243, respectively, with resolution higher than ours. For the other 
cases, detailed photometric optical/NIR information and higher-resolution 
spectroscopy is mandatory to refine the spectral classification and 
properties of these objects.

\begin{table*}
\caption[]{Main results for the X--ray binaries (see Fig. 10) identified 
in the present sample of {\it INTEGRAL} sources. The upper part of the 
table deals with the HMXBs identified here, whereas the lower part reports 
on the single LMXB of our sample.} 
\hspace{-1.2cm}
\scriptsize
\vspace{-.5cm}
\begin{center}
\begin{tabular}{lccccccccr}
\noalign{\smallskip}
\hline
\hline
\noalign{\smallskip}
\multicolumn{1}{c}{Object} & \multicolumn{2}{c}{H$_\alpha$} & 
\multicolumn{2}{c}{H$_\beta$} &
 $R$ & $A_V$ & $d$ & Spectral & \multicolumn{1}{c}{$L_{\rm X}$} \\
\cline{2-5}
\noalign{\smallskip} 
 & EW & Flux & EW & Flux & mag & (mag) & (kpc) & type & \\

\noalign{\smallskip}
\hline
\noalign{\smallskip}

IGR J00515$-$7328 & 4.5$\pm$0.6 & 8.4$\pm$1.0 & $<$1.1 & $<$3.4 & 
15.18$^{\rm a}$ & $\sim$1.1$^{\rm b}$ & 60$^{\rm c}$ & O8\,V or B0\,III & 0.5$^{\rm d}$ (0.1--2; {\it R}) \\
 & & & & & & & & & 1.3 (0.16--3.5; {\it E}) \\
 & & & & & & & & & 8--11$^{\rm d,e}$ (0.2--10; {\it X}) \\
 & & & & & & & & & $<$0.027 (0.2--10; {\it N}) \\
 & & & & & & & & & 15--30 (0.3--10; {\it X}) \\

 & & & & & & & & & \\

RX J0101.8$-$7223 & 28.4$\pm$1.4 & 46$\pm$2 & 2.4$\pm$0.5 & 10$\pm$2 &
 $\sim$14.8$^{\rm f}$ & $\sim$1.5 & 60$^{\rm c}$ & O9--B0\,III & 2.3--5.8$^{\rm g}$ (0.2--10; {\it A, R, N}) \\
 & & & & & & & & & 3.9$^{\rm f}$ (0.2--10; {\it N}) \\
 & & & & & & & & & 36.5$^{\rm h}$ (0.2--10; {\it N}) \\

 & & & & & & & & & \\

IGR J22534+6243 & 31.8$\pm$1.6 & 81.0$\pm$0.4 & 3.4$\pm$0.5 & 3.3$\pm$0.5 &
 13.7 & $\sim$6.7 & $\sim$4.4 & early B\,III & 0.003 (0.1--2.4; {\it R}) \\
 & & & & & & & & & 0.012--0.10 (0.5--10; {\it R, X, C}) \\
 & & & & & & & & & 0.14 (17--60; {\it I}) \\

\noalign{\smallskip}
\hline
\noalign{\smallskip}

IGR J18256$-$1035 & 5.5$\pm$0.5 & 0.11$\pm$0.03 & $<$10 & $<$0.03 &
 --- & --- & 8$^{\rm i}$ & --- & 0.22 (0.3--10; {\it C}) \\
 & & & & & & & & &    0.15 (2--10; {\it X}) \\
 & & & & & & & & &    0.47 (20--40; {\it I}) \\
 & & & & & & & & & $<$0.21 (40--100; {\it I}) \\
 & & & & & & & & &    0.56 (17--60; {\it I}) \\

\noalign{\smallskip} 
\hline
\noalign{\smallskip}
\multicolumn{10}{l}{Note: EWs are expressed in \AA, line fluxes are
in units of 10$^{-15}$ erg cm$^{-2}$ s$^{-1}$, X--ray luminosities
are in units of} \\
\multicolumn{10}{l}{10$^{35}$ erg s$^{-1}$, and the reference band 
(between round brackets) is expressed in keV.} \\
\multicolumn{10}{l}{In the last column, the upper case letter indicates the satellite 
and/or the instrument with which the} \\
\multicolumn{10}{l}{corresponding X--ray flux measurement was obtained (see text).} \\
\multicolumn{10}{l}{$^{\rm a}$: from Massey (2002); $^{\rm b}$: only the Galactic 
reddening along the line of sight was assumed;} \\
\multicolumn{10}{l}{$^{\rm c}$: from Harries et al. (2003); $^{\rm d}$: from Coe et al. (2010);
$^{\rm e}$: from Sturm et al. 2011; } \\
\multicolumn{10}{l}{$^{\rm f}$: from Haberl et al. (2008); $^{\rm g}$: from Haberl \& Pietsch (2004);} \\
\multicolumn{10}{l}{$^{\rm h}$: from Townsend et al. (2013); $^{\rm i}$: assumed (see text).} \\
\noalign{\smallskip}
\hline
\hline
\end{tabular} 
\end{center} 
\end{table*}

To conclude this section, we mention that none of these objects has any 
known radio source positionally associated with them. This indicates that 
all these X--ray binaries are unlikely to produce collimated (jet-like) 
outflows; that is, their microquasar nature can be ruled out.

\subsection{CVs}

\begin{figure*}[th!]
\mbox{\psfig{file=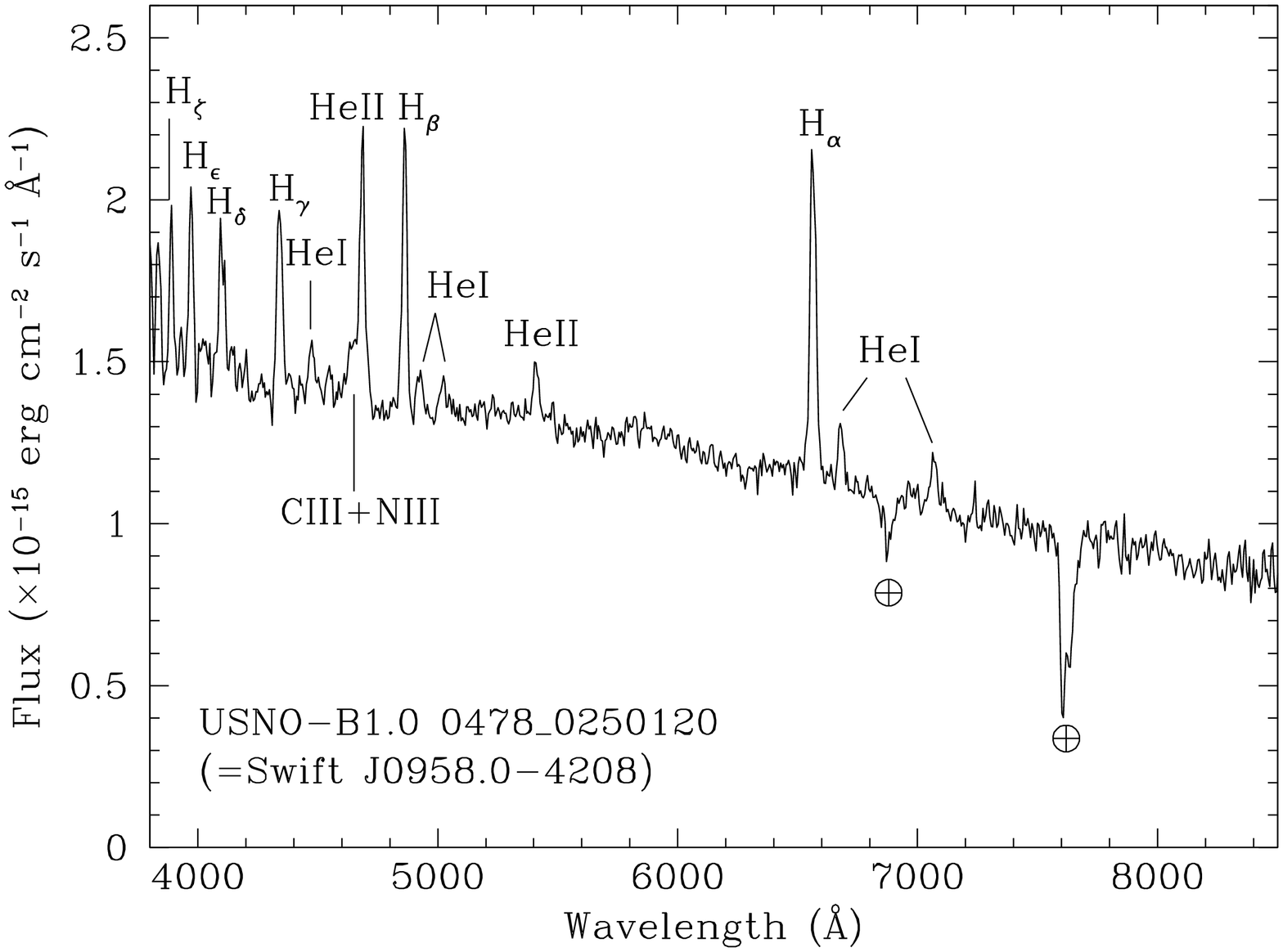,width=9cm,angle=270}}
\mbox{\psfig{file=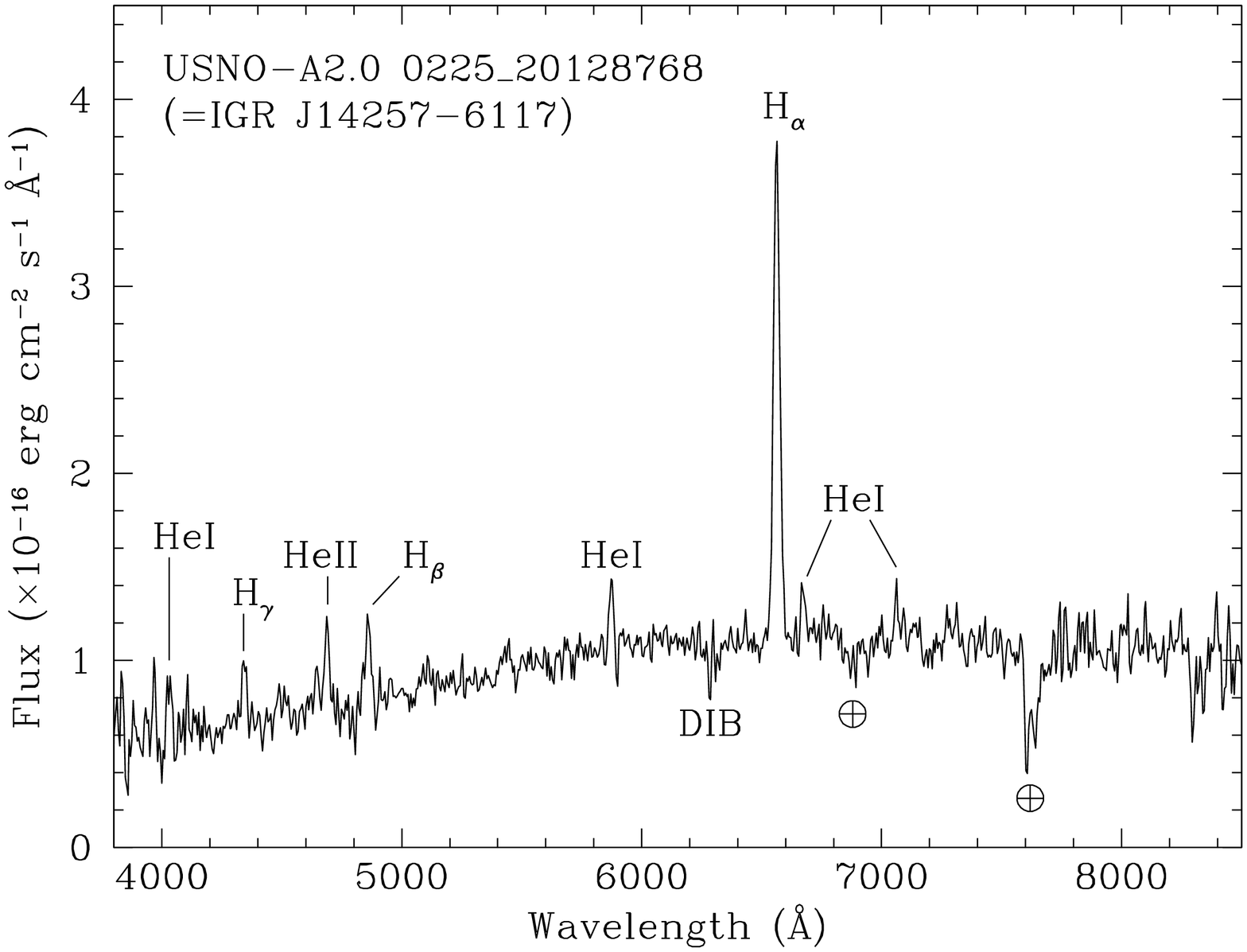,width=9cm,angle=270}}

\vspace{-.9cm}
\mbox{\psfig{file=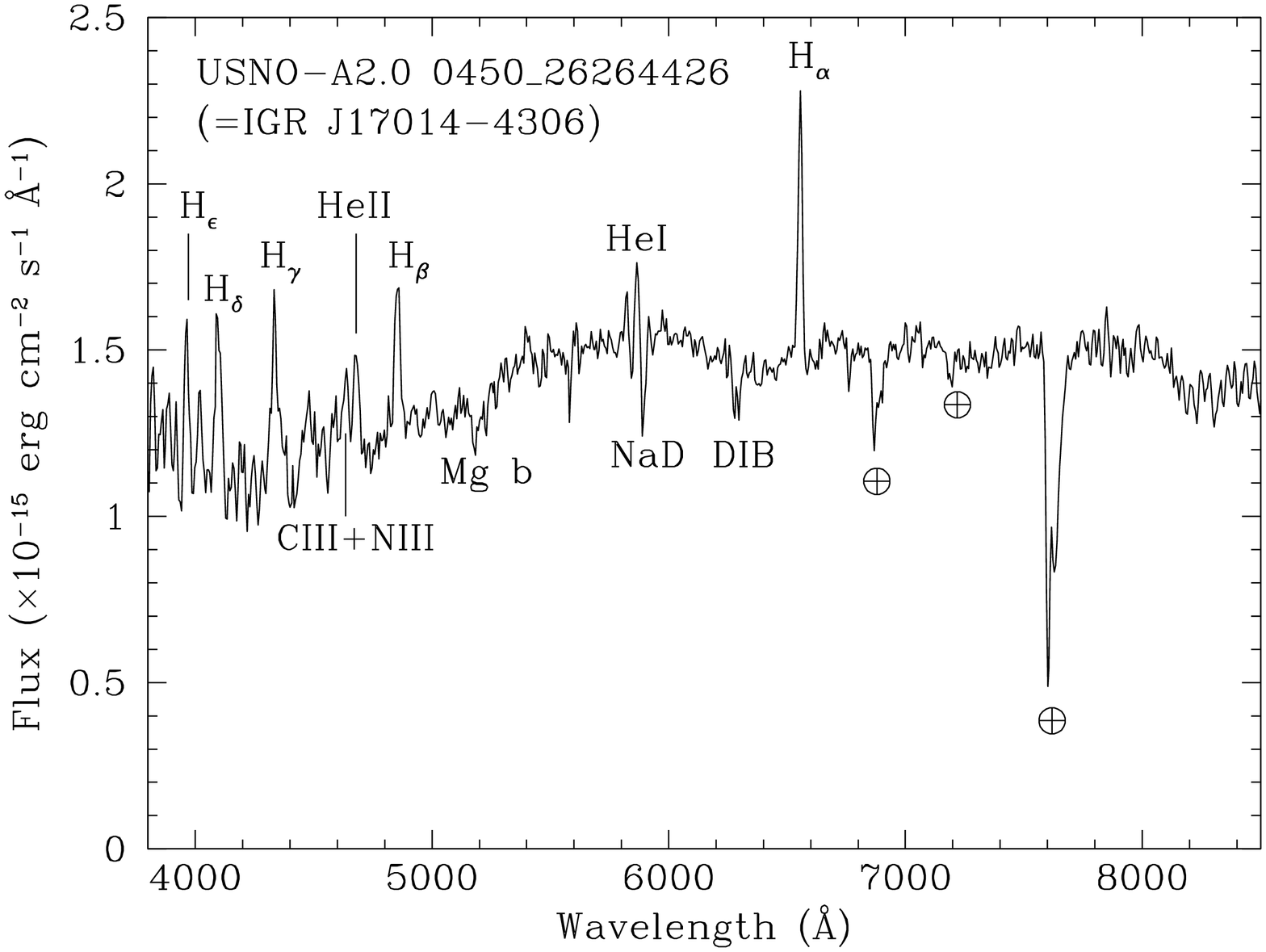,width=9cm,angle=270}}
\mbox{\psfig{file=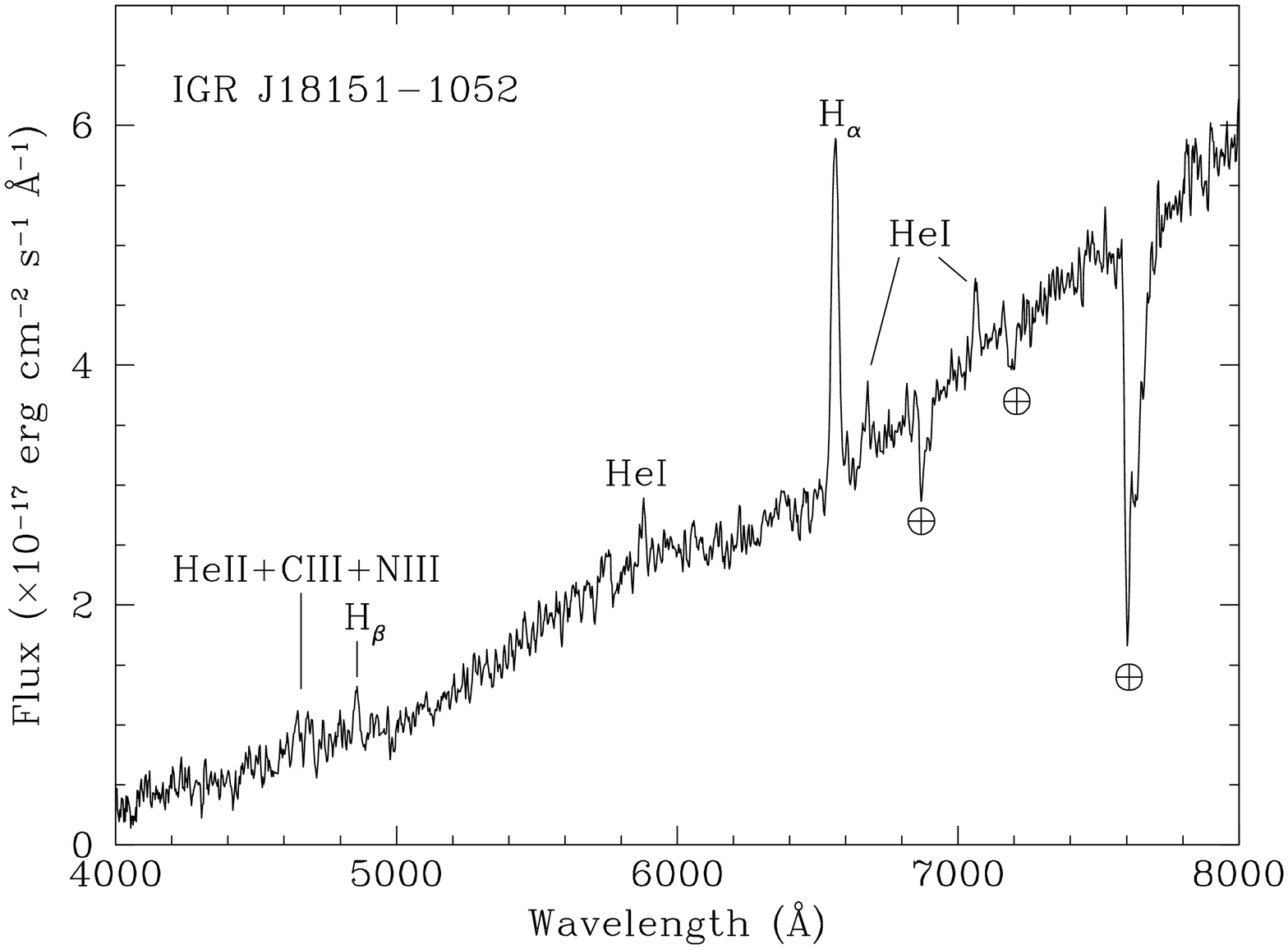,width=9cm,angle=270}}

\vspace{-.9cm}
\parbox{9.5cm}{
\psfig{file=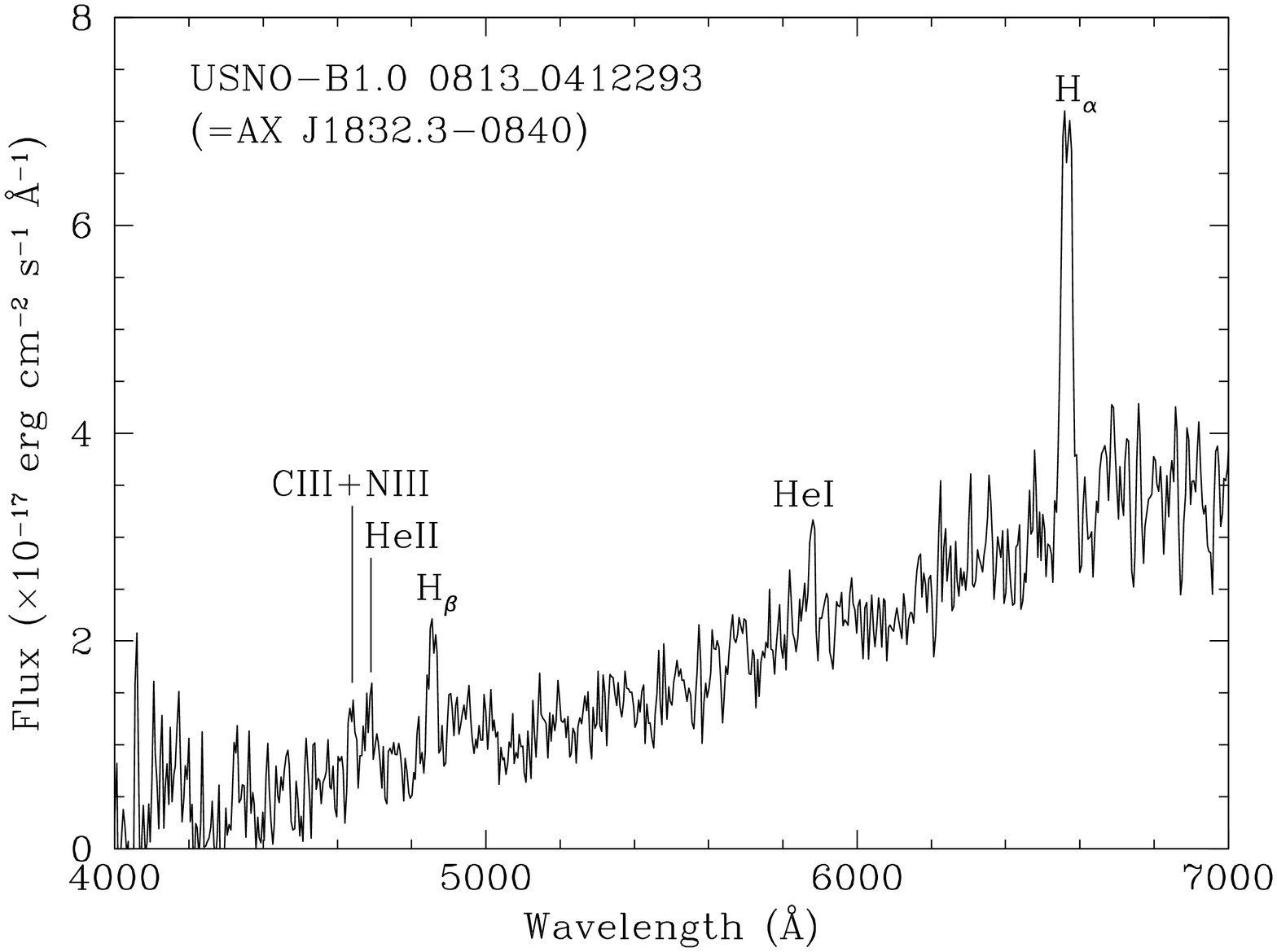,width=9cm,angle=270}
}
\hspace{0.8cm}
\parbox{8cm}{
\vspace{-.5cm}
\caption{Spectra (not corrected for the intervening Galactic absorption) 
of the optical counterparts of the five CVs belonging to the sample of 
{\it INTEGRAL} sources presented in this paper. For each spectrum, 
the main spectral features are labelled. The symbol $\oplus$ indicates 
atmospheric telluric absorption bands.}}
\end{figure*}

Spectroscopy of the five sources classified as CVs in our present sample 
(Fig. 11) indicates the typical characteristics of objects belonging to this 
class --- that is, Balmer (up to H$_\zeta$ for the optical counterpart of Swift 
J0958.0$-$4208) and helium emission lines. In all cases these spectral
features are at redshift $z$ = 0, which of course means that these objects
lie within the Galaxy. Our spectroscopic results for IGR J18151$-$1052 are 
consistent with those reported by Lutovinov et al. (2012) for this source.

The main spectroscopic results and the main astrophysical parametres, which 
can be inferred from the available observational data, are reported in Table 8. 
The X--ray luminosities in this table were obtained using the fluxes of 
Sugizaki et al. (2000), Watson et al. (2009), Cusumano et al. (2010),
Kaur et al. (2010), Scaringi et al. (2010), Krivonos et al. (2010, 2012),
Landi et al. (2012a, 2013), Lutovinov et al. (2012), and Baumgartner et 
al. (2013).

It can be noted from Table 8 that Swift J0958.0$-$4208 shows an He {\sc 
ii} $\lambda$4686 / H$_\beta$ equivalent width (EW) ratio that is larger 
than 1; moreover, this same parametre approaches unity in sources IGR 
J14257$-$6117, IGR J17014$-$4306 and AX J1832.3$-$0840. Besides, in nearly 
all CVs of our sample the EWs of these two emission lines 
are larger than 10 \AA. This strongly indicates that these sources are 
possibly harboring a magnetized white dwarf (WD) and thus may be polar 
or intermediate polar (IP) CVs (see e.g. Cropper 1990 and Warner 
1995). This further supports the IP CV classification suggested for 
AX J1832.3$-$0840 from the X--ray data analysis (Kaur et al. 2010 and 
references therein).

We however recall that this proposed classification for the other four 
sources needs an independent confirmation through the measurement 
of both the orbital and the WD spin periods, given that optical 
spectroscopy alone is sometimes insufficient to determine the magnetic 
nature of CVs (see e.g. Pretorius 2009 and de Martino et al. 2010).

It should be noted that three of these objects with the exception of 
Swift J0958.0$-$4208 and IGR J17014$-$4306 show the presence of reddening 
along the line of sight. This can be stated from the continuum shape and 
the H$_\alpha$/H$_\beta$ flux ratio. Using this parametre we estimate the 
value for the absorption in the optical $V$ band along the line of sight 
to these sources (reported in Table 8). Because of the uncertainties tied 
to the distance determination for absorbed CVs (see e.g. Paper IX), we do 
not attempt such an evaluation and we assume a distance of 1 kpc for these 
objects. For these three cases we also note that the $V$-band absorption 
inferred from the Balmer line ratio is always (and, in some cases, 
remarkably) lower than the Galactic one along their direction (that is, 
$\sim$33, $\sim$4.4, and $\sim$60 mag for IGR J14257$-$6117, IGR 
J18151$-$1052, and AX J1832.3$-$0840, respectively, according to Schlegel 
et al. 1998). This suggests that they are located in the near side of the 
Galaxy. 

We conclude this section by recalling that two soft X--ray sources (\#1 
and \#2 in Landi et al. 2013) are found within the hard X--ray error 
circle of Swift J0958.0$-$4208 with the stronger one (\#1) positionally 
associated with the CV identified in this work. If we follow the 
prescription of Tomsick et al. (2012), consider the X--ray fluxes 
reported in Landi et al. (2013), and use the conservative error radius 
of Baumgartner et al. (2013) for Swift J0958.0$-$4208, we find that the 
positional chance coincidence probabilities for sources \#1 and \#2 of 
Landi et al. (2013) of being associated with this hard X--ray object are 
1\% and 9\%, respectively. This result suggests that, indeed, the CV 
identified here has the largest probability of being physically associated 
with Swift J0958.0$-$4208; however, we cannot exclude a contribution from 
the other object (\#2) to the total high-energy flux of this source 
detected with {\it INTEGRAL}. More observations are thus needed to 
clarify this issue and the nature of source \#2 of Landi et al. (2013).

\begin{table*}
\caption[]{Synoptic table containing the main results concerning the 
five CVs identified in the present sample of {\it INTEGRAL} sources (see 
Fig. 11).}
\scriptsize
\vspace{-.3cm}
\begin{center}
\begin{tabular}{lcccccccccr}
\noalign{\smallskip}
\hline
\hline
\noalign{\smallskip}
\multicolumn{1}{c}{Object} & \multicolumn{2}{c}{H$_\alpha$} & 
\multicolumn{2}{c}{H$_\beta$} & \multicolumn{2}{c}{He {\sc ii} $\lambda$4686} & 
$R$ & $A_V$ & $d$ & \multicolumn{1}{c}{$L_{\rm X}$} \\
\cline{2-7}
\noalign{\smallskip} 
 & EW & Flux & EW & Flux & EW & Flux & mag & (mag) & (pc) & \\

\noalign{\smallskip}
\hline
\noalign{\smallskip}

Swift J0958.0$-$4208 & 30.6$\pm$1.5 & 34.6$\pm$1.7 & 16.1$\pm$0.8 & 21.6$\pm$1.1 & 19.6$\pm$1.0 & 26.5$\pm$1.3 & 
 15.5$^{\rm a}$ & $\sim$0 & $\sim$200 & 2.2 (2--10; {\it X}) \\
 & & & & & & & & & & 6.7 (17--60; {\it I}) \\
 & & & & & & & & & & 4.0 (15--150; {\it B}) \\
 & & & & & & & & & & 4.4 (14--195; {\it B}) \\

& & & & & & & & & & \\ 

IGR J14257$-$6117 & 75$\pm$5 & 8.4$\pm$0.6 & 18$\pm$3 & 1.4$\pm$0.2 & 16$\pm$2 & 1.19$\pm$0.18 & 
 17.2 & $\sim$2.3 & 1000$^{\rm b}$ & 53 (2--10; {\it X}) \\
 & & & & & & & & & & 87 (17--60; {\it I}) \\
 & & & & & & & & & & 130 (14--195; {\it B}) \\

& & & & & & & & & & \\ 

IGR J17014$-$4306 & 11.1$\pm$0.6 & 16.4$\pm$0.8 & 13.2$\pm$1.3 & 16.1$\pm$1.6 & 10.4$\pm$1.6 & 11.9$\pm$1.8 & 
 15.1 & $\sim$0 & $\sim$170 & 0.90 (0.2--12; {\it N}) \\
 & & & & & & & & & & 2.5 (17--60; {\it I}) \\
 & & & & & & & & & & 3.0 (15--150; {\it B}) \\
 & & & & & & & & & & 5.5 (14--195; {\it B}) \\

& & & & & & & & & & \\ 

IGR J18151$-$1052& 29$\pm$2 & 0.89$\pm$0.06 & 12$\pm$4 & 0.10$\pm$0.03 & $<$10 & $<$0.08 & 
 $\approx$18$^{\rm c}$ & $\sim$3.6 & 1000$^{\rm b}$ & 48 (2--10; {\it X}) \\
 & & & & & & & & & & 110 (17--60; {\it I}) \\
 & & & & & & & & & & 110 (15--150; {\it B}) \\

& & & & & & & & & & \\ 

AX J1832.3$-$0840 & 50$\pm$5 & 1.50$\pm$0.15 & 43$\pm$9 & 0.37$\pm$0.07 & 36$\pm$12 & 0.15$\pm$0.05 & 
 19.6$^{\rm c}$ & $\sim$1.1 & 1000$^{\rm b}$ & 110 (0.2--12; {\it N}) \\
 & & & & & & & & & & 130 (0.7--10; {\it A}) \\
 & & & & & & & & & &  84 (2--10; {\it N}) \\
 & & & & & & & & & &  74 (17--60; {\it I}) \\
 & & & & & & & & & &  45 (20--100; {\it I}) \\

\noalign{\smallskip} 
\hline
\noalign{\smallskip} 
\multicolumn{11}{l}{Note: EWs are expressed in \AA, line fluxes are
in units of 10$^{-15}$ erg cm$^{-2}$ s$^{-1}$, X--ray luminosities
are in units of} \\
\multicolumn{11}{l}{10$^{31}$ erg s$^{-1}$, and the reference band (between 
round brackets) is expressed in keV.} \\
\multicolumn{11}{l}{In the last column, the upper case letter indicates the satellite and/or the 
instrument with which the} \\
\multicolumn{11}{l}{corresponding X--ray flux measurement was obtained (see text).} \\
\multicolumn{11}{l}{$^{\rm a}$: from Monet et al. (2003); $^{\rm b}$: assumed (see text); $^{\rm c}$:
from Lutovinov et al. (2012).} \\
\noalign{\smallskip} 
\hline
\hline
\noalign{\smallskip} 
\end{tabular} 
\end{center}
\end{table*}

\subsection{Active stars}

\begin{figure}
\psfig{file=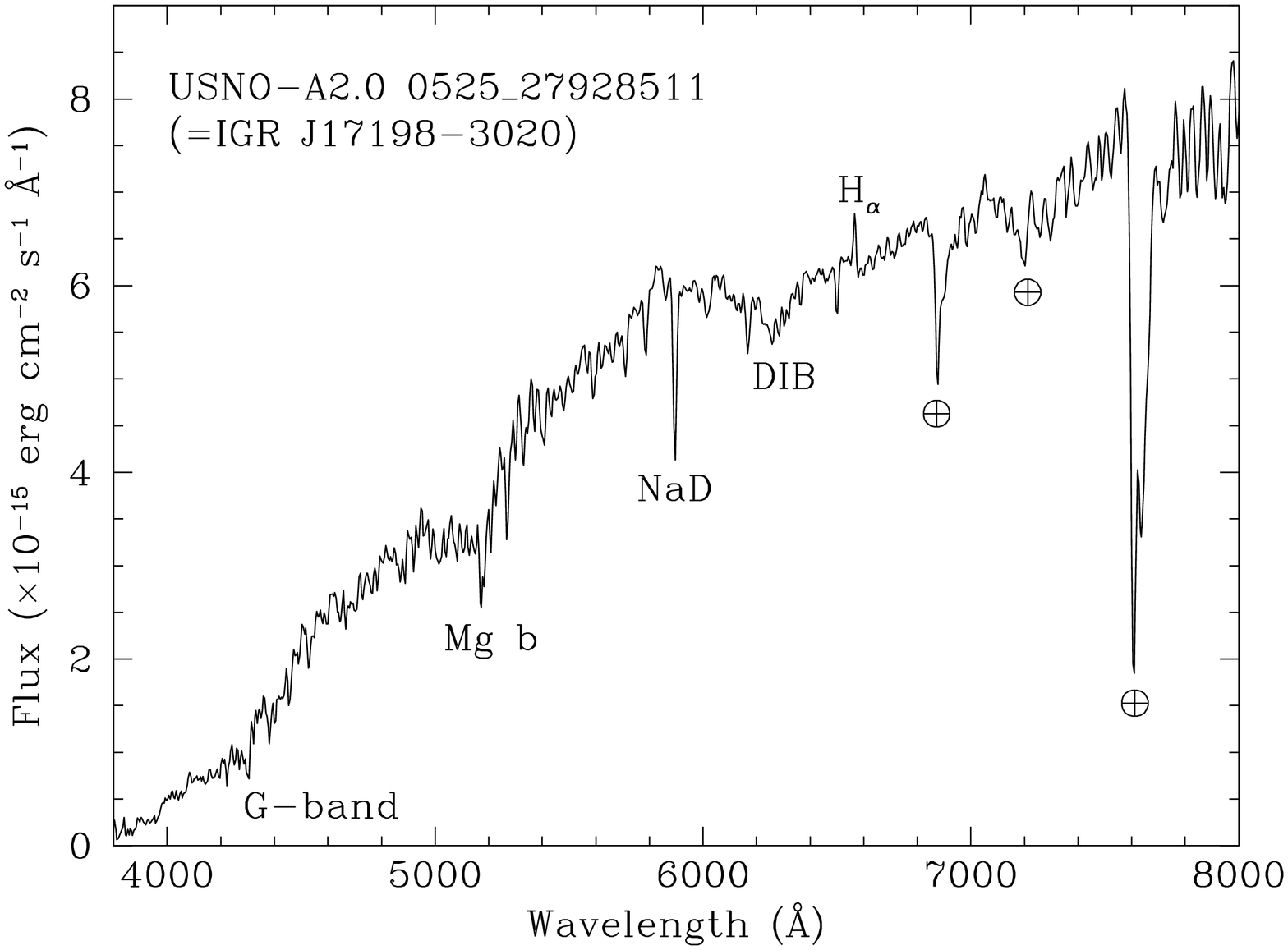,width=9cm,angle=270}
\caption{Spectrum (not corrected for the intervening Galactic absorption) 
of the optical counterpart of the chromospherically active star IGR 
J17198$-$3020 belonging to the sample of {\it INTEGRAL} sources presented 
in this paper. The main spectral features are labelled. The symbol 
$\oplus$ indicates atmospheric telluric absorption bands.}
\end{figure}

\begin{table*}
\caption[]{Main observational results for the active star IGR 
J17198$-$3020 (see Fig. 12) as identified in the present sample of {\it 
INTEGRAL} sources.}
\hspace{-1.2cm}
\scriptsize
\vspace{-.5cm}
\begin{center}
\begin{tabular}{lcccccr}
\noalign{\smallskip}
\hline
\hline
\noalign{\smallskip}
\multicolumn{1}{c}{Object} & \multicolumn{2}{c}{H$_\alpha$} & 
$R$ & $A_V$ & $d$ & \multicolumn{1}{c}{$L_{\rm X}$} \\
\cline{2-3}
\noalign{\smallskip} 
 & EW & Flux & mag & (mag) & (pc) & \\

\noalign{\smallskip}
\hline
\noalign{\smallskip}

IGR J17198$-$3020 & 1.02$\pm$0.15 & 6.3$\pm$0.9 & 12.3 & $\sim$0.9 & $\sim$350 & 0.83 (0.3--10; {\it X}) \\
 & & & & & & 10 (20--100; {\it I}) \\

\noalign{\smallskip} 
\hline
\noalign{\smallskip}
\multicolumn{7}{l}{Note: EWs are expressed in \AA, line fluxes are
in units of 10$^{-15}$ erg cm$^{-2}$ s$^{-1}$, X--ray luminosities} \\
\multicolumn{7}{l}{are in units of 10$^{31}$ erg s$^{-1}$, and the 
reference band (between round brackets) is expressed in keV.} \\
\multicolumn{7}{l}{In the last column, the upper case letter indicates the satellite 
and/or the instrument} \\
\multicolumn{7}{l}{with which the corresponding X--ray flux measurement 
was obtained (see text).} \\
\noalign{\smallskip}
\hline
\hline
\end{tabular} 
\end{center} 
\end{table*}

As seen in Fig. 12, the optical counterpart of source IGR J17198$-$3020
shows a star-like continuum typical of late-G/early-K type stars with a faint 
but nevertheless evident $H_\alpha$ emission at $z$ = 0.
This is comparable to what was found in other cases that were tentatively 
identified as RS CVn stars (Paper VI; Paper IX). We thus suggest that 
this source can be classified as a chromospherically active star as well.
This conclusion is also supported by the apparent X--ray variability pointed
out for this object by Bird et al. (2010; see also Section 2).

The main observed and inferred parametres for this object are listed in 
Table 9. Luminosities are computed using the X--ray fluxes given in
Bird et al. (2010) and Luna et al. (2012).

Assuming a similarity with the active star II Peg (with magnitude $R \sim$ 
6.9 and distance 40 pc; see Monet et al. 2003, van Leeuwen 2007 and Paper IX) 
as suggested in Rodriguez et al. (2010), we obtain a distance of 480 pc for 
IGR J17198$-$3020. However, this figure should strictly be considered as an 
upper limit, given that no correction for the interstellar absorption was 
applied. This should actually be present, due to the reddened appearance 
of the optical spectrum of this object when compared to similar sources in 
Papers VI and IX. If we correct the $B-R$ color index (2.9 mag) of the 
optical counterpart of IGR J17198$-$3020 so to bring it equal to the one 
of II Peg (1.5 mag; Monet et al. 2003), we derive an absorption 
$A_V \sim$ 0.9 mag and in turn a distance $d \sim$ 350 pc for the source. 
An independent support to this approach comes from the fact that this value 
for $A_V$ is compatible with the hydrogen column density N$_{\rm H}$ derived 
by Luna et al. (2012), if we assume the empirical formula of Predehl \& 
Schmitt (1995). We thus used this latter value for the distance to determine 
the X--ray luminosities reported in Table 9. Clearly, a deeper multiwavelength 
follow-up is vividly advised to better characterize this active star 
identification also in this case.

\subsection{Statistics}

\begin{figure}[t!]
\hspace{-0.5cm}
\psfig{file=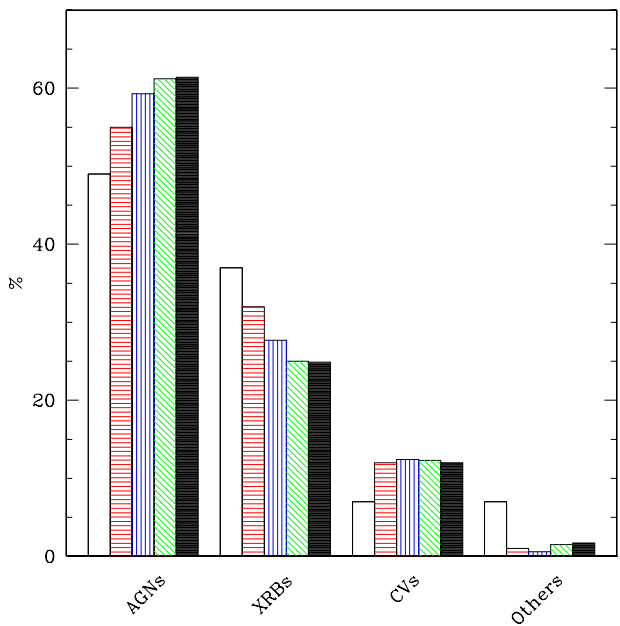,width=9.5cm}
\vspace{-0.7cm}
\caption{Histogram, subdivided into source types, showing the percentage 
of {\it INTEGRAL} objects of known nature in the fourth IBIS survey
(Bird et al. 2010; open columns) and of {\it INTEGRAL} sources from various 
surveys and identified through optical or NIR spectroscopy at the time of
acceptance of papers VII (November 2008; horizontally dashed columns), 
VIII (May 2010; vertically dashed columns), and IX (December 2011; diagonally 
dashed columns) along with those in the present one (filled columns).}
\end{figure}

As done in our previous papers, we give an update of the 
statistics concerning the identifications of new hard X--ray {\it 
INTEGRAL} sources including the results of this work and those in 
Lutovinov et al. (2012), Smith et al. (2012), Sturm et al. (2012), and 
Ratti et al. (2013).
Up to now, 240 {\it INTEGRAL} objects have been identified through optical 
or NIR spectroscopy; their separation into the main classes reported in 
Papers VI-IX is the following: 147 (61.2\%) are AGNs, 60 (25.0\%) are 
X--ray binaries, 29 (12.1\%) are CVs, and 4 cases (1.7\%) are likely 
identified as active stars. This seems to confirm a possible observational 
bias towards the former group of sources (see the discussion in Papers 
VIII and IX). We also keep detecting (likely magnetic) CVs within the 
considered putative counterparts of hard X--ray sources.

By examining the breakdown of low-redshift ($z <$ 0.5) AGN subclasses, we 
see that 68 objects (that is, 55\% of the AGN identifications) are Seyfert 
1 galaxies and 54 (44\%) are narrow-line AGNs (including 50 Seyfert 2 
galaxies and 4 LINERs), while the high-redshift ($z>$ 0.5) QSOs, 
the XBONGs, and the BL Lacs amount to 16, 6, and 3 objects (7\%, 2\%, and 
1\% of the total of identified {\it INTEGRAL} sources), respectively.

It is moreover confirmed that a substantial number of high-redshift QSOs 
can be found with the use of medium-sized telescopes (such as TNG in the 
present study). With this paper, we increase the number of newly-identified 
{\it INTEGRAL} sources at $z >$ 0.5 by more than 50\% (that is, from 10 to 16), 
allowing a deeper mapping of the distribution and properties of these high-$z$, 
high-energy sources.

When we compare the average AGN redshift of the known 
extragalactic sources in the IBIS surveys ($<$$z$$> \sim$ 0.14: Bird et 
al. 2010; Krivonos et al. 2010, 2012) to that of the sample of the 
present paper ($<$$z$$>$ = 0.404), we see that this value, albeit slightly 
lower than that of the sample in Paper IX, corroborates the results of 
that work. In particular, it indicates on average that we can explore the 
hard X--ray sky at distances that are more than three times larger than 
those of the known emitting sources belonging to the {\it INTEGRAL} surveys. 
Moreover, nearly half (11 out of 23) of the AGNs identified here lie 
farther than the above mentioned average redshift of known catalogued 
IBIS objects. We also remark that a non-negligible contribution to this 
search is afforded by 2-metre class telescopes, as two of the six high-$z$ 
QSOs of our sample have been identified with this type of facility.

We also find that the large majority of the high-redshift QSOs found in 
our identification program are of Type 1 (the only Type 2 case being IGR 
J19295$-$0919, as reported in Sect. 4.1). This is reasonably explained by 
the evidence that soft X--rays, needed for a precise localization of the 
source, are not easily detected from Type 2 QSOs due to local absorption 
in these sources, which adds to the faintness tied to their distance.

Concerning the Galactic objects, we find that 45 and 14 objects (i.e. 75\% and 
23\% of the X--ray binary identifications) are HMXBs and LMXBs, respectively,
whereas a single one is classified as an intermediate-mass X--ray binary (see
Ratti et al. 2013). In addition, most of the sources hosting an accreting WD 
(24, that is 83\% of them) are definite or likely dwarf novae (eminently of 
magnetic type), and the remaining 5 are symbiotic stars.

A comparison of the above figures with those in the largest {\it INTEGRAL} 
survey of Bird et al. (2010) and those of our previous Papers VII-IX (see 
the histograms in Fig. 13) indicates no significant changes in the 
relative weights among the various classes; rather, a saturation of the 
identification percentages seems to have been reached.
Finally, we remark that 206 of the 240 optical and NIR spectroscopic 
identifications examined in this section (i.e., more than 85\% of the 
total) were obtained within the framework of our spectroscopic follow-up 
program originally started in 2004 (Papers I-IX, the present work, and 
references therein).

\section{Conclusions}

We presented further results from our ongoing identification program of 
{\it INTEGRAL} sources through optical spectroscopy (Papers I-IX) at 
various telescopes worldwide. In the present work, we identified and 
characterized 33 objects (two of which contribute to the hard X--ray 
emission labelled as source IGR J16058$-$7253) with an unknown or poorly 
explored nature, which are listed in hard X--ray sky surveys from this 
satellite. This was accomplished with the use of six telescopes of 
different apertures from 1.5 to 3.6 metres and of archival data from one 
spectroscopic survey.

Our main results are as follows.

\begin{itemize}

\item
The majority of identifications is made of AGNs. Most of them (13) are of 
Type 1, eight cases are of Type 2, and two are XBONGs. This confirms the 
trend of our previous findings that optical spectroscopy preferentially 
allows the identification of hard X--ray emitting AGNs.

\item
Six of these AGNs lie at high redshift ($z >$ 0.5), and five lie at $z >$ 1.
This allowed the detections of high-$z$, high-energy emitting QSOs to 
increase by 50\%. We also show that the black hole mass estimates for 
these sources support the idea that hard X--ray surveys may efficiently 
identify powerful AGNs in the distant Universe.

\item
The only LINER found in the present sample, IGR J02045$-$1156, had an 
intermediate Seyfert 1 appearance in an optical spectrum acquired in the 
late 1980s. This indicates that a change in the AGN environment structure
occurred for this source.

\item
Ten objects lie in the local Universe, eight in the Galaxy, and two in the 
SMC. Of them, five are CVs, three are Be/X HMXBs (two of them belong to the 
SMC), one is a LMXB, and one is possibly a flare star of RS CVn type. This 
finding again supports that a non-negligible percentage of hard X--ray 
CVs is routinely found in optical spectroscopic searches for unidentified 
high-energy sources.

\end{itemize}

We also report to the best of our knowledge the discovery of the 
farthest hard X--ray emitting XBONG (IGR J10200$-$1436, at $z$=0.391) and 
of the farthest Type 2 QSO (IGR J19295$-$0919, at $z$=0.741) within the 
{\it INTEGRAL} surveys. This confirms the importance of this 
identification work on catalogued but unidentified high-energy sources, 
because peculiar objects can be found within the considered samples (see, 
for instance, Paper IX; Masetti et al. 2007, 2008b; Bassani et al. 2012; 
de Martino et al. 2010, 2013).
With the present data, we also correct two results given in Paper IX,
namely the actual redshift ($z$=2.02) of the Type 1 QSO IGR J16388+3557
and the correct counterpart of source IGR J06293$-$1359.
We again stress that our findings indicate the validity of this approach, 
which combines X--ray catalogue cross-correlation, follow-up observations 
with soft X--ray satellites capable of providing arcsec-sized error boxes 
(such as {\it Chandra}, {\it XMM-Newton}, {\it Swift} or {\it NuSTAR}), 
and optical spectroscopy to pinpoint the nature of still unidentified or 
poorly known {\it INTEGRAL} sources.

Present and future surveys at optical and NIR wavelengths, such as the 
ongoing Vista Variables in the V\'{i}a L\'actea (VVV: Minniti et al. 2010; 
Saito et al. 2012) public NIR survey, will permit the identification of 
variable sources in the fields of the objects detected in published and 
forthcoming {\it INTEGRAL} catalogues. This will facilitate the detection 
of putative NIR counterparts for these high-energy sources by means of 
accurate positional information and/or variability studies. This approach 
was already tested, providing encouraging results and allowing constraints 
on the nature of transient and persistent hard X--ray sources, especially 
in crowded fields such as those along the Galactic Plane and Bulge (Rojas 
et al. 2012ab, 2013; Rojas et al., in preparation).

\begin{acknowledgements}

We thank Silvia Galleti for Service Mode observations at the Loiano 
telescope, and Roberto Gualandi for night assistance; Giorgio 
Martorana and Mauro Rebeschini for Service Mode observations at the Asiago 
telescope and Luciano Traverso for coordinating them; Aldo Fiorenzano,
Vania Lorenzi and Walter Boschin for Service Mode observations at the TNG; 
Manuel Hern\'andez, Rodrigo Hern\'andez, Alberto \'Alvarez and Alberto
Miranda for Service Mode observations at the CTIO telescope and Fred 
Walter for coordinating them. NM thanks Daniel Stern for useful discussions.
We also thank the anonymous referee for the high consideration expressed 
about this work.
This research has made use of the ASI Science Data Center Multimission 
Archive; it also used the NASA Astrophysics Data System Abstract Service, 
the NASA/IPAC Extragalactic Database (NED), and the NASA/IPAC Infrared 
Science Archive, which are operated by the Jet Propulsion Laboratory, 
California Institute of Technology, under contract with the National 
Aeronautics and Space Administration.
This publication made use of data products from the Two Micron All 
Sky Survey (2MASS), which is a joint project of the University of 
Massachusetts and the Infrared Processing and Analysis Center/California 
Institute of Technology, funded by the National Aeronautics and Space 
Administration and the National Science Foundation.
This research has also made use of data extracted from the Six-degree 
Field Galaxy Survey archive; it has also made use of the SIMBAD and VIZIER
databases operated at CDS, Strasbourg, France, and of the HyperLeda 
catalogue operated at the Observatoire de Lyon, France.
NM acknowledges financial support via ASI-INAF agreement No. I/009/10/0 
and thanks the Departamento de Astronom\'{i}a y Astrof\'{i}sica of the
Pontificia Universidad Cat\'olica de Chile in Santiago for the warm
hospitality during the preparation of this paper.
PP and RL are supported by the ASI-INAF agreement No. I/033/10/0.
DM is supported by the BASAL CATA PFB-06 and FONDECYT No. 1130196 grants.
LM acknowledges financial support from the University of Padua through 
grant No. CPS0204. 
\end{acknowledgements}

\end{document}